\documentclass[11pt]{article}

\pdfmapfile{+cm.map}
\pdfmapfile{+cmextra.map}
\pdfmapfile{+symbols.map}

\usepackage[margin=1in]{geometry}
\usepackage{setspace}
\usepackage{booktabs}
\usepackage{longtable}
\usepackage{threeparttablex}
\usepackage{array}
\usepackage{adjustbox}
\usepackage{graphicx}
\usepackage{caption}
\usepackage{subcaption}
\usepackage{amsmath,amssymb}
\usepackage{pdflscape}
\usepackage{enumitem}
\usepackage{xcolor}
\usepackage{microtype}
\usepackage{verbatim}
\usepackage{natbib}
\bibpunct[, ]{(}{)}{,}{a}{}{,}
\def\bibfont{\small}

\usepackage{bibunits}

\defaultbibliographystyle{informs2014}
\defaultbibliography{refs}

\usepackage{hyperref}
\usepackage{cleveref}
\usepackage{float}
\usepackage{placeins}

\hypersetup{colorlinks=true,citecolor=blue!55!black,linkcolor=blue!55!black,urlcolor=blue!55!black}
\microtypesetup{expansion=false}
\newcommand{\placeholder}[1]{\textcolor{red!70!black}{\textbf{[PLACEHOLDER: #1]}}}
\newcommand{\HR}{\mathrm{HR}}
\newcommand{\SMD}{\mathrm{SMD}}

\newcommand{\balancefigure}[3]{%
  \begin{figure}[p]
    \centering
    \includegraphics[
      width=0.94\textwidth,
      height=0.80\textheight,
      keepaspectratio,
      trim=1mm 1mm 1mm 1mm,
      clip
    ]{#1}
    \vspace{-2mm}
    \caption{Covariate balance for #2. Points report absolute standardized
    mean differences before and after matching. The dashed reference line
    marks the prespecified threshold of 0.10. Variables for which the
    standardized mean difference is undefined are omitted.}
    \label{#3}
  \end{figure}
  \clearpage
}

\title{\textbf{Classification as Search Infrastructure: How Category Creation, Addition and Cleanup Shape Knowledge Retrieval}}

\author{Kerstin Hötte,$^{1}$\thanks{Corresponding author: kerstin.hotte@kedgebs.com 
} 
Nicolò Barbieri,$^{23}$ %
Su Jung Jee$^{4}$ %
\\[0.5em]
\small $^{1}$KEDGE Business School, France\\
\small $^{2}$Department of Economics and Management, University of Ferrara, Italy\\
\small $^{3}$SEEDS -- Sustainability, Environmental Economics and Dynamics Studies, Italy\\
\small $^{4}$Leeds University Business School, University of Leeds, United Kingdom
}

\date{\today}

\begin{document}
\begin{bibunit}
\maketitle

\begin{abstract}
Classification systems shape how searchers find relevant objects. We study how revisions to classification architecture affect retrieval by changing the routes through which objects enter consideration. We theorize that creating a cross-cutting category can reduce classificatory distance and translation costs, whereas adding routes to an established system can increase route-selection and updating costs; category cleanup can restore contrast. Empirically, we exploit the 2013 creation and 2020 revision of Y02/Y04 green-technology codes in the Cooperative Patent Classification. We match treated patents to unchanged green patents within technology--cohort cells and estimate examiner- and applicant-specific effects on first-citation timing and citation volume. Category creation is associated with a 25\% higher examiner first-citation hazard. In the mature system, additions are associated with 13--15\% lower examiner hazards, while cleanup is associated with a 14\% higher hazard and 12\% higher citation volume. Effects are generally stronger for examiners, whose search task is more tightly coupled to patent classification. Timing and volume need not move together: creation and green-domain entry shift timing without detectable changes in volume, while cleanup increases both speed and recorded reach. Classification infrastructure may therefore change when relevant objects are found even when overall recorded attention is unchanged, and greater classificatory detail need not improve search efficiency.

\end{abstract}

\newpage
\section{Introduction}

Categories help audiences compare and evaluate objects \citep{hsu2009, negro2010, DurandPaolella2013,kovacs2021}. 
When organized into classification systems, they also influence which objects become visible and enter consideration before evaluation begins.
In complex information environments, searchers cannot inspect every potentially relevant object and rely on labels, hierarchical branches, cross-references, and co-classifications to narrow where to look and which alternatives to inspect. 
Classification systems can thus operate as external search infrastructure, making some retrieval routes more salient and accessible than others \citep{bowker_star_1999}. 

Most category research begins after an object has entered an audience's consideration set and examines how categorical fit, ambiguity, or spanning affects its evaluation \citep{hsu2009,negro2010,DurandPaolella2013,kovacs2021,cudennec2023}. Much less is known about how category systems shape the earlier process through which objects are found and enter consideration, or how revisions to those systems alter retrieval. 
This gap matters because classification systems are deliberately maintained infrastructures: as domains evolve, categories are introduced, removed, and reorganized \citep{lafondkim2019,navisglynn2010,barbieri2025evolving}. Such dynamics can change retrieval routes without changing the underlying objects themselves. 

In this paper, we ask how changes in category architecture affect whether and when relevant objects are found. 
We develop a theory of \emph{classification-mediated search} in which category architecture shapes the retrieval routes that searchers recognize and use, thereby affecting search costs and entry into consideration sets. 
Category architecture can therefore affect search efficiency by changing how quickly relevant objects are retrieved. 

Our theoretical framework distinguishes substantive technological distance from \emph{classificatory distance}. Technological distance concerns the relatedness of the underlying knowledge components \citep{fleming2001,fleming_sorenson_2004,rosenkopfnerkar2001}. 
Classificatory distance instead is the retrieval separation induced by how a prevailing taxonomy represents their relation. 
Because classification systems themselves change, an object can remain technologically unchanged while becoming classificatorily closer to, or more distant from, a search problem \citep{lafondkim2019,barbieri2025evolving}.

Classificatory distance affects search costs because the taxonomy can represent technological relations more or less directly. 
This can be associated with translation costs when no recognizable common route connects a search problem to a relevant object and searchers must translate alternative labels into their context. Route-selection costs arise when several plausible routes compete for attention. 
Revisions can additionally create updating costs by changing previously learned mappings between categories and objects. 
Category creation may therefore lower translation costs by providing a missing shared route under an interpretable label, while incremental label additions to a mature system may increase route-selection and updating costs. Category cleanup through label deletion may reduce route-selection costs by removing redundant routes and restoring category contrast. We expect these effects to be stronger when search tasks rely more directly on the classification infrastructure.  

Patent classification systems provide empirical leverage for studying these theoretical arguments. 
They represent institutionalized infrastructures explicitly designed to support prior-art search across millions of technically heterogeneous documents \citep[sec.~904]{usptompep904}. Revisions of these systems can change the classificatory representation of already-public patents without changing the underlying inventions. We study the Y02/Y04 green-technology layer of the Cooperative Patent Classification (CPC), a nonexclusive, cross-cutting classification that links climate-related technologies across otherwise distinct technical branches \citep{veefkind2012,angelucci2018}. 

We exploit two system-level changes to this infrastructure. The first is the January 2013 launch of the green-technology layer in the CPC environment at the USPTO. Because this event coincided with the broader introduction of the CPC, our design identifies differential exposure to the new cross-cutting green layer but not the effect of CPC adoption itself, comparing launch-family patents with other green patents exposed to the same institutional transition. 
The second is a comprehensive revision of the now-mature green classification in August 2020, which added new codes, replaced existing ones, and deleted others while in most cases preserving a patent's membership in the broader green domain through another assignment. Together, these events allow us to distinguish category creation from code addition, replacement, and cleanup within an established hierarchical system that permits co-classification. 

Using USPTO patent and examination data, we study how these revisions affect observable search outcomes for existing patents. 
Our primary outcome is the time from the classification event to a patent's first post-event citation. We interpret this as a proxy for system-level retrieval efficiency: it captures when a patent is first recorded as cited in a later application, rather than the time spent on an individual query. Citation counts over corresponding time windows provide a secondary outcome capturing the overall volume of recorded attention. 
Methodologically, we develop novel document-based citation-timing measures from patent prosecution histories, allowing us to locate citation-related activity within the examination process instead of dating citations by the citing patent's eventual publication or grant.

We analyze examiner- and applicant-added citations separately because their search objectives and processes differ. 
Examiners perform a formal, documented and accountable prior-art search as part of patent examination, including a systematic search across relevant and analogous technological areas. Applicant-side citations instead arise from a more heterogeneous combination of prior knowledge and technological search, disclosure obligations, attorney practice, and strategic considerations \citep{sampat2010, cotropia2013, alcacergittelman2006, alcacer2009}. 
This distinction allows us to examine whether classification changes have a stronger impact when search tasks are more tightly coupled to the classification infrastructure. 


Our analysis uses event-specific matching among patents publicly available for at least three years before each event \citep{stuart2010,ho2011}. Within publication-cohort-by-technology cells, we match treated patents to unused current-green controls with no relevant classification change around the event, using propensity scores based on pre-event covariates and citation histories. 
We estimate matched-pair-stratified Cox models for the time to the first post-event citation and matched-pair fixed-effect Poisson difference-in-differences (DiD) models for citation counts in symmetric 24-month pre- and post-event windows. 
To assess the plausibility of our identifying assumptions, we examine remaining pre-event treated--control differences using pseudo-policy DiDs for citation counts and Cox placebo tests for citation timing. 
These diagnostics are most favorable for examiner-side creation and additions but generally weaker for applicants. Replacement and cleanup show some pre-event differences concentrated in more remote timing windows.  

Three principal findings emerge. 
First, patents assigned to a green CPC code created in 2013 have a 25\% higher examiner first-citation hazard, without a statistically detectable change in volume. 
Second, green code additions during the 2020 revision are associated with 13\% lower examiner first-citation hazards, and 15\% lower when the addition creates a patent's first recorded route into the green domain. 
Addition also shows an approximately 11\% lower examiner citation volume, whereas the count effect of green domain entry is indistinguishable from zero. 
Third, cleanup deletions that remove a green CPC route while retaining green-domain membership are associated with a 14\% higher examiner first-citation rate and 12\% higher examiner citation volume. 
Replacement shows no detectable effect. 
Effects are generally stronger and more precisely estimated for examiners, particularly for creation and cleanup. Mature-system additions, however, slow first citation for both audiences. 

Creation and additions with green domain entry alter first-citation timing without detectable changes in examiner citation volume. By contrast, cleanup increases speed and volume, accompanied by a broadened reach across examination units. Classification architecture can therefore alter search efficiency without necessarily changing overall retrieval volume. 
The core examiner timing findings are robust across alternative comparison groups and exact-cell designs, outcome horizons and implementation lags, citation-date proxies, matching specifications, and alternative CPC assignment and cited-record definitions, while citation-count estimates show some sensitivity to measurement choices. 

The paper makes three contributions. First, it extends category research from evaluation to search processes. Existing research largely examines how categorical fit, ambiguity, and spanning affect audience evaluations once an object has entered consideration \citep{hsu2009,negro2010,DurandPaolella2013,kovacs2021}. We develop a theory of \emph{classification-mediated search} in which category architecture shapes which objects enter consideration sets and how quickly they are found. This connects research on categories with theories of situated attention and information infrastructure \citep{ocasio1997,bowker_star_1999}.

Second, we develop a lifecycle account of category architecture. 
We theorize that category creation, addition, and cleanup affect different search costs: creation can reduce translation costs, mature-system additions can increase route-selection and updating costs, and cleanup can restore contrast by removing overlapping routes.
Category maintenance is therefore an organizational intervention that changes the search environment faced by users.
This lifecycle perspective extends previous work on category emergence and evolution, category differentiation, and historical change in classification systems \citep{navisglynn2010,DurandPaolella2013,Mohlin2014,lafondkim2019}.

Third, we distinguish \emph{technological distance} from \emph{classificatory distance}. Research commonly uses patent-class relationships to measure technological novelty, relatedness, and boundary spanning \citep{rosenkopfnerkar2001, fleming2001, benner2008close, mcnamee2013can}. Yet classification systems are mutable: creating a cross-cutting category can absorb an administrative boundary without changing the invention's underlying knowledge combination. 
Measures based on contemporaneous classifications can therefore confound substantive recombination with historically contingent taxonomy design \citep{lafondkim2019, barbieri2025evolving}. 
Because classification systems co-evolve with the domains they represent, a new cross-cutting category can make unchanged inventions appear less boundary spanning while changing the routes through which searchers find them. Observed classification-based novelty and search behavior may therefore reflect technological development, revisions to institutional representation, or both \citep{lafondkim2019,barbieri2025evolving}.
Our results demonstrate the relevance of this distinction for behavioral outcomes: revisions to the classificatory representation of otherwise unchanged inventions can affect when they are retrieved.
Whether such revisions also facilitate a transition in technological search from more curiosity-driven exploration toward efficiency-oriented exploitation, or merely affect classification-based measures of these behaviors, remains beyond our empirical design.

Together, these contributions position classification maintenance as an information-infrastructure design problem that can influence the efficiency of organizational search. Our results show that greater classificatory detail is not uniformly associated with faster retrieval. 
The consequences depend on the balance among access, route multiplicity, and the costs of updating learned mappings. 
For classification maintainers, our results highlight retrieval speed, organizational reach, and adjustment costs as relevant dimensions of category-system performance, particularly when revisions alter established retrieval routes. 
For technology managers and researchers, they also underscore that patent landscapes and longitudinal measures reflect not only changes in the underlying technological domain but also changes in the classificatory infrastructure through which technology is represented and retrieved.


\FloatBarrier
\section{Theory and hypotheses}
\subsection{Classifications as information infrastructure}


Classification systems codify how large bodies of information are partitioned and who is expected to search where \citep{garicano2000,starruhleder1996}. 
These structures guide attention by encoding expectations about where relevant objects are likely to be found \citep{Mohlin2014}: labels signal properties of category members, nested branches delimit search neighborhoods, and co-classifications connect otherwise separate locations.

Classification system architecture influences search behavior through several cognitive processes that provide the behavioral foundations for the concept of search costs developed below.
First, a category label affects \emph{cue salience}: an interpretable label makes a relation easier to recognize and provides stronger ``information scent'' about the likely value of following a route \citep{pirollicard1999,zunino2019}. 
Second, hierarchies in classification systems influence \emph{consideration-set construction}. Because inspecting additional alternatives is costly, decision makers evaluate only a subset of the available objects. The hierarchical organization of the search environment affects which objects enter that subset and in what order \citep{hauserwernerfelt1990}. 
Third, classification architecture affects \emph{search and stopping rules}. Boundedly rational searchers do not inspect every possible location but stop when an aspiration is met \citep{simon1955}. 
A classification with clear contrast can therefore concentrate search and support effective stopping \citep{kovacs2021}. 


Taken together, classification architecture affects the sequence of attention and therefore when and whether relevant objects enter consideration. We call this \emph{classification-mediated search}. Empirically, we examine whether changing a patent's external classification alters recorded search outcomes while the underlying technological landscape remains fixed.

\subsection{Classificatory distance and search costs} 

To study the impact of category changes, we conceptualize \emph{classificatory distance} as the retrieval separation induced by the taxonomy between a search problem and a relevant object. 
It is lower when shared or nearby categories provide a recognizable common route and higher when search requires traversing distant branches or translating across classificatory vocabularies. The hierarchical structure therefore matters because nearby branches imply a shorter and more recognizable route than locations separated higher in the taxonomy.

It is conceptually distinct from technological distance that reflects the substantive distance between knowledge components. Greater technological distance can make unfamiliar knowledge more difficult to understand, evaluate, and recombine \citep{fleming_sorenson_2004,fleming2001}. 
Classificatory distance instead concerns how these relations are represented within the prevailing classification system, which is historically contingent \citep{lafondkim2019}.
Two patents may be technologically close yet classificatorily distant when the taxonomy provides no common or recognizable route between them. Conversely, a shared classification code can place technologically heterogeneous patents on the same retrieval path \citep{bowker_star_1999, uspto_mpep902}. 
Because patent classifications are frequently used to measure technological relatedness \citep{benner2008close, mcnamee2013can}, the two forms of distance can be difficult to distinguish empirically. 

We distinguish two mechanisms through which classificatory distance generates retrieval frictions. 
\emph{Translation costs} arise when the taxonomy provides no recognizable route connecting the search problem to a relevant object, requiring the searcher to translate the problem across classificatory vocabularies \citep{bowker_star_1999}. 
\emph{Route-selection costs} arise when several plausible or overlapping routes require additional effort to determine where to search \citep{knudsen2007}. 
Changes in categorization systems can further create \emph{updating costs}, which are transitional costs that arise when a revision changes previously learned mappings between categories and objects. 
A classification revision can therefore reduce one retrieval friction while increasing another or temporarily imposing updating costs. 

Building on these cognitive retrieval frictions, we develop hypotheses about the impact of the creation and subsequent revision of the Y02/Y04 green-technology layer in the CPC system.

\subsection{Category creation}

Before the creation of a dedicated cross-cutting category for green technology, searchers had to anticipate which established functional branches might contain relevant green inventions and translate a climate-related search task across multiple classification vocabularies \citep{veefkind2012}. A new cross-cutting category can \textit{absorb} such boundaries by institutionalizing a relation that previously required cross-boundary search. 

The introduction of the Y02/Y04 infrastructure did not make relevant patents technologically closer, but may have reduced their classificatory distance by providing a common cross-cutting retrieval route. This route reduces the need to translate across otherwise separate classificatory vocabularies, lowering translation costs and bringing previously dispersed technological knowledge into consideration sets sooner.

\begin{quote}
\textbf{Hypothesis 1 (category creation for a new domain).}
Assignment to a newly created cross-cutting category reduces classificatory distance by providing a shared retrieval route, thereby lowering translation costs and reducing the time to first subsequent citation relative to otherwise comparable patents without the new retrieval route.
\end{quote}

We use the 2013 launch of the Y02/Y04 green-classification layer to test Hypothesis~1.


\subsection{Mature-system overhaul}

Classification design involves a trade-off between grouping objects broadly enough to make meaningful similarities visible and differentiating them finely enough to preserve contrast among individual members \citep{Mohlin2014}. 
Too few routes can create high translation costs because relevant objects remain dispersed across classificatory locations.
Too many overlapping routes can instead create high route-selection costs because several routes appear plausible and no single route provides a sufficiently strong cue.

This trade-off depends on the lifecycle of the classification system. Creation of a new category in an under-differentiated domain can supply a missing route and reduce classificatory distance. 
However, as such a category matures, additional routes can impose updating costs by altering established mappings, introducing alternative search paths, and requiring searchers to revise where they expect relevant objects to be found. 
Additional routes may also expand the set of plausible locations, increasing route-selection costs because searchers must determine which route to follow \citep{knudsen2007}. 

Therefore, the addition of a subcategory in a mature classification system can increase search costs. By contrast, removing redundant and overlapping routes can reduce route-selection costs by leaving fewer, more distinctive alternatives in the consideration set and concentrating attention on routes with stronger category contrast. 

\begin{quote}
\textbf{Hypothesis 2a (mature-system subcategory addition).} In an established classification infrastructure, adding category routes increases route-selection and updating costs during the adjustment period, thereby decreasing the rate of first subsequent citation relative to unchanged patents in the same domain.
\end{quote}

\begin{quote}
\textbf{Hypothesis 2b (mature-system subcategory cleanup).} In an established classification infrastructure, deleting a category route while preserving another domain route reduces route-selection costs, thereby increasing the rate of first subsequent citation relative to unchanged patents in that domain.
\end{quote}

We use the August 2020 overhaul of Y02/Y04 hierarchical branches to test Hypotheses~2a and 2b. Some patents receiving newly added Y02/Y04 codes were already members of the green domain, whereas others entered the green domain through the newly added codes for the first time. 
Comparing the two groups distinguishes the general effect of adding routes from the narrower effect of creating a patent's first recorded route into the domain. 
Similarly, some patents lost a granular Y02/Y04 code yet retained green-domain membership through a previous code assignment.

\subsection{Audience--task coupling}
The primary actors using patent classifications for information search are inventors, applicants and their representatives, and patent examiners, whose search objectives and routines differ. We define \emph{audience--task coupling} as the degree to which an audience's search task is formally organized around the institutional classification infrastructure. The same change in classification architecture should have larger behavioral consequences when this coupling is stronger.

Applicant-added citations arise from a heterogeneous process combining inventors' prior knowledge and search, disclosure obligations, applicant-side search and disclosure practices, and strategic considerations \citep{sampat2010,cotropia2013,roachcohen2013,alcacer2009}. Inventors begin from the technological problem and knowledge used to develop the invention, while applicant-side references may also be disclosed because they are already known, considered legally relevant, or strategically useful. Applicant-side citations therefore need not result from a search organized around the patent classification system. 

Examiners, by contrast, follow a more standardized institutional retrieval process in which classification forms part of the information-organizing infrastructure \citep{alcacergittelman2006,alcacer2009,kovacs2021}. They define the field of search, consult relevant and analogous arts, and document the search performed \citep[sec.~904]{usptompep904}. Examiners are technically specialized even within art units \citep{righisimcoe2019},  yet the examination task requires them to search beyond the application's most immediate domain \citep[sec.~904]{usptompep904}. Their specialization often makes cross-domain routing and evaluation difficult \citep{fergusoncarnabuci2017}.

Classification architecture may also affect the breadth of examiner search. By providing a recognizable route across otherwise specialized technological domains, a cross-cutting category can reduce the classificatory translation required to reach relevant prior art outside an examiner's familiar domain \citep{fergusoncarnabuci2017}. 
Classification changes may therefore influence not only how quickly relevant patents are found, but also whether relevant patents are reached from examination units outside the focal patent's immediate technological location. 

Examiner search is therefore more tightly coupled to the classification infrastructure than applicant-side search. USPTO search guidance explicitly requires examiners to define a field of search and consult relevant and analogous art \citep[sec.~904]{usptompep904}. 
Classification helps organize this formal search routine even for technically specialized examiners who must search beyond their immediate domain. 
Changes in classificatory distance should consequently have more direct effects on the sequencing and breadth of examiner consideration sets than on applicant-side search. More generally, audience--task coupling moderates the effect of classification architecture: changes in translation, route-selection, or updating costs should produce larger observable responses when the search task relies more strongly on the revised taxonomy.

\begin{quote}
\textbf{Hypothesis 3 (audience--task coupling).} The effects of category creation, mature-system addition, and cleanup on search outcomes are larger in their predicted direction for audiences whose search tasks are more tightly coupled to the classification infrastructure.
\end{quote}

In our setting, this implies larger effects for examiner-added than for applicant-added citations.

\FloatBarrier

\section{Data and methods}
\label{sec:empirical-setting}
\FloatBarrier
\subsection{Empirical context: CPC codes for green technology}
\label{subsec:greenCPC-methods}
The CPC is a patent classification system maintained jointly by the European Patent Office (EPO) and the USPTO. Patents can be classified by multiple CPC codes, allowing search to draw on both the system's hierarchical structure and links created through co-classification.

We study two system-level classification events. For category creation, we use January 1, 2013, when the EPO and USPTO launched the CPC and the USPTO began transitioning from the United States Patent Classification (USPC) to the CPC \citep{epo_cpc_history,uspto_mpep902}. 
The transition was gradual: published applications and patent documents continued to receive USPC classifications through December 2014, and we do not observe when individual documents became searchable under particular CPC codes \citep{uspto_mpep902}.
Our focal green layer comprises Y02 (climate-change mitigation and adaptation) and Y04S (smart-grid) codes, which are cross-cutting classifications that link climate-related technologies across conventional technical branches, such as energy, transport, buildings, chemistry, and information technology, without replacing those classifications \citep{veefkind2012,angelucci2018,epo_y02_updates}. Because CPC adoption was system-wide, our 2013 design identifies differential exposure to this green layer among patents subject to the same broader institutional transition. 
We therefore use January 1, 2013, as the institutional launch date and examine different treatment lags in the robustness checks to account for possible administrative delays.

For the mature-system overhaul, we use August 1, 2020, when a major revision of the established Y02/Y04 green-classification layer entered into force. CPC Notices of Changes (NoCs) provide the official record of additions, deletions, and other revisions to individual CPC codes \citep{usptonoc}. We scraped the NoCs, coded the symbols added or deleted in the August 2020 version, and validated them against official EPO and USPTO records \citep{epo_y02_updates,usptonoc}. 
A separate USPTO automated application-routing reform followed in October 2020 \citep{pairolerodegrazia2026}. We address concerns related to this contemporaneous institutional change through robustness checks that vary the timing of treatment and outcome measurement. 

We distinguish the following treatments: 
\begin{description}[style=nextline,leftmargin=0.5cm]

\item[2013 category creation (H1).]
A patent is treated if it belongs to the family of green CPC codes introduced with the 2013 launch.

\item[2020 category addition (H2a).]
A patent is treated if it receives a green code introduced in August 2020 without a simultaneous green-code deletion.

\item[2020 strict category addition (supplementary, H2a).]
A patent is treated if it was not classified as green before August 2020 and enters the green domain at that date through a newly added code.

\item[2020 category replacement (supplementary, H2a and H2b).]
A patent is treated if one new green code is added and another deleted in August 2020, while green-domain membership is retained.

\item[2020 category cleanup (H2b).]
A patent is treated if a pre-event green code is deleted, no other code is added, and green-domain membership is retained through another pre-existing code.

\end{description} 
Green-domain exit events when patents lose their green domain membership through deletion are too rare to yield an eligible treated sample and cannot be estimated. 

\subsection{Data}
\label{subsec:data-methods}

Our analysis combines three data sources. PatentsView provides CPC assignments observed at patent issue and in current records, citation links to patent grants and pre-grant publications, and patent characteristics including assignees, inventors, claims, and backward references \citep{patentsview}. At-issue CPC assignments reflect historical classifications at the date of patent issue, while current records capture classifications today. Our treatment and control group definitions use this distinction. 
The 2022 USPTO Patent Examination Research Dataset (PatEx) provides Case Management System (CMS) document histories for patent applications \citep{USPTOPatEx}. Finally, CPC NoCs provide the effective dates at which CPC symbols were added to or deleted from the classification system \citep{usptonoc}. We merge these code-level changes to the CPC symbols assigned to individual patents to construct the event-specific treatment indicators described in Section~\ref{subsec:greenCPC-methods}.

After linking these sources, the analysis contains 4,097,057 granted patents with a pre-grant publication after January 2005. 
For each event, we restrict the sample to patents with an earliest pre-grant publication at least 36 months before the event date, providing three complete years of observable pre-event history for matching. This leaves 951,990 patents for the 2013 event and 2,837,191 for the 2020 event. 


Our primary outcome is the time from the classification event to the first subsequent citation, which we use as a proxy for system-level search efficiency. Rather than dating citations by the filing or publication of the citing patent, we use citation-related documents recorded in the citing application's CMS prosecution history. These documents are filed by applicants or generated during examination and therefore locate citation activity within the prosecution process. 
For applicant-added citations, the baseline proxy is the earliest Information Disclosure Statement (IDS/SB08) or USPTO Form PTO-1449 reference-list event. For examiner-added citations, it is the earliest USPTO Form PTO-892 examiner reference list or, when no PTO-892 is observed, the earliest nonfinal or final office-action package (CTNF/CTFR). These dates identify when citation-related documentation enters the prosecution record, not when an individual reference was first discovered. 
Appendix~\ref{app:citation-timing} provides the complete code map and timing proxies. 

We complement the timing outcome with the number of examiner- and applicant-added forward citations within a 24-month window (Section~\ref{subsec:citation-volume}). 
Because the PatEx CMS document histories end in early 2023, we use 24 months as the common headline follow-up period to measure the outcome variables for both events.

\FloatBarrier

\subsection{Matching design}
\label{subsec:matching-methods}
Classification revisions are not random: classification authorities may revise technological domains that are expanding, maturing, or already experiencing changes in attention \citep{lafondkim2019}. Because such pre-event dynamics may predict both treatment and subsequent citations, we use matching to construct treated--control comparisons with similar observable characteristics and citation trajectories \citep{ho2011,stuart2010}. 

To isolate from simultaneous reclassification activities, we require controls not to experience any CPC revision during the six months before or after the focal event. In our headline analyses, we further restrict controls to current green patents, keeping treated and control patents within the green technology domain to avoid conflation with broader differences between green and non-green technologies \citep{veefkind2012,angelucci2018}. 
We assess the alternative of using broad-patent controls in Section~\ref{sec:exact-cell-robustness}.

Among the patents with the 36 months of pre-event observability described above, we use exact technology--cohort cell matching, whereby each treated patent can be matched only to a control from the same technology--cohort cell. 
Our baseline design uses earliest-publication-year by four-digit examiner art unit (year--AU4) for 2013 and earliest-publication-year by two-digit technology center (year--TC2) for 2020. 
These specifications provide acceptable balance for the respective headline designs and, for 2020, common support across all four treatment branches. 
Alternative cell definitions are reported in the robustness checks (Section~\ref{sec:exact-cell-robustness}).

Within these exact cells, we estimate a propensity score by logistic regression using pre-event patent characteristics and citation histories. The covariates capture invention scope, prior-art intensity, organizational and procedural characteristics, technological maturity, and prior examiner- and applicant-citation trajectories (see Appendix~\ref{app:matching-covariates} for details). 
Our primary matching procedure restricts potential pairs using a standardized propensity-score caliper and, within this restriction, uses Mahalanobis distance to prioritize similarity in the three annual logged citation levels and two adjacent-year changes for each audience. 
We assess post-match covariate balance using standardized mean differences (SMD) and use a maximum absolute SMD below 0.10 as our design target. Implementation details are reported in Appendix~\ref{app:matching-implementation}.

We then perform greedy one-to-one nearest-neighbor matching without replacement, ordered from the closest eligible pair to the most distant \citep{rosenbaumrubin1985,ho2011,stuart2010}. 


Because matching addresses observed pre-event differences and cannot capture unobserved shocks, we examine treated--control differences before each event using the falsification tests described in Section~\ref{subsec:robustness-methods}. 
Matching within common technology--cohort environments also limits differential exposure to broader institutional changes, but cannot eliminate contemporaneous shocks that affect treated patents differently after the event (see Section~\ref{sec:limitations}).

\FloatBarrier 
\subsection{Econometric specification}
\subsubsection{Search efficiency}
\label{subsec:search-efficiency}

We operationalize search efficiency as the time from the classification event to the first recorded subsequent citation: a shorter interval indicates that the focal patent enters the searcher's recorded consideration set sooner, reflecting faster retrieval. 
We estimate the time to first citation separately for examiner- and applicant-added citations using Cox proportional hazards models \citep{cox1972}, using a right-censored follow-up of 24 months. For audience $a$ and matched pair $m$, our headline model is

\begin{equation}
 h^{a}_{im}(t)=h^{a}_{0m}(t)\exp\left(\theta^{a}D_i+Z_i'\kappa^{a}\right),
 \label{eq:cox}
\end{equation}
where $t$ denotes time since the classification event, $D_i$ indicates whether patent $i$ is treated, $h^{a}_{0m}(t)$ is the matched-pair-specific baseline hazard, and $Z_i$ contains the patent-level controls together including annual logged examiner and applicant pre-event citations to adjust for residual observed differences after matching (Appendix~\ref{app:outcome-controls}).

We estimate matched-pair-stratified Cox proportional hazards models by partial likelihood, with robust sandwich standard errors clustered by matched pair \citep{cox1972} and report $\exp(\widehat{\theta}^a)$ as the hazard ratio (HR), with $\HR>1$ indicating an earlier and $\HR<1$ a later first citation. 
We conduct proportional-hazards tests and calculate restricted mean survival time (RMST) contrasts, indicating the number of citation-free days, as supplementary diagnostics (see Appendix~\ref{app:survival-diagnostics}).

\subsubsection{Citation volume}
\label{subsec:citation-volume}

As a secondary outcome, we assess whether classification changes alter the volume of recorded attention by comparing examiner- and applicant-added citation counts across symmetric 24-month pre- and post-event windows. 
Given the skewed nature of citation counts and prevalence of zeros, we estimate a stacked two-period matched-pair Poisson DiD model \citep{santossilva2006}, where each patent contributes one observation per window: 
\begin{equation}
\mathbb{E}[Y^a_{it}\mid D_i,Post_t,Z_i,m]
=\exp\left(\alpha^a_m+\lambda^a Post_t+\beta^a D_i
+\delta^a(D_i\times Post_t)+Z_i'\gamma^a
\right),
\label{eq:did}
\end{equation}
where $Y^a_{it}$ is the citation count for patent $i$, audience $a$, and period $t$; $D_i$ indicates treatment and $Post_t$ the post-event period; $\alpha^a_m$ denotes matched-pair fixed effects and $Z_i$ contains the patent-level controls, including the pre-event annual logged examiner and applicant citation counts.

The coefficient $\lambda^a Post_t$ is the common pre--post period effect and $\beta^a$ captures the conditional pre-event treated--control difference. 
The interaction coefficient $\delta^a$ captures the differential pre--post change associated with treatment and $\exp(\delta^a)$ is the corresponding multiplicative DiD effect, comparing the proportional pre--post change among treated and controls patents. 
We report it as percentage change $100[\exp(\widehat{\delta}^a)-1]$. Standard errors are clustered by matched pair \citep{cameronmiller2015}. 

\subsubsection{Audience--task coupling}
\label{subsec:audience-methods}
To assess audience--task coupling, we compare the examiner- and applicant-specific estimates from the headline models above. Because the Cox models are estimated separately by audience, differences in first-citation effects provide descriptive comparisons. 
For citation volume, we additionally estimate a pooled model that formally tests the examiner--applicant difference. 
We pool both audiences and extend Equation~\ref{eq:did} to a triple-difference model:
\begin{align}
\mathbb{E}[Y_{iat}\mid D_i,Post_t,E_a,Z_i,m]
&=\exp\bigl(\alpha_m+\lambda Post_t+\mu E_a+\beta D_i
+\phi(D_i\times E_a) \nonumber\\
&+\psi(Post_t\times E_a)+\delta(D_i\times Post_t) \nonumber\\
&+\tau(D_i\times Post_t\times E_a)+Z_i'\gamma\bigr),
\label{eq:audience-ddd}
\end{align}
where $Y_{iat}$ is the citation count for patent $i$, audience $a$, and period $t$, and $E_a$ equals one for examiner-added and zero for applicant-added citations. 
We stack the audience-specific two-period panels that each patent contributes pre- and post-event observations for both audiences. 
The applicant DiD treatment effect is $\delta$, the examiner effect is $\delta+\tau$, and the triple interaction $\tau$ tests whether treatment effects differ between audiences. 
On the multiplicative scale, $\exp(\widehat{\tau})$ compares the examiner and applicant DiD incidence-rate ratios: values above one indicate a more positive and values below one a more negative examiner response. Standard errors are clustered by matched pair. 
As a functional-form check, we also estimate an analogous linear model in raw citation counts, where the triple interaction provides an additive examiner--applicant treatment contrast that is directly interpretable in citation-count units. 

\subsection{Identification and robustness}
\label{subsec:robustness-methods}

Because matching cannot rule out differential pre-event trajectories, we use falsification tests to assess whether treated and matched control patents were already evolving differently before the classification events.

For citation volume, we estimate adjacent-year pseudo-policy DiDs comparing treated--control changes from event years $-3$ to $-2$ and from years $-2$ to $-1$, treatment coefficients for years $-2$ and $-1$ relative to year $-3$, and test their joint significance. These diagnostics are estimated using the primary Poisson specification and repeated linearly in raw citation counts. 

For citation timing, we estimate matched Cox placebo models over the preceding 12, 24, and 36 months and a separate fourth-year pre-event window. Rejection indicates a pre-existing treated--control difference in citation volume or timing and therefore weakens the identifying interpretation of the corresponding post-event estimate. Appendix~\ref{app:pretrend-construction} provides the complete construction of these diagnostics.

We further assess robustness to four broad sources of design dependence.
First, we vary the control group by including non-green patents and modifying the exact technology--cohort cells, control-exclusion window, and matching implementation to assess dependence on the construction of the matched counterfactual. 
Second, we vary outcome horizons, implementation lags, and citation-date proxies to assess sensitivity to outcome timing and possible administrative delays in the implementation of revised CPC codes at the patent level. 
Third, we vary whether treatment and citation outcomes are constructed from grant- or pre-grant-publication records to assess sensitivity to data-source and focal unit definition. 
Finally, we compare the Poisson models with alternative functional forms. Appendix~\ref{app:robustness-specifications} reports the complete specifications and implementation details.

We additionally conduct two exploratory analyses to inform interpretation of the results. 
First, we examine whether classification revisions alter the breadth of examiner search, including the number of distinct examiners and technological units reaching a focal patent. 
Second, we explore whether treatment effects vary with characteristics of the technological and classificatory environment, organizational and application characteristics, and prior exposure to the green-classification infrastructure. These analyses use the headline matched samples and are interpreted as supplementary rather than confirmatory. These analyses additionally report Benjamini--Hochberg-adjusted $q$-values for multiple testing. Appendix~\ref{app:exploratory-methods} provides further details.

\FloatBarrier
\section{Results}
\label{sec:results}

\subsection{Matching performance}
\label{subsec:matching-results}

\IfFileExists{generated/tables/table_matching_performance.tex}{
  \begin{table}[htbp]
\centering
\caption{Sample construction and matching performance}
\label{tab:matching}
\small
\begin{adjustbox}{max width=\linewidth,max totalheight=0.82\textheight,keepaspectratio,center}
\begin{threeparttable}
\begin{tabular}{lrrrrrrrrr}
\toprule
\shortstack{Exact-cell\\design} & \shortstack{Eligible\\treated} & \shortstack{Eligible\\controls} & \shortstack{Supported\\treated} & \shortstack{Supported\\controls} & \shortstack{Matched\\pairs} & Cells & \shortstack{Treated\\retained} & \shortstack{Max\\SMD} & \shortstack{Pretrend\\SMD} \\
\midrule
\multicolumn{10}{l}{\textit{Creation}} \\[2pt]
Pub. year $\times$ AU4 & 39,201 & 21,806 & 34,990 & 18,288 & 6,419 & 927 & 18.3\% & 0.089 & 0.089 \\
\addlinespace
\multicolumn{10}{l}{\textit{Addition}} \\[2pt]
Pub. year $\times$ TC2 & 17,080 & 149,563 & 17,080 & 105,473 & 9,556 & 106 & 55.9\% & 0.074 & 0.074 \\
\addlinespace
\multicolumn{10}{l}{\textit{Strict addition}} \\[2pt]
Pub. year $\times$ TC2 & 14,283 & 149,563 & 14,283 & 88,015 & 7,478 & 105 & 52.4\% & 0.097 & 0.097 \\
\addlinespace
\multicolumn{10}{l}{\textit{Replacement}} \\[2pt]
Pub. year $\times$ TC2 & 2,966 & 149,563 & 2,966 & 54,027 & 2,336 & 71 & 78.8\% & 0.087 & 0.085 \\
\addlinespace
\multicolumn{10}{l}{\textit{Cleanup}} \\[2pt]
Pub. year $\times$ TC2 & 55,623 & 149,563 & 55,623 & 128,324 & 46,420 & 102 & 83.5\% & 0.095 & 0.035 \\
\addlinespace
\bottomrule
\end{tabular}
\begin{tablenotes}[flushleft]\footnotesize
\item[]
The compiled grant-level universes before event-specific restrictions are 2013: 4,097,057; 2020: 4,097,057 . Exact-cell design identifies the publication-cohort-by-technology restriction; Cells is the number of exact cells represented in the final match. Eligible patents satisfy the 36-month public-history requirement, have a nonmissing exact-cell identifier, and have complete three-year pre-event citation histories. Supported candidates satisfy the exact-cell requirement and the within-cell control-pool restriction. Matched pairs are the unique treated--control pairs retained by one-to-one matching without replacement. Treated retained is the percentage of supported treated patents retained in the final match. Max SMD is the largest absolute post-match standardized mean difference across all diagnostic covariates. Pretrend SMD is the largest absolute post-match standardized mean difference across the six raw annual pre-event citation measures: three examiner-citation counts and three applicant-citation counts. The 2020 treatment branches overlap and should not be added.
\end{tablenotes}
\end{threeparttable}
\end{adjustbox}
\end{table}

}{\placeholder{Run the R script to generate the matching-performance table.}}

Table~\ref{tab:matching} summarizes sample attrition, matching retention, and post-match balance for the baseline analyses.
For 2013, 61,007 patents meet the eligibility criteria. Exact-cell support and the within-cell control-pool restriction leave 53,278 candidates, including 34,990 of the 39,201 eligible treated patents. After one-to-one matching, the sample retains 6,419 treated patents and 6,419 unique controls in 927 publication-year--AU4 cells, with a maximum post-match SMD of 0.089.
For 2020, exact-cell support does not remove any eligible treated patents. One-to-one matching retains 9,556 of 17,080 addition patents, 7,478 of 14,283 strict-addition patents, 2,336 of 2,966 replacement patents, and 46,420 of 55,623 cleanup patents. Maximum post-match SMDs range from 0.074 to 0.097. The replacement analysis is based on a substantially smaller matched sample compared to the other treatments. 

\begin{figure}[!htbp]
\centering
\includegraphics[
width=\linewidth,
trim=1mm 1mm 1mm 1mm,
clip
]{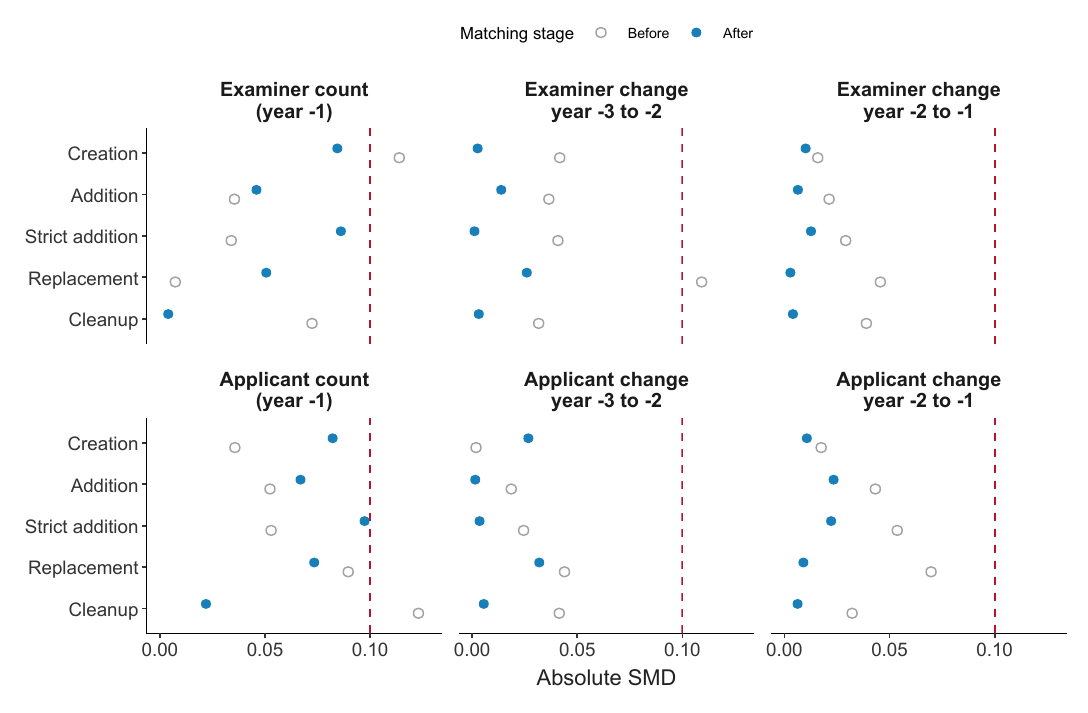}
\caption{Balance in pre-event citation counts and slopes. The panels report absolute standardized mean differences before and after matching for the most recent annual logged examiner- and applicant-citation levels and the adjacent-year changes in those logged citation histories. The dashed line denotes the prespecified $|\SMD|=0.10$ balance threshold.}
\label{fig:smd-headline}
\end{figure}

Figure~\ref{fig:smd-headline} shows the pre- and post-match balance in pre-event citation counts and trends. Appendix Figures~\ref{fig:smd-creation}--\ref{fig:smd-cleanup} and Appendix Table~\ref{tab:smd-full} provide the complete covariate-level balance audit. All five headline designs remain below the 0.10 balance criterion after matching. 
Appendix Table~\ref{tab:outcome-descriptives} reports citation count descriptives for each treatment sample after matching.  

\FloatBarrier
\subsection{Pre-event diagnostics}
\label{subsec:pretrend-results}

We next evaluate the identifying assumptions of our design by asking whether treated and matched control patents exhibited differential citation trajectories before the reclassification events. The null hypothesis states that, before reclassification, treatment and control patents do not exhibit differential changes in citation patterns. 

\IfFileExists{generated/tables/table_pretrend_diagnostics.tex}{
\begin{table}[htbp]
\centering
\caption{Pre-event diagnostics}
\label{tab:pretrends}
\small
\begin{adjustbox}{max width=\linewidth,max totalheight=0.82\textheight,keepaspectratio,center}
\begin{threeparttable}
\begin{tabular}{rrr@{\hspace{4pt}}rrr@{\hspace{6pt}}rr@{\hspace{4pt}}rr}
\toprule
\multicolumn{6}{c}{Count diagnostics} & \multicolumn{4}{c}{Timing diagnostics} \\
\cmidrule(lr){1-6}\cmidrule(lr){7-10}
\multicolumn{3}{c}{Examiner} & \multicolumn{3}{c}{Applicant} & \multicolumn{2}{c}{Examiner} & \multicolumn{2}{c}{Applicant} \\
\cmidrule(lr){1-3}\cmidrule(lr){4-6}\cmidrule(lr){7-8}\cmidrule(lr){9-10}
\shortstack{Min\\$p$} & $N_{p<.05}$ & \shortstack{Joint\\$p$} & \shortstack{Min\\$p$} & $N_{p<.05}$ & \shortstack{Joint\\$p$} & \shortstack{Min\\$p$} & $N_{p<.05}$ & \shortstack{Min\\$p$} & $N_{p<.05}$ \\
\midrule
\multicolumn{10}{l}{\textit{Creation}} \\[2pt]
$.065$ & 0/10 & $.158$ & $.006$ & 5/10 & $.098$ & $.383$ & 0/4 & $.071$ & 0/4 \\
\addlinespace
\multicolumn{10}{l}{\textit{Addition}} \\[2pt]
$.046$ & 2/10 & $.320$ & $.035$ & 2/10 & $.107$ & $.369$ & 0/4 & $.142$ & 0/4 \\
\addlinespace
\multicolumn{10}{l}{\textit{Strict addition}} \\[2pt]
$.018$ & 1/10 & $.061$ & $.023$ & 4/10 & $.067$ & $.300$ & 0/4 & $.007$ & 1/4 \\
\addlinespace
\multicolumn{10}{l}{\textit{Replacement}} \\[2pt]
$.111$ & 0/10 & $.705$ & $.016$ & 1/10 & $.261$ & $.028$ & 2/4 & $.268$ & 0/4 \\
\addlinespace
\multicolumn{10}{l}{\textit{Cleanup}} \\[2pt]
$.019$ & 2/10 & $.055$ & $<.001$ & 10/10 & $<.001$ & $.001$ & 1/4 & $<.001$ & 2/4 \\
\addlinespace
\bottomrule
\end{tabular}
\begin{tablenotes}[flushleft]\footnotesize
\item[]
Each treatment separator is followed by examiner and applicant diagnostics. Count diagnostics comprise ten tests in total: five Poisson tests (two adjacent-year pseudo-policy DiDs, two annual treatment coefficients relative to the oldest pre-event year, and one joint Wald test) plus the same five tests estimated with a linear model in raw citation counts. Min $p$ is the smallest unadjusted $p$-value; $N_{p<.05}$ reports the number of rejecting tests divided by the number reported. Joint $p$ tests whether the annual Poisson pre-event treatment coefficients are jointly zero. Timing diagnostics comprise four Cox falsification tests covering the preceding 12, 24, and 36 months and the separate fourth pre-event year. These are falsification diagnostics, not treatment-effect estimates. Exact count estimates appear in Appendix Figure~\ref{fig:count-placebos-appendix}; Figure~\ref{fig:timing-placebos} shows the 36-, 24-, and 12-month Cox placebos, while Appendix Table~\ref{tab:timing-pretrend-details} reports all four timing placebos, including the separate fourth pre-event year.
\end{tablenotes}
\end{threeparttable}
\end{adjustbox}
\end{table}

}{\placeholder{Run the R script to generate the diagnostic table.}}

\begin{figure}[htbp]
\centering
\includegraphics[
width=0.75\linewidth,
trim=1mm 1mm 1mm 1mm,
clip
]{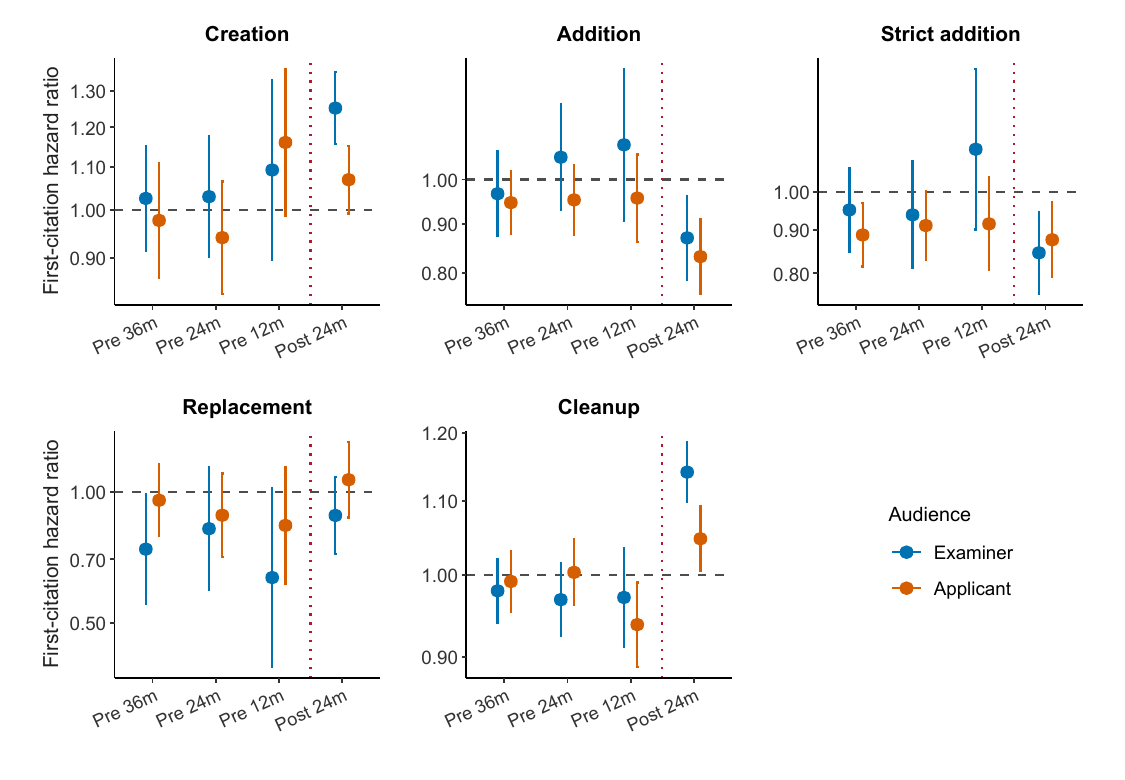}
\caption{Pre-event timing-placebo estimates and the headline post-event estimate. Each treatment panel reports Cox hazard ratios for the overlapping 36-, 24-, and 12-month pre-event risk windows, together with the 24-month post-event estimate. Points are estimates and whiskers are 95\% confidence intervals. 
The dashed horizontal line marks one and the vertical line separates pre- and post-event estimates. 
}
\label{fig:timing-placebos}
\end{figure}

Table~\ref{tab:pretrends} summarizes rejection counts, minimum $p$-values, and the joint count test, while Appendix Tables~\ref{tab:count-pretrend-details} and~\ref{tab:timing-pretrend-details} report individual estimates. Figure~\ref{fig:timing-placebos} presents the Cox timing placebos, with corresponding count estimates in Appendix Figure~\ref{fig:count-placebos-appendix}.

The examiner diagnostics for 2013 creation are favorable: none of the ten count diagnostics or four timing placebos rejects at 5\%, with minimum $p$-values of $.065$ and $.383$, respectively, and the Poisson joint test of $p=.158$.
For applicants, the timing tests also do not reject the null, but five of the ten applicant count tests reject (minimum $p=.006$).
Thus, the examiner diagnostics support the identifying interpretation more strongly than the applicant count diagnostics.

For 2020 addition, no examiner or applicant timing placebo rejects, and the examiner joint count test does not reject ($p=.320$). The two individual count rejections concern the same year-$-3$-to-$-2$ contrast in the linear model. 
For strict addition, examiner timing diagnostics are also favorable and the joint count test is marginal ($p=.061$), with one individual rejection in the count diagnostics.
Applicant diagnostics are weaker: four count diagnostics and the 36-month timing placebo ($\HR=.888$, $p=.007$) reject the null, whereas the shorter 24- and 12-month windows closer to the event do not. 
Overall, timing diagnostics are stronger than count diagnostics, particularly for examiners.

Replacement has favorable examiner count diagnostics, with none of the ten tests rejecting. Timing diagnostics are weaker: the separate fourth-year and 36-month placebos reject ($HR=.591$, $p=.028$; $HR=.739$, $p=.042$), whereas the 24- and 12-month windows do not ($p=.246$ and $.060$). 
For applicants, no timing placebo rejects; the only count rejection is the linear year-$-1$ versus year-$-3$ contrast ($p=.016$).

For cleanup, examiner diagnostics vary by pre-event window. The two count rejections concern the same year-$-1$ versus year-$-3$ contrast under alternative functional forms. 
Neither adjacent-year pseudo-policy DiD rejects, and the Poisson joint test is marginal ($p=.055$). Of the four examiner timing placebos, only the separate fourth-year window rejects ($HR=.916$, $p=.001$). 
Because this interval lies outside the 36-month history used for matching, it provides less direct evidence of a differential trajectory approaching the event.
Applicant diagnostics are again much weaker: all ten count tests and two of four timing tests reject. We therefore interpret the examiner cleanup result as a robust association, but not the applicant estimates causally. 

\FloatBarrier
\subsection{Search efficiency and citation volume}
\label{subsec:headline-results}
\IfFileExists{generated/tables/table_headline_effects.tex}{
  \begin{landscape}
\begin{table}[htbp]
\centering
\caption{Baseline results for search efficiency and citation volumes}
\label{tab:headline}
\small
\setlength{\tabcolsep}{4pt}
\renewcommand{\arraystretch}{1.15}
\begin{tabular*}{\linewidth}{@{\extracolsep{\fill}}lcc@{\hspace{5pt}}cc@{\hspace{5pt}}cc@{\hspace{5pt}}cc@{\hspace{5pt}}cc@{}}
\toprule
 & \multicolumn{2}{c}{2013: Creation} & \multicolumn{2}{c}{2020: Addition} & \multicolumn{2}{c}{2020: Strict addition} & \multicolumn{2}{c}{2020: Replacement} & \multicolumn{2}{c}{2020: Cleanup} \\[2pt]
\cmidrule(lr){2-3}\cmidrule(lr){4-5}\cmidrule(lr){6-7}\cmidrule(lr){8-9}\cmidrule(lr){10-11}
 & Examiner & Applicant & Examiner & Applicant & Examiner & Applicant & Examiner & Applicant & Examiner & Applicant \\
\midrule
\multicolumn{11}{l}{\textit{Panel A. Time to first citation: Cox models}} \\[2pt]
Treated & $0.225^{***}$ & $0.067^{*}$ & $-0.139^{***}$ & $-0.184^{***}$ & $-0.167^{***}$ & $-0.131^{**}$ & $-0.123$ & $0.066$ & $0.132^{***}$ & $0.047^{**}$ \\[-1pt]
 & $(0.040)$ & $(0.038)$ & $(0.051)$ & $(0.045)$ & $(0.057)$ & $(0.052)$ & $(0.104)$ & $(0.103)$ & $(0.020)$ & $(0.021)$ \\
HR & 1.252 & 1.069 & 0.870 & 0.832 & 0.846 & 0.877 & 0.884 & 1.069 & 1.141 & 1.048 \\
\addlinespace[2pt]
$N$ (patents) & 12,838 & 12,838 & 19,112 & 19,112 & 14,956 & 14,956 & 4,672 & 4,672 & 92,840 & 92,840 \\
Events & 3,302 & 4,495 & 1,864 & 2,973 & 1,525 & 2,390 & 483 & 548 & 11,915 & 11,223 \\
Concordance & 0.622 & 0.693 & 0.614 & 0.679 & 0.611 & 0.718 & 0.645 & 0.664 & 0.582 & 0.626 \\
Treatm. Schoenfeld $p$ & $.124$ & $.044$ & $.596$ & $.017$ & $.718$ & $.011$ & $.230$ & $.341$ & $.006$ & $.968$ \\
Global Schoenfeld $p$ & $.174$ & $.006$ & $.229$ & $.004$ & $.784$ & $.029$ & $.124$ & $.088$ & $.010$ & $.157$ \\
RMST diff., 12m & -9.7 & -3.7 & 2.5 & 4.9 & 3.6 & 6.3 & -1.9 & -3.2 & -3.0 & -1.5 \\[-1pt]
\quad 95\% CI & [-12.7, -6.6] & [-7.0, -0.5] & [1.0, 3.9] & [2.6, 6.9] & [1.7, 5.4] & [3.6, 8.8] & [-4.7, 1.2] & [-6.7, 0.6] & [-3.7, -2.2] & [-2.3, -0.6] \\
RMST diff., 24m & -26.1 & -14.2 & 8.1 & 14.4 & 10.3 & 18.0 & -1.2 & -8.5 & -8.4 & -4.0 \\[-1pt]
\quad 95\% CI & [-33.8, -18.9] & [-23.2, -6.3] & [4.0, 11.6] & [9.2, 19.6] & [6.1, 14.8] & [12.6, 23.8] & [-8.6, 6.2] & [-18.0, -0.0] & [-10.5, -6.5] & [-6.2, -1.9] \\
\addlinespace
\multicolumn{11}{l}{\textit{Panel B. Citation counts: Poisson DiD}} \\[2pt]
Treated $\times$ Post & $0.049$ & $-0.178^{***}$ & $-0.111^{**}$ & $-0.018$ & $0.001$ & $-0.022$ & $-0.243^{**}$ & $-0.108$ & $0.117^{***}$ & $-0.063^{**}$ \\[-1pt]
 & $(0.041)$ & $(0.048)$ & $(0.050)$ & $(0.058)$ & $(0.057)$ & $(0.062)$ & $(0.106)$ & $(0.134)$ & $(0.019)$ & $(0.032)$ \\
IRR & 1.050 & 0.837 & 0.895 & 0.982 & 1.001 & 0.978 & 0.784 & 0.897 & 1.124 & 0.939 \\
\addlinespace[2pt]
$N$ (patent-periods) & 15,068 & 16,524 & 11,004 & 15,140 & 8,948 & 12,004 & 2,784 & 3,136 & 61,728 & 58,960 \\
\bottomrule
\end{tabular*}
\renewcommand{\arraystretch}{1}
\begin{minipage}{0.98\linewidth}\footnotesize
Panel A reports Cox log-hazard coefficients and SEs; HR is $\exp(\widehat\theta)$. Models are stratified by matched pair and follow patents for 24m. Panel B reports Poisson DiD log-incidence-rate coefficients and SEs; IRR is $\exp(\widehat\delta)$. Models include pair fixed effects and compare symmetric 24m pre- and post-event windows. Matched-pairs with no positive counts for the analyzed audience and window do not contribute to fixed-effect Poisson estimation; the reported number of observations for Poisson regressions can therefore be smaller than twice the matched-patent count. Model SEs are clustered by matched pair. For the Cox estimates, $SE[\exp(\widehat\beta)]=\exp(\widehat\beta)SE(\widehat\beta)$. For count effects, $SE\{100[\exp(\widehat\delta)-1]\}=100\exp(\widehat\delta)SE(\widehat\delta)$. Schoenfeld $p$-values test proportional hazards; small values indicate nonproportionality. RMST is the treated-minus-control difference in citation-free days; CIs use 499 matched-pair bootstrap replications, and negative values indicate earlier treated-patent citation. Concordance is Harrell's rank-discrimination statistic. Post-match balance is reported in Table~\ref{tab:matching}. All models use the narrow citation-date proxy. $^{***}p<.01$, $^{**}p<.05$, $^{*}p<.10$.
\end{minipage}
\end{table}
\end{landscape}

}{\placeholder{Run the R script to generate the headline-effects table.}}
Table~\ref{tab:headline} shows our main results. 

\paragraph{2013 category creation.} 
Membership in the green category created in 2013 is associated with a 25.2\% higher examiner first-citation hazard ($\HR=1.252$, 95\% confidence interval (CI) [1.157, 1.355], $p<.001$), observed in a sample of 12,838 matched patents and 3,302 first-citation events during the 24 months after January 2013. 
Neither the treatment-specific nor the global Schoenfeld test rejects proportional hazards ($p=.124$ and $p=.174$), meaning that the data do not provide evidence that the examiner HR changes systematically over the 24-month follow-up period. 
The RMST contrast indicates that treated patents receive their first examiner citation approximately 26.1 days earlier than controls (95\% CI [$-33.8$, $-18.9$]). 
The applicant HR is smaller and less precisely estimated ($\HR=1.069$, 95\% CI [0.993, 1.151], $p=.078$). Proportional hazards are rejected for the applicant model, so its HR is interpreted as an average over the follow-up period. The corresponding RMST difference is $-14.2$ days (95\% CI [$-23.2$, $-6.3$]), indicating an approximately two-week earlier first citation among treated patents.

The examiner count DiD is 5.0\% (95\% CI [$-3.1$\%, 13.7\%], $p=.233$), whereas the applicant count estimate is $-16.3$\% (95\% CI [$-23.8$\%, $-8.0$\%], $p<.001$). Together with the favorable examiner pre-event diagnostics, these results provide support for Hypothesis~1 on the examiner timing margin. However, they do not indicate a corresponding general increase in examiner citation volume. For applicants, the weaker count diagnostics limit a causal interpretation as treatment appears to correlate with a reduction in applicant citations.

\paragraph{2020 mature-system additions.}
For additions we observe a 13.0\% lower examiner first-citation hazard ($\HR=0.870$, 95\% CI [0.787, 0.962], $p=.007$; 19,112 patents and 1,864 events) and a 16.8\% lower applicant hazard ($\HR=0.832$, 95\% CI [0.762, 0.909], $p<.001$; 2,973 events). 
The examiner proportional-hazards tests do not reject, and treated patents are cited about 8.1 days later (95\% CI [4.0, 11.6]). 
The applicant model rejects proportional hazards, but its RMST estimate points in the same direction, with a 14.4 days later first citation (95\% CI [9.2, 19.6]). The examiner count estimate suggests $-10.5$\% fewer citations during the 24 months post-event window (95\% CI [$-18.9$\%, $-1.3$\%], $p=.026$), whereas the applicant count estimate is $-1.8$\% and not significant ($p=.751$).

The timing pattern is similar for strict additions: the examiner first-citation hazard is 15.4\% lower ($\HR=0.846$, 95\% CI [0.757, 0.946], $p=.003$), as is the applicant hazard, by 12.3\% ($\HR=0.877$, 95\% CI [0.792, 0.971], $p=.012$).
The examiner proportional-hazards tests again do not reject, and the 24-month examiner RMST difference indicates 10.3 additional citation-free days (95\% CI [6.1, 14.8]). For applicants, the proportional-hazards restriction is rejected, but its RMST difference, which does not require proportional hazards, is likewise positive at 18.0 days (95\% CI [12.6, 23.8]). 
For both audiences, count estimates are indistinguishable from zero. 

The slower examiner citation in both addition treatment groups, together with favorable examiner pre-event timing diagnostics, supports Hypothesis~2a. 

\paragraph{2020 cleanup and replacement.}

Category cleanup is associated with a 14.1\% higher examiner first-citation hazard ($\HR=1.141$, 95\% CI [1.098, 1.187], $p<.001$; 92,840 patents and 11,915 events) and 12.4\% greater examiner citation volume (95\% CI [8.2\%, 16.7\%], $p<.001$). The examiner proportional-hazards tests reject and the HR reflects an average over the follow-up period. The RMST indicates 8.4 fewer days before the first examiner citation arrives (95\% CI [$-10.5$, $-6.5$]). 
The applicant hazard also increases modestly by 4.8\% ($\HR=1.048$, 95\% CI [1.005, 1.092], $p=.030$), with an RMST difference of $-4.0$ days (95\% CI [$-6.2$, $-1.9$]), while applicant citation volume decreases by 6.1\% ($p=.049$). 

The direction of the examiner cleanup estimates is consistent with Hypothesis~2b, and their larger magnitude relative to the applicant estimates is consistent with Hypothesis~3. The examiner pre-event evidence is mixed but considerably more favorable than the applicant evidence: the examiner joint count test is marginal, with two of ten count diagnostics and one of four timing diagnostics rejecting, whereas the applicant diagnostics reject for all ten count tests and two timing windows. We therefore treat the examiner cleanup result as a robust association and avoid a causal interpretation of the applicant cleanup estimates.

The effects of label replacement remain inconclusive. The examiner first-citation hazard is 11.6\% lower ($\HR=0.884$, 95\% CI [0.722, 1.083], $p=.233$), while the applicant hazard is 6.9\% higher ($\HR=1.069$, 95\% CI [0.874, 1.306], $p=.518$). The examiner count estimate is $-21.6$\% ($p=.022$), but it is not accompanied by a timing effect and is based on a substantially smaller matched sample. 
The examiner count placebos are favorable, with none of ten tests rejecting the hypothesis of equal pre-event citation trajectories among treated and untreated patents, but two of four examiner timing placebos reject. On the applicant side, one count test and no timing tests reject. The inconclusive and poorly significant outcomes, limited support, and unfavorable examiner timing diagnostics do not allow us to draw any conclusions about the impact of internal category relabeling.

\subsection{Audience--task coupling}
\label{subsec:audience-results}
The distinct patterns observed for the timing of examiner and applicant citations already provide descriptive support for Hypothesis~3. 
Following the 2013 creation event, the examiner timing response is larger and more precisely estimated than the applicant response. 
The cleanup treatment also produces larger examiner effects, although the pre-event diagnostics demand greater caution regarding a causal interpretation. 
Mature-system additions delay the time of first citation for both audiences, whereby the effect of base additions appears greater on applicants, while strict addition with green domain entry produces larger examiner effects. 
Because the Cox models are estimated separately by audience, these differences remain descriptive comparisons and do not provide formal tests on the differences of applicant and examiner coefficients.

\IfFileExists{generated/tables/table_audience_ddd.tex}{
  \begin{table}[htbp]
\centering
\caption{Examiner--applicant contrasts in citation volume}
\label{tab:audience-ddd}
\small
\begin{adjustbox}{max width=\linewidth,max totalheight=0.82\textheight,keepaspectratio,center}
\begin{threeparttable}
\begin{tabular}{lrrrrr}
\toprule
Model & Estimate & SE & 95\% CI & $p$ & $N$ \\
\midrule
\multicolumn{6}{l}{\textit{Creation}} \\[2pt]
Poisson & $1.254^{***}$ & 0.078 & [1.110, 1.417] & $<.001$ & 40,928 \\
Linear & $0.091$ & 0.065 & [-0.038, 0.219] & $.166$ & 51,352 \\
\addlinespace
\multicolumn{6}{l}{\textit{Addition}} \\[2pt]
Poisson & $0.911$ & 0.069 & [0.785, 1.058] & $.222$ & 38,792 \\
Linear & $-0.123^{*}$ & 0.065 & [-0.250, 0.003] & $.057$ & 76,448 \\
\addlinespace
\multicolumn{6}{l}{\textit{Strict addition}} \\[2pt]
Poisson & $1.023$ & 0.086 & [0.867, 1.206] & $.788$ & 30,888 \\
Linear & $-0.177^{***}$ & 0.068 & [-0.310, -0.043] & $.010$ & 59,824 \\
\addlinespace
\multicolumn{6}{l}{\textit{Replacement}} \\[2pt]
Poisson & $0.874$ & 0.146 & [0.629, 1.214] & $.422$ & 9,016 \\
Linear & $0.060$ & 0.039 & [-0.017, 0.136] & $.127$ & 18,688 \\
\addlinespace
\multicolumn{6}{l}{\textit{Cleanup}} \\[2pt]
Poisson & $1.197^{***}$ & 0.045 & [1.113, 1.288] & $<.001$ & 182,152 \\
Linear & $-0.024$ & 0.025 & [-0.072, 0.024] & $.324$ & 371,360 \\
\addlinespace
\bottomrule
\end{tabular}
\begin{tablenotes}[flushleft]\footnotesize
\item[]
Poisson estimates are ratios of examiner and applicant incidence-rate ratios for the treatment-by-post-by-examiner interaction. Their displayed standard errors are delta-method standard errors on the ratio scale. Linear estimates are examiner-minus-applicant triple-difference coefficients in citation counts. Both estimators include matched-pair, post-period, and audience fixed effects. Standard errors are clustered by matched pair. $^{***}p<.01$, $^{**}p<.05$, $^{*}p<.10$.
\end{tablenotes}
\end{threeparttable}
\end{adjustbox}
\end{table}

}{\placeholder{Run the R script to generate the audience triple-difference table.}}

\begin{figure}[htbp]
  \centering
  \includegraphics[
    width=\linewidth,
    trim=1mm 1mm 1mm 1mm,
    clip
  ]{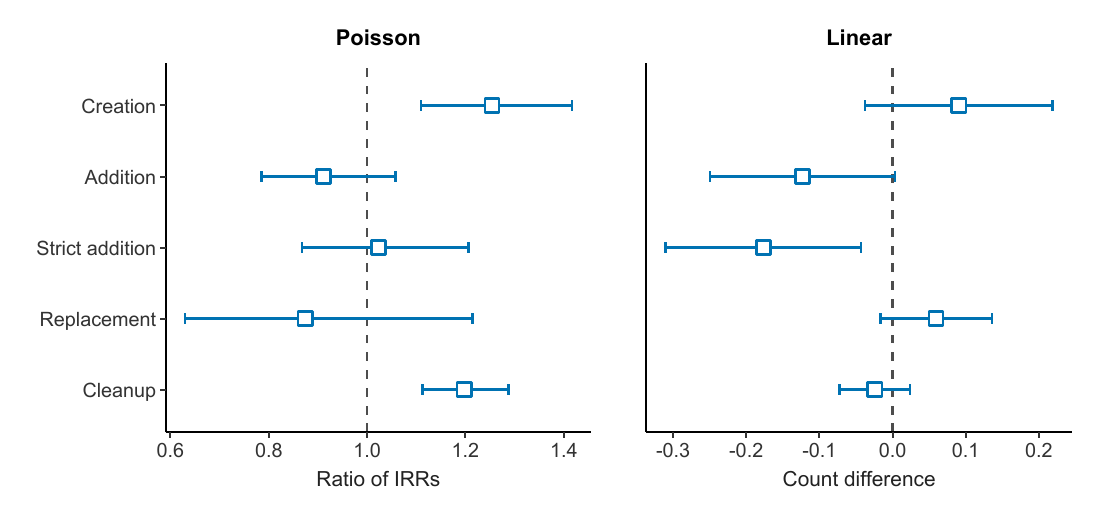}
  \caption{Examiner--applicant contrasts in citation volume. The Poisson panel reports the ratio of examiner and applicant incidence-rate ratios. One denotes equal proportional treatment effects across audiences. The linear panel reports the examiner-minus-applicant triple difference in citation counts. Zero denotes equal additive treatment effects. Squares are point estimates and whiskers show 95\% confidence intervals.}
  \label{fig:audience-ddd}
\end{figure}

Table~\ref{tab:audience-ddd} and Figure~\ref{fig:audience-ddd} report the formal examiner--applicant comparisons for citation counts. In 2013, the Poisson triple difference implies an examiner-to-applicant ratio of incidence-rate ratios of 1.254 (95\% CI [1.110, 1.417], $p<.001$). For cleanup, the corresponding ratio is 1.197 (95\% CI [1.113, 1.288], $p<.001$). The remaining contrasts are statistically indistinguishable from one, suggesting the absence of any significant examiner--applicant difference.

To obtain the contrasts in an additive form, we estimate linear models on raw citation counts in levels. 
They do not reproduce statistically meaningful creation and cleanup contrasts ($p=.166$ and $p=.324$), while strict addition produces a significant linear but not Poisson contrast with significantly fewer post-treatment citations from applicants rather than examiners. 
Because the Poisson model is better suited to the skewed, zero-heavy count outcomes, it remains our primary model. 
Despite the statistical limitations of our audience contrast analyses, the results provide partial support for Hypothesis~3, concentrated in the creation and cleanup count and timing estimates, while the audience contrast for addition treatments remains less conclusive. 

\FloatBarrier
\subsection{Supplementary analyses}
\label{subsec:supplementary-results}

\subsubsection{Breadth of examiner search}
\label{subsec:breadth-results}

\IfFileExists{generated/tables/table_breadth_absolute.tex}{
  \begin{table}[htbp]
\centering
\caption{Category revision effects on the breadth of examiner search}
\label{tab:breadth-absolute}
\small
\begin{adjustbox}{max width=\linewidth,max totalheight=0.82\textheight,keepaspectratio,center}
\begin{threeparttable}
\begin{tabular}{lccccc}
\toprule
Measure & Creation & Addition & \shortstack{Strict\\addition} & Replacement & Cleanup \\
\midrule
Distinct examiners & $6.6$\% & $-8.8^{*}$\% & $-0.2$\% & $-20.4^{**}$\% & $12.2^{***}$\% \\
 & {\scriptsize [-1.3, 15.2]} & {\scriptsize [-17.0, 0.2]} & {\scriptsize [-10.4, 11.2]} & {\scriptsize [-35.3, -2.0]} & {\scriptsize [8.2, 16.4]} \\
 & {\scriptsize $p=.101$} & {\scriptsize $p=.056$} & {\scriptsize $p=.975$} & {\scriptsize $p=.032$} & {\scriptsize $p=<.001$} \\
\addlinespace[6pt]Distinct AU4s & $8.3^{**}$\% & $-8.4^{*}$\% & $-0.3$\% & $-20.6^{**}$\% & $11.0^{***}$\% \\
 & {\scriptsize [0.6, 16.6]} & {\scriptsize [-16.6, 0.7]} & {\scriptsize [-10.5, 11.0]} & {\scriptsize [-35.3, -2.6]} & {\scriptsize [7.1, 15.0]} \\
 & {\scriptsize $p=.035$} & {\scriptsize $p=.070$} & {\scriptsize $p=.953$} & {\scriptsize $p=.027$} & {\scriptsize $p=<.001$} \\
\addlinespace[6pt]Distinct TC2s & $6.8^{*}$\% & $-4.4$\% & $2.7$\% & $-19.0^{**}$\% & $10.6^{***}$\% \\
 & {\scriptsize [-0.5, 14.6]} & {\scriptsize [-12.7, 4.6]} & {\scriptsize [-7.3, 13.9]} & {\scriptsize [-33.2, -1.8]} & {\scriptsize [6.9, 14.5]} \\
 & {\scriptsize $p=.071$} & {\scriptsize $p=.327$} & {\scriptsize $p=.606$} & {\scriptsize $p=.032$} & {\scriptsize $p=<.001$} \\
\bottomrule
\end{tabular}
\begin{tablenotes}[flushleft]\footnotesize
\item[]
The first row for each measure reports the percentage effect; the following rows report the 95\% confidence interval and two-sided $p$-value. Effects are implied by matched-pair Poisson difference-in-differences models for distinct sources of examiner citations in symmetric 24-month pre- and post-event windows. Model-specific observation counts and supplementary compositional-reach measures are reported in Appendix Table~\ref{tab:breadth-deep}. $^{***}p<.01$, $^{**}p<.05$, $^{*}p<.10$.
\end{tablenotes}
\end{threeparttable}
\end{adjustbox}
\end{table}

}{\placeholder{Run the R script to generate the absolute-reach breadth table.}}

Classification changes may affect not only how quickly a patent is retrieved but also how broadly it is reached across distinct examination units at the USPTO. 
We operationalize this organizational reach as the numbers of distinct examiners, AU4s, and TC2s citing the focal patent. Table~\ref{tab:breadth-absolute} summarizes the estimates and Appendix Table~\ref{tab:breadth-deep} reports supplementary compositional-reach measures.

For 2013 category creation, the estimated numbers of distinct citing examiners, AU4s, and TC2s increase by 6.6\% ($p=.101$), 8.3\% ($p=.035$), and 6.8\% ($p=.071$), respectively, suggesting at most a modest expansion in examiner reach. 
Creation appears to primarily accelerate first examiner citation timing without a clearly detectable pattern of increased organizational reach.

Addition provides similarly limited support for changes in the breadth of search. The numbers of distinct citing examiners and AU4s decline by 8.8\% ($p=.056$) and 8.4\% ($p=.070$), respectively, while the TC2 estimate is smaller and insignificant. The strict addition treatment shows no detectable change. 
Changes in citation timing are not accompanied by robust changes in organizational reach, although we observe some patterns of decline. 
Category replacement is associated with narrower examiner reach: the numbers of distinct citing examiners, AU4s, and TC2s decline by approximately 19--21\% ($p<.05$ for each). 
Given weaker matching support and unfavorable timing diagnostics, we consider the replacement estimates as descriptive associations. 
Cleanup, by contrast, consistently expands absolute examiner reach. It increases the numbers of distinct citing examiners by 12.2\% ($p<.001$), citing AU4s by 11.0\% ($p<.001$), and citing TC2s by 10.6\% ($p<.001$). These increases closely match the 12.4\% increase in total examiner citation volume. 

Taken together, the breadth results sharpen the distinction between the timing and scale of retrieval. Creation and strict addition shift first-citation timing without clear changes in citation volume or organizational reach, whereas base addition combines slower retrieval with some evidence of reduced volume and reach. 
Cleanup shows the clearest scaling response, with higher citation volume and more distinct technological units reaching the patent.

\FloatBarrier
\subsubsection{Exploratory heterogeneity}
\label{subsec:heterogeneity-results}

Appendix Table~\ref{tab:moderators-bh} examines moderators capturing organizational context (entity and assignee status), the technological and classificatory search environment (technological maturity and CPC scope), application lineage, and prior membership of the green domain. These characteristics probe whether classification effects depend on features of the invention, the applicant, or familiarity with established classificatory routes. 
After adjusting for multiple testing, few effects remain statistically significant. Thus, within the observable variation remaining after matching, classification effects appear to be only weakly conditioned by the tested characteristics.

The clearest heterogeneity patterns can be observed for category cleanup. For applicant citation volume, persistent membership in the 2013 green-code family is associated with a stronger cleanup effect (ratio of incidence-rate ratios $=1.50$, $q<.001$). 
This is consistent with cleanup being more beneficial when it removes redundant routes while preserving an established green retrieval route. 
Persistent membership is itself nonrandom, however, so this interaction cannot distinguish familiarity with an established route created during the 2013 launch event from other differences between persistent and newer green patents. 
For examiner citation volume, broad CPC scope attenuates the cleanup effect (ratio of incidence-rate ratios $=.85$, $q=.038$), consistent with removal of one route providing less simplification when many alternative classificatory routes remain. Neither corresponding timing interaction remains significant after multiplicity adjustment.

\FloatBarrier
\subsection{Robustness}
\label{subsec:results_robustness}

\subsubsection{Control groups and matching}
\label{sec:exact-cell-robustness}

Appendix Tables~\ref{tab:robustness-summary} and~\ref{tab:control-pool-headline} report the robustness checks for variation in comparison group and exact-cell definitions, crossing unchanged-green versus broad-patent controls with alternative technology--cohort cells. 

For the 2013 creation event, the examiner hazard remains significantly above one in all six designs. Estimates are similar in direction and magnitude across green and broad controls, but the broad-control matches are less well balanced. For example, under year--AU4 cells the maximum SMD increases from .089 with green controls to .144 with broad controls. 
This greater residual imbalance reinforces our preference for green controls, which improve comparability and isolate variation within the green-technology domain.

For 2020, the effects of addition and strict-addition on examiner hazards are significantly below one in all six designs and cleanup hazards are significantly above one in all six, whereas replacement remains mixed and inconclusive.
We therefore use the green-control year--TC2 design for all 2020 treatments because it preserves the within-green-domain contrast and satisfies the balance criterion across all branches. 

Our main results are also stable to nine alternative matching implementations (Appendix Table~\ref{tab:matching-sensitivity}). Examiner timing effects remain significant in all nine specifications for creation, cleanup, and strict addition, and in seven of nine for addition. 
Cleanup also increases examiner citation volume by 9.7--12.1\% in every specification. Replacement is distinctly less stable, with acceptable balance in only four of nine specifications and no statistically significant timing estimate.
The results are likewise unchanged when we impose a stricter 12-month exclusion window around CPC revisions for control patents.

\FloatBarrier
\subsubsection{Outcome horizons, implementation lags, and functional form}
Examiner timing results remain directionally stable across alternative outcome horizons, implementation lags, and CMS-based citation-date proxies (Appendix Table~\ref{tab:outcome-robustness-details}). 
Across these checks, examiner HRs range from 1.126 to 1.290 for creation, from 0.809 to 0.865 for addition, from 0.816 to 0.835 for strict addition, and from 1.088 to 1.175 for cleanup. All remain statistically significant in their predicted direction. 
Applicant creation HRs remain close to one and statistically insignificant. Applicant addition effects remain below one and generally become stronger over longer or delayed outcome measurement windows. For strict addition, the 12-month estimate is not significant, whereas the 24- and 36-month and delayed-window estimates are. 
Applicant cleanup HRs are smaller, around 1.04--1.06, and less stable under delayed windows. Replacement remains imprecise: examiner HRs are generally below one but rarely significant, while applicant HRs remain close to one throughout.

Citation-volume results are more specification dependent. The 2013 examiner effect attenuates from 10.7\% over 12 months to 6.1\% over 24 months and 0.8\% over 36 months, but none of these estimates reaches conventional significance. 
For addition, examiner Poisson estimates remain negative at roughly 7--10\% across windows, lags, and citation-date proxies, but alternative functional forms are essentially null. Strict-addition count effects are likewise small and insignificant. Thus, neither addition branch shows a systematic horizon-dependent volume response.

Cleanup remains an exception. Its examiner count effect is positive and significant throughout, declining from 17.2\% over 12 months to 12.1\% over 24 months and to 10.3\% and 9.4\% under three- and six-month implementation lags. Alternative functional forms using linear counts and $\log(1+y)$ are also positive and significant. A linear probability model for receiving any citation, capturing the extensive margin, is also positive and significant. 
Replacement examiner count estimates remain negative across several specifications, but weaker balance, smaller samples, and unfavorable timing diagnostics preclude a causal interpretation.

These timing variations reduce sensitivity to uncertainty about implementation dates and broader institutional changes at the USPTO. For 2013, the persistence of the examiner result under delayed windows suggests that it does not depend on the exact date at which patents became searchable under CPC during the USPTO transition. For 2020, estimates are likewise stable across implementation lags, although the design cannot separate the August CPC revision from the October USPTO routing reform.

\FloatBarrier
\subsubsection{Cited-record and CPC-definition checks}

Appendix Table~\ref{tab:routes} independently varies the CPC record used to define treatment—grant versus pre-grant publication—and the record on which subsequent citations are observed. 
For category creation, the examiner timing result is stable across all four combinations ($\HR=1.25$--$1.29$, all $p<.001$), with satisfactory balance.
When cleanup treatment is defined from grant-level CPC assignments, examiner retrieval remains faster, regardless whether citations are observed to the focal grant ($\HR=1.13$, $p<.001$) or its pre-grant publication ($\HR=1.17$, $p<.001$). 
In contrast, defining cleanup from pre-grant CPC assignments leaves only 27--28 treated patents and severe imbalance (maximum SMD .392--.400). Cox estimates are therefore unavailable and the remaining count estimates are too imprecise to interpret.

The 2020 addition results are more sensitive to the cited record than to the CPC basis used to define treatment. 
When citations are observed to grants, examiner hazards remain below one whether treatment is defined from grant or pre-grant CPC assignment ($\HR=.82$, $p=.002$ and $\HR=.81$, $p<.001$, respectively). When citations are instead observed to the pre-grant publication, the corresponding HRs are approximately one and statistically insignificant. These alternative matched samples do not meet our SMD balance target (maximum SMD .129--.138) and cannot be interpreted as clean causal contrasts.

Applicant estimates are generally more sensitive to the cited record, but the interpretative strength of our design differs across treatments. 
For creation, all four matches satisfy the balance criterion. The effect on applicant timing is close to one and not significant when citations are observed to grants. However, it rises to approximately 1.14--1.15 and becomes significant when citations are observed to pre-grant publications. 
For addition, the applicant timing estimate changes its direction with the cited record: when citations are observed to grants, the HR is below one ($\HR=.79$ or $.90$), whereas for citations observed to pre-grant publications it rises above one ($\HR=1.08$ or $1.14$). All four matches remain above the balance target and these observations provide weaker evidence. 
Grant-defined cleanup retains good balance and yields faster applicant timing for both cited records ($\HR=1.09$ and $1.11$), whereas the applicant citation-volume effect reverses from $+11.7\%$ for citations to grants to $-5.9\%$ for citations to pre-grant publications. Both count estimates are statistically significant.
Replacement remains imbalanced across the available variants. 
Overall, the CPC record primarily changes treatment membership and common support, whereas the cited record can change measured outcomes holding treatment definition fixed. Pre-grant CPC definitions do not systematically retain more or fewer patents across treatments. 

\FloatBarrier
\section{Discussion}

\subsection{Categories change retrieval before they change evaluation}
Much category research has examined how categorical fit, spanning, and ambiguity shape evaluation after an object has entered an audience's consideration set \citep{hsu2009,negro2010,cudennec2023}. Recent work has expanded this agenda to attention and search. For example, \citet{kovacs2021} show that category contrast influences whether inventions are noticed and cited. We extend this research by treating classification architecture as mutable. 
Rather than asking how audiences respond to a given categorical position, we show that changing the category system itself can alter when an object enters consideration. 

Our distinction between timing and volume adds a further dimension to research on access to patented knowledge. Earlier research has found that earlier disclosure can lead to more subsequent patent citations \citep{baruffaldi2020patents} and that machine translation supports citations to previously difficult-to-access foreign patents \citep{buettner2022}. These studies argue that reducing barriers to patented information can facilitate knowledge diffusion. 
Our setting instead concerns the findability of patents that are already available within a complex technological landscape. We show that the design of search routes can affect when patents are retrieved without uniformly affecting citation volume. 

The breadth results provide suggestive evidence about when faster retrieval also expands recorded attention. Creation accelerates retrieval without clearly increasing the number of distinct citing examiners or technological units, whereas cleanup shows higher citation volumes and organizational reach. 
Thus, volume increases coincide with broader reach, whereas timing shifts can occur without a corresponding expansion of the citing audience. 

\subsection{Classification architecture across stages of maturity}

The contrast between category creation and mature-system revision suggests that the consequences of classification changes depend on the existing architecture.  
A missing cross-cutting route can impose translation costs because relevant objects remain dispersed across classificatory vocabularies, and creating that route can make retrieval easier. 
Once users have developed routines around an established system, however, additional routes may raise route-selection and updating costs, whereas removing a route while preserving domain access may simplify search. 
A related tension appears in environmental labeling of consumption goods, where labels can reduce information problems but their proliferation can also reduce informativeness and create confusion among users \citep{brecard2014}.
This resembles the granularity trade-off in \citet{Mohlin2014}, in which finer categories reduce within-category heterogeneity but leave fewer observations on which to base predictions. 
Our research adds a search-cost dimension to this trade-off: too little structure can make relevant objects difficult to locate, while excessive route multiplicity can make the resulting system costly to navigate. 

A related stage dependence appears in research on the emergence of new market categories. 
As a new market category emerges and becomes legitimated, attention shifts from establishing the category as a whole toward differentiation within it \citep{navisglynn2010}.
Our setting concerns formal classification rather than producer self-categorization, but it points to a similar change over different stages in the category lifetime: broad connection can be valuable when a domain is emerging, whereas differentiation among routes becomes increasingly important once the system is established. 
Empirically, collapsing additions, replacements, cleanup, and domain exit into a single reclassification indicator can therefore combine changes operating through different mechanisms and potentially opposite signs \citep{barbieri2025evolving}.

\subsection{Classificatory distance, technological search, and measurement}

Classificatory distance separates technological boundary spanning from the retrieval costs created by its representation. Research on technological search treats unfamiliar recombination and boundary-spanning exploration as sources of novelty, uncertainty, and technological impact \citep{fleming2001,rosenkopfnerkar2001}. Our argument adds a representational layer: the same technological relation can become easier to retrieve when the classification system supplies a shared route. 
Formal classification provides an institutional map of where relevant objects are likely to be found. This parallels \citet{fleming_sorenson_2004} at the level of search guidance: scientific knowledge helps inventors identify useful technological combinations, whereas classification encodes expectations about where existing objects with particular attributes can be located. Both can make search more directed by structuring complex technological search spaces. 

The distinction between classificatory and technological distance is important because technologies and categories co-evolve. Research on industry emergence shows that technological designs and the categories used to represent them develop jointly \citep{grodal2015coevolution,zunino2019}, and patent classifications are repeatedly revised as technological domains change \citep{lafondkim2019,barbieri2025evolving}. 
As technological trajectories develop, classification systems are revised to reflect emerging and increasingly established technological relations. 
Hence, ex post classificatory distance can partly reflect which technological relations became institutionally recognized rather than only how novel or distant they were ex ante. 
Reclassification can consequently make the same knowledge combination appear closer, more conventional, or less boundary spanning in class-based measures than before the update. Longitudinal research based on patent classes therefore risks attributing changes in the institutional representation of technology to changes in inventive behavior unless classification versions are preserved or technological distance is measured independently of the mutable taxonomy.

Coevolution can also have a behavioral implication. Once a previously dispersed technological relation is institutionalized through a recognizable category, subsequent search may become more directed even though the underlying knowledge relation has not changed. 
A revision can therefore make a search path appear less exploratory in classification-based measures while simultaneously making actual search processes more routinized. 

\subsection{Managerial and policy implications}

Classification systems are information infrastructures and their design matters for organizations that must search large information spaces under limited attention and search capacity \citep{ocasio1997,garicano2000,starruhleder1996}. 
Recognizable categories that provide retrieval routes can reduce the costs to locate relevant objects. 
Our results are consistent with a central trade-off: additional structure may reduce translation costs when useful routes are missing while increasing route-selection and updating costs once routes overlap or learned mappings change. 
For patent offices and other organizations maintaining large classification systems, greater classificatory detail is therefore not synonymous with greater search efficiency, and performance depends on how the architecture structures retrieval routes for its users \citep{Mohlin2014,lafondkim2019}. 

Our audience results add a user-side dimension to this design problem. Classification changes have their clearest consequences when the search task is formally organized around the taxonomy, as in patent examination. This suggests that the performance of an information infrastructure cannot be separated from the tasks and routines of its users: the same revision may matter greatly for institutionally coupled search while playing little role for users who rely primarily on prior knowledge or retrieval practices outside the formal taxonomy.

For technology managers and patent analysts, taxonomy revisions can alter both an invention's visibility and its measured position in technological space without changing its content. This matters for patent landscapes and competition analyses that infer technological proximity, novelty, or movement from classification codes: changes in class-based distance can reflect revisions to the taxonomy as well as changes in the technologies being mapped \citep{lafondkim2019,barbieri2025evolving}. 
In other mutable information infrastructures---such as product catalogs, scientific taxonomies, or organizational knowledge repositories---changes in category architecture can likewise alter both measured similarity among objects or competitors, and the costs of identifying them. 

\section{Limitations}
\label{sec:limitations}

First, exposure to classification revision is not randomly assigned. Exact-cell matching, pre-event citation histories, DiD estimation, and falsification diagnostics reduce observable differences, but classification authorities may revise precisely those domains that are already expanding, maturing, or changing in attention. 

Second, neither event is a fully isolated patent-specific shock: the 2013 green-layer launch coincided with the system-wide CPC adoption and gradual USPC phase-out, while the August 2020 revision preceded the October USPTO routing reform \citep{uspto_mpep902, pairolerodegrazia2026}. Moreover, our treatment dates are defined at the system level and we do not exactly measure when individual patents became searchable under the revised codes. 
Comparing treated and control patents within common institutional environments, matching on pre-event characteristics, and using placebo and timing robustness checks reduce sensitivity to these contemporaneous changes, but cannot rule out heterogeneous effects on treated and control patents.

Third, citation timing is measured from application-level prosecution events rather than from the moment when a reference was discovered. This limitation is particularly important for applicants. Their first observable citation-related event often occurs near the application's entry into the patent system, while the inventive process, prior-art search, and accumulation of relevant knowledge may have begun earlier. Applicant-side timing therefore indicates when prior art becomes recorded in prosecution rather than when an inventor or representative first encountered it. 
Nevertheless, whether a patent appears in citations across subsequent applications remains informative about its recorded visibility, and implementation lags capture delayed responses to treatment. Examiner-side timing proxies are more closely tied to their search processes because applicants face less systematic documentation requirements, although neither measure reveals reference-specific discovery times.

Fourth, our design focuses on patents that already existed before the classification change. This allows us to hold technological content fixed, but does not identify how newly filed inventions respond to an already available category structure, for example through changes in search, positioning, or subsequent impact.

Finally, external validity is bounded by patent classifications explicitly built for prior-art search and by the green-technology setting. 
Patent examination is highly formalized and tightly coupled to classification, which may make infrastructure effects more visible than in less codified search environments. 
Whether similar mechanisms operate in product catalogs, scientific taxonomies, digital platforms, or organizational knowledge repositories is therefore an empirical question likely to depend on the search task, user expertise, and the extent to which retrieval relies on the taxonomy.

\section{Conclusion}
Classification systems form part of the infrastructure through which organizations search for relevant information and objects. 
The creation of a cross-cutting green-technology classification is associated with faster examiner retrieval, additions to the mature system with slower retrieval, and cleanup with faster retrieval alongside broader recorded attention and organizational reach. 
These patterns are consistent with a trade-off between translation costs, route multiplicity, and the costs of updating learned mappings, and they are clearest for examiners whose formal search task is tightly coupled to patent classification. 
By separating technological distance from classificatory distance, we theorize how revisions to the institutional representation of technology can alter retrieval frictions and show empirically that these revisions can affect realized search outcomes. Changing the map can therefore change what is found and how quickly it is reached, even when the territory itself remains fixed.

\section*{Acknowledgments}
The authors thank François Lafond for comments and discussions. 
CRediT author contributions statement: KH: Conceptualization, methodology, formal analysis, data curation, investigation, visualization, writing (original draft, review \& editing), project administration. NB: Conceptualization, writing (review \& editing). SJ: Conceptualization, validation, writing (review \& editing). 
Generative AI tools (ChatGPT, Claude, and Gemini) were used under author supervision for coding and editorial assistance. All AI-assisted outputs were reviewed and revised before use; the authors take full responsibility for the research.

\clearpage
\putbib
\end{bibunit}

\clearpage
\thispagestyle{empty}

\vspace*{0.25\textheight}

\begin{center}
{\Large\bfseries Online Appendix\par}

\vspace{1.5em}

{\large
Classification as Search Infrastructure:\\
How Category Creation, Addition and Cleanup Shape Knowledge Retrieval
\par}

\vspace{2em}

{\normalsize Supplementary Materials\par}
\end{center}

\vfill
\clearpage

\appendix
\begin{bibunit}
The appendix provides further detail on data construction, technical implementation and methods, as well as supplementary diagnostics and results. Appendix~\ref{app:reproducibility} records the computational implementation of the analysis. Appendix~\ref{app:data-construction} documents the citation-timing measures used in Section~\ref{sec:empirical-setting}. Appendix~\ref{app:matching-diagnostics} expands the matching and matched-sample diagnostics summarized in Section~\ref{subsec:matching-results}. Appendix~\ref{app:estimator-details} provides additional details for the estimators introduced in Section~\ref{sec:empirical-setting}. Finally, Appendix~\ref{app:supp-results} reports the specification-level falsification, robustness, moderator, and breadth results summarized in Section~\ref{sec:results}.

\section{Software and computational reproducibility}
\label{app:reproducibility}
All data construction and statistical analyses were performed in R version 4.4.1 \citep{RCoreTeam2024}. Data preparation primarily used \texttt{data.table} version 1.16.2 \citep{barrett2024datatable}, together with \texttt{lubridate} version 1.9.3 and \texttt{stringr} version 1.5.1. One-to-one matching was implemented with \texttt{MatchIt} version 4.7.2 \citep{ho2011}. Cox models, Schoenfeld-residual diagnostics, and survival-function calculations used \texttt{survival} version 3.6-4 \citep{therneau2024survival,therneaugrambsch2000}. Fixed-effect Poisson and linear models were estimated with \texttt{fixest} version 0.12.1 \citep{berge2018}. Excel workbooks were written with \texttt{openxlsx} version 4.2.8.1. The analyses were run under 64-bit Windows 11. 

\section{Data construction and variable definitions}
\label{app:data-construction}

\setcounter{table}{0}
\renewcommand{\thetable}{B\arabic{table}}
\setcounter{figure}{0}
\renewcommand{\thefigure}{B\arabic{figure}}

Our data pipeline links PatentsView patents and citation edges to their PatEx application histories, constructs the audience-specific citation-date proxies, combines grant- and publication-level CPC assignments with dated CPC revisions, and reduces these data to the event-specific products used by the matching and outcome models. The sections below document core variables that enter the reported analyses. 
Upon final publication of this article, code and data will be made available under an open license in a public repository.

\subsection{Treatment definitions}
\label{app:treatment-construction}

Historical patent-specific CPC assignment dates are unavailable, so our treatment variables identify exposure to dated changes in the classification system rather than the exact date on which an individual patent received or lost a code. We distinguish CPC assignments observed at issue from assignments present in the current CPC snapshot. At-issue records capture the classification attached to a patent when it issued, whereas current records indicate whether a code remains assigned subsequently.

For 2013, treatment identifies patents carrying codes from the launch family described in Section~\ref{subsec:greenCPC-methods}. Operationally, the launch family comprises focal Y02/Y04S symbols whose recorded CPC establishment date is no later than January 1, 2013, or for which no later establishment date is available. We validated the resulting code list against the January 2013 CPC scheme and official EPO and USPTO documentation \citep{epo_y02_updates,usptonoc}. 

For 2020, we link code additions and deletions recorded in the CPC NoCs to patent-level CPC assignments and classify the resulting patterns as addition, strict addition, replacement, or cleanup. Current green status, as recorded in current CPC assignments, is used to define the headline green control pool and to distinguish replacement or cleanup that preserves green-domain membership from domain exit; strict addition additionally requires that the patent was not green before the event. The headline treatment definitions use grant-level CPC records and exclude simultaneous CPC changes outside the affected green branches. As a measurement robustness check, we reconstruct the corresponding treatment definitions from pre-grant-publication CPC records, obtained from Patents View.

\FloatBarrier

\subsection{Citation timing code map}
\label{app:citation-timing} 
PatEx records when a citation-related document was submitted to the CMS and entered into the citing application’s administrative record. Although these dates do not reveal when an individual reference was cognitively discovered, they provide a closer proxy for the timing of the search and discovery process than conventional measures based on patent publication dates.
The underlying dates come from the PatEx \texttt{cms\_documents.csv} table, using \texttt{document\_code} and \texttt{mailroom\_date} columns, with code definitions obtained from \texttt{cms\_document\_codes.csv} \citep{USPTOPatEx}.

For examiner- and applicant-added citations, we construct three proxies (Table~\ref{tab:timing-map}): the ``narrow'' proxy used in the baseline analysis and the broader ``earliest'' and ``latest'' proxies used for sensitivity checks with broader timing bounds. The ``narrow'' proxy uses the earliest Information Disclosure Statement (IDS/SB08) or PTO-1449 date, the latter identifying a list of references cited by the applicant and considered by the examiner. 
For examiner-added citations, it uses the earliest PTO-892 list of references cited by the examiner and, when PTO-892 is absent, the earliest CTNF or CTFR office-action package, denoting a nonfinal or final rejection. 
The broader ``earliest'' and ``latest'' proxies, respectively, capture the earliest and latest administrative events at which a citation may have entered the application record through examiner or applicant actions. 



Because CMS dates are observed at the citing-application rather than the individual-reference level, eligible citation links associated with the same application inherit the corresponding audience-specific proxy date.

\begin{table}[H]
\centering
\caption{Construction of citation-date proxies}
\label{tab:timing-map}
\small
\begin{tabular}{
p{0.11\textwidth}
p{0.14\textwidth}
p{0.30\textwidth}
p{0.35\textwidth}}
\toprule
Audience & Proxy & Eligible CMS codes & Temporal rule \\
\midrule

Applicant &
Narrow (base) &
\texttt{IDS}, \texttt{1449} &
Earliest \texttt{mailroom\_date} among the two codes. \\

Applicant &
Earliest &
\texttt{IDS}, \texttt{1449}, \texttt{FOR},
\texttt{REF.OTHER}, \texttt{NTC.IDS.CONS} &
Earliest \texttt{mailroom\_date} among all five codes. \\

Applicant &
Latest &
\texttt{IDS}, \texttt{1449}, \texttt{FOR},
\texttt{REF.OTHER}, \texttt{NTC.IDS.CONS} &
Latest \texttt{mailroom\_date} among all five codes that is no later
than the patent issue date, where an issue date is available; otherwise
the latest observed eligible date. \\

\midrule

Examiner &
Narrow (base) &
Primary: \texttt{892}; fallback: \texttt{CTNF}, \texttt{CTFR} &
Use the earliest \texttt{892} date whenever any \texttt{892} record
exists for the citing application. Only when no \texttt{892} is observed,
use the earliest \texttt{CTNF}/\texttt{CTFR} date. \\

Examiner &
Earliest &
\texttt{892}, \texttt{CTNF}, \texttt{CTFR},
\texttt{NTC.CITE.IMP} &
Earliest \texttt{mailroom\_date} among all four codes. \\

Examiner &
Latest &
\texttt{892}, \texttt{CTNF}, \texttt{CTFR} &
Latest eligible \texttt{mailroom\_date} no later than the patent issue
date, where an issue date is available; otherwise the latest observed
eligible date. \\

\bottomrule
\end{tabular}

\begin{minipage}{0.92\textwidth}
\footnotesize
\textit{Notes:}
All dates are document-event proxies for the citing application. \texttt{IDS} is an Information Disclosure Statement (SB08); \texttt{1449} is the legacy list of references cited by the applicant and considered by the examiner; \texttt{FOR} denotes a foreign-reference record; \texttt{REF.OTHER} denotes another patent, application, or search reference document; and \texttt{NTC.IDS.CONS} is a notice of consideration for imported citations. On the examiner side, \texttt{892} is the list of references cited by the examiner, \texttt{CTNF} and \texttt{CTFR} are non-final and final rejection packages, and \texttt{NTC.CITE.IMP} is a notice of imported citations. 
After these application-level proxy dates are linked to the PatentsView citation edges, we deduplicate citation events to one focal-patent--citing-patent pair per audience and proxy, retaining the earliest eligible event date.
\end{minipage}
\end{table}


\FloatBarrier

\subsection{Matching covariates}
\label{app:matching-covariates}

This section provides additional detail on the matching variables. The covariates address four sources of observable differences between treated and control patents: prior visibility and citation trajectories; invention scope and prior-art intensity; organizational and procedural characteristics; and the maturity of the technological domain.

Table~\ref{tab:covariate-definitions} documents the variables that enter the matching specifications and their construction. Patent characteristics and citation links are drawn primarily from PatentsView \citep{patentsview}; application-level entity, examination, and continuity information is obtained from PatEx \citep{USPTOPatEx}. Additional variables used in exploratory specifications are not listed.

\footnotesize
\begin{longtable}{
  p{0.19\textwidth}
  p{0.35\textwidth}
  p{0.17\textwidth}
  p{0.19\textwidth}}

\caption{Covariates used in the matching design}
\label{tab:covariate-definitions}\\

\toprule
Construct & Definition and transformation & Data source & Rationale \\
\midrule
\endfirsthead

\multicolumn{4}{c}{\tablename\ \thetable\ -- continued} \\
\toprule
Construct & Definition and transformation & Primary source & Matching rationale \\
\midrule
\endhead

Examiner citations &
Numbers of examiner-added citing patents in each of the three annual
pre-event periods; entered as $\log(1+x)$ levels, zero indicators, and
two adjacent-year changes in the logged counts. &
PatentsView citations + PatEx timing &
Prior visibility and trajectory of recorded examiner attention. \\

Applicant citations &
Numbers of applicant-added citing patents in each of the three annual
pre-event periods; entered as $\log(1+x)$ levels, zero indicators, and
two adjacent-year changes in the logged counts. &
PatentsView citations + PatEx timing &
Prior applicant-side visibility and citation trajectory. \\

Claims &
Number of claims, transformed by $\log(1+x)$; missing values are
median-imputed with a missingness indicator. &
PatentsView &
Scope and complexity of the focal invention. \\

Inventors &
Number of distinct inventors, transformed by $\log(1+x)$; missing
values are median-imputed with a missingness indicator. &
PatentsView &
Team size and organizational complexity of inventive activity. \\

Backward U.S. patent citations &
Number of backward references to U.S. patent grants,
transformed by $\log(1+x)$. Missing counts are coded as zero. &
PatentsView &
Prior-art intensity and embeddedness in patented knowledge. \\

Backward foreign-patent citations &
Number of backward references to foreign patents,
transformed by $\log(1+x)$. Missing counts are coded as zero. &
PatentsView &
International prior-art intensity. \\

Non-patent-literature references &
Number of backward references to non-patent literature, excluding
patent-office notices, transformed by $\log(1+x)$. Missing counts are
coded as zero. &
PatentsView &
Embeddedness in scientific and other non-patent knowledge. \\

Small/micro entity &
Indicator for small- or micro-entity status. &
PatEx &
Organizational resources and applicant characteristics. \\

U.S. assignee &
Indicator equal to one when the focal patent's assignee country is the
United States. &
PatentsView &
Geographic and institutional location of the assignee. \\

Continuation &
Indicator that the application is recorded as a continuation. &
PatEx continuity records &
Application lineage and prior procedural history. \\

Foreign priority and priority age &
Categorical measure distinguishing no foreign priority, unavailable or
invalid priority age, and foreign-priority ages of 0--3, 4--12,
13--24, and more than 24 months. &
PatentsView priority records &
Prior application history and time over which related knowledge may
have entered prosecution records. \\

Technological maturity &
Cumulative number of patents published before the event in the focal
patent's retained USPC subclass; transformed by $\log(1+x)$ and
standardized within the event data. Missing values are median-imputed
with a missingness indicator. &
Derived from patent publication records and retained USPC subclass &
Density of existing knowledge and baseline search opportunities in the
technological domain. \\

\bottomrule
\end{longtable}

\noindent\footnotesize

All variables are measured before the focal classification event. The propensity-score model uses the complete set above. For the three-year citation histories, Mahalanobis distance additionally prioritizes the annual logged levels and adjacent-year logged changes. The outcome models retain the patent-level controls and annual logged citation levels, but do not separately re-enter the zero indicators or citation slopes; see Appendix~\ref{app:outcome-controls}.

\normalsize

\FloatBarrier
\section{Matching details}
\label{app:matching}
\subsection{Matching implementation}
\label{app:matching-implementation}

Exact technology--cohort matching is a binding restriction: treated and control patents must share the same exact cell, and cells without both treatment states do not contribute to the match. A candidate treatment specification is attempted only when the resulting candidate pool contains at least 25 treated patents and 50 controls overall; these thresholds do not apply separately to each exact cell. 

To select the exact technology--cohort cells for the headline analyses, we evaluate three definitions---year--AU4, year--TC2, and quarter--TC2---within the preferred green-control design and repeat these specifications with broad-patent controls as robustness checks. AU4 provides the finer examination-environment comparison, TC2 uses a coarser technology grouping, and quarter--TC2 uses a finer publication cohort. For 2013, the green-control year--AU4 design is the only design in the six-specification grid that satisfies the maximum-SMD balance criterion. For 2020, the green-control year--TC2 design satisfies the balance criterion across all four treatment branches while retaining greater common support than the finer AU4 specification. We therefore use these event-specific exact-cell definitions in the headline analyses and report the remaining combinations as robustness checks.

The primary procedure uses greedy one-to-one nearest-neighbor propensity-score matching without replacement, beginning with a standardized caliper of 0.10 and using Mahalanobis distance to prioritize similarity in the pre-event citation histories. If the maximum post-match absolute SMD remains above 0.10, the algorithm also evaluates a tighter 0.05 caliper. A wider 0.20 caliper is attempted only when the tighter alternatives fail the hard matched-support floors of at least 20 treated patents, 10 unique controls, and a control effective sample size of at least 10. If supported propensity-score matches remain unavailable, or if their best balance remains above 0.10, the procedure additionally evaluates exact-cell nearest-neighbor Mahalanobis matching without a propensity-score caliper. Among supported candidates, it retains the match with the lowest maximum absolute SMD. 
A pure-Mahalanobis fallback is not selected for any headline sample. The 2013 creation match and the 2020 addition, strict-addition, and cleanup matches use the initial 0.10 caliper, whereas the 2020 replacement match uses the tighter 0.05-caliper specification. Matching-sensitivity analyses vary these implementation choices separately. 

Before matching, we cap the number of eligible controls retained within each exact cell to limit computationally excessive candidate pools: the cap is 50 controls per treated patent for the creation and addition branches and 75 for the 2020 replacement and cleanup branches. When the eligible control pool exceeds the cap, controls are sampled using a fixed random seed. The matching-sensitivity analyses vary the caliper specification, the control-pool cap, and the random seed to assess dependence on these implementation choices.

For covariate $k$, balance is summarized by
\begin{equation}
 \SMD_k=
 \frac{\bar X_{Tk}-\bar X_{Ck}}
 {\sqrt{(s_{Tk}^{2}+s_{Ck}^{2})/2}},
\end{equation}
where $\bar X_{Tk}$ and $\bar X_{Ck}$ denote the weighted treated and control means and $s_{Tk}^{2}$ and $s_{Ck}^{2}$ their corresponding weighted variances. We use a maximum absolute SMD below 0.10 across the diagnostic variables as the headline balance criterion.

\subsection{Matching diagnostics}
\label{app:matching-diagnostics}

\setcounter{table}{0}
\renewcommand{\thetable}{C\arabic{table}}
\setcounter{figure}{0}
\renewcommand{\thefigure}{C\arabic{figure}}

Section~\ref{subsec:matching-results} reports sample retention and the maximum post-match SMD for each headline treatment. This section expands that audit. Table~\ref{tab:outcome-descriptives} first describes the raw citation outcomes in the final matched samples, while Table~\ref{tab:smd-full} and Figures~\ref{fig:smd-creation}--\ref{fig:smd-cleanup} report balance for the complete set of matching variables.
\FloatBarrier

\subsection{Covariate balance}
\label{app:balance-diagnostics}

\IfFileExists{generated/tables/table_full_smd_after.tex}{\scriptsize
\setlength{\tabcolsep}{3pt}
\begin{ThreePartTable}
\begin{TableNotes}[flushleft]\footnotesize
\item[]
Entries are absolute standardized mean differences after matching. The design criterion is $|\mathrm{SMD}|<0.10$. Each column reports one event-specific treatment definition.
\end{TableNotes}
\begin{longtable}{lccccc}
\caption{Covariate-level post-match standardized mean differences}\label{tab:smd-full}\\
\toprule
 & \multicolumn{1}{c}{2013} & \multicolumn{4}{c}{2020} \\
\cmidrule(lr){2-2}\cmidrule(lr){3-6}
Matching variable & Creation & Addition & \shortstack{Strict\\addition} & Replacement & Cleanup \\
\midrule
\endfirsthead
\multicolumn{6}{c}{\tablename\ \thetable\ -- continued} \\
\toprule
 & \multicolumn{1}{c}{2013} & \multicolumn{4}{c}{2020} \\
\cmidrule(lr){2-2}\cmidrule(lr){3-6}
Matching variable & Creation & Addition & \shortstack{Strict\\addition} & Replacement & Cleanup \\
\midrule
\endhead
\midrule
\multicolumn{6}{r}{\footnotesize Continued on next page} \\
\endfoot
\bottomrule
\insertTableNotes
\endlastfoot
Applicant citations, pre-year 1 (log1p) & 0.082 & 0.067 & 0.097 & 0.073 & 0.022 \\
No applicant citation, pre-year 1 & 0.070 & 0.074 & 0.095 & 0.044 & 0.017 \\
Priority age: no priority & 0.047 & 0.025 & 0.018 & 0.018 & 0.095 \\
Examiner citations, pre-year 3 (raw) & 0.089 & 0.051 & 0.063 & 0.035 & 0.013 \\
Applicant citations, pre-year 2 (log1p) & 0.087 & 0.050 & 0.083 & 0.076 & 0.028 \\
Applicant citations, pre-year 1 (raw) & 0.087 & 0.029 & 0.049 & 0.085 & 0.018 \\
Non-patent literature citations (log1p) & 0.068 & 0.003 & 0.000 & 0.087 & 0.041 \\
Examiner citations, pre-year 3 (log1p) & 0.086 & 0.050 & 0.063 & 0.032 & 0.013 \\
Examiner citations, pre-year 1 (log1p) & 0.084 & 0.046 & 0.086 & 0.051 & 0.004 \\
Examiner citations, pre-year 1 (raw) & 0.085 & 0.048 & 0.085 & 0.051 & 0.004 \\
US assignee & 0.010 & 0.034 & 0.036 & 0.031 & 0.083 \\
No examiner citation, pre-year 1 & 0.076 & 0.041 & 0.082 & 0.049 & 0.003 \\
No applicant citation, pre-year 3 & 0.067 & 0.051 & 0.082 & 0.017 & 0.025 \\
Applicant citations, pre-year 3 (log1p) & 0.081 & 0.046 & 0.075 & 0.047 & 0.035 \\
Backward US patent citations (log1p) & 0.080 & 0.028 & 0.041 & 0.080 & 0.010 \\
No applicant citation, pre-year 2 & 0.078 & 0.052 & 0.077 & 0.061 & 0.021 \\
Examiner citations, pre-year 2 (log1p) & 0.077 & 0.037 & 0.070 & 0.058 & 0.009 \\
Applicant citations, pre-year 2 (raw) & 0.075 & 0.030 & 0.045 & 0.076 & 0.014 \\
No examiner citation, pre-year 3 & 0.075 & 0.048 & 0.059 & 0.028 & 0.010 \\
Technology stock (standardized log) & 0.062 & 0.060 & 0.074 & 0.002 & 0.040 \\
Examiner citations, pre-year 2 (raw) & 0.072 & 0.035 & 0.067 & 0.060 & 0.010 \\
No examiner citation, pre-year 2 & 0.070 & 0.037 & 0.071 & 0.056 & 0.008 \\
Inventors (log1p) & 0.071 & 0.030 & 0.057 & 0.070 & 0.013 \\
Applicant citations, pre-year 3 (raw) & 0.069 & 0.016 & 0.026 & 0.054 & 0.005 \\
Priority age: over 24 months & 0.015 & 0.046 & 0.069 & 0.053 & 0.062 \\
Continuation application & 0.019 & 0.009 & 0.008 & 0.057 & 0.069 \\
Priority age: 4--12 months & 0.031 & 0.019 & 0.055 & 0.068 & 0.053 \\
Backward foreign patent citations (log1p) & 0.047 & 0.034 & 0.027 & 0.063 & 0.063 \\
Small or micro entity & 0.059 & 0.051 & 0.035 & 0.026 & 0.048 \\
Claims (log1p) & 0.008 & 0.030 & 0.048 & 0.012 & 0.000 \\
Priority age: 0--3 months & 0.002 & 0.004 & 0.010 & 0.035 & 0.025 \\
Priority age: 13--24 months & 0.033 & 0.016 & 0.013 & 0.030 & 0.012 \\
Applicant citation slope, years 2--3 & 0.027 & 0.001 & 0.004 & 0.032 & 0.006 \\
Examiner citation slope, years 2--3 & 0.003 & 0.014 & 0.001 & 0.026 & 0.003 \\
Applicant citation slope, years 1--2 & 0.011 & 0.023 & 0.022 & 0.009 & 0.006 \\
Examiner citation slope, years 1--2 & 0.010 & 0.006 & 0.013 & 0.003 & 0.004 \\
\end{longtable}
\end{ThreePartTable}
\normalsize
}{\placeholder{Run the R script to generate the full SMD table.}}

\balancefigure
  {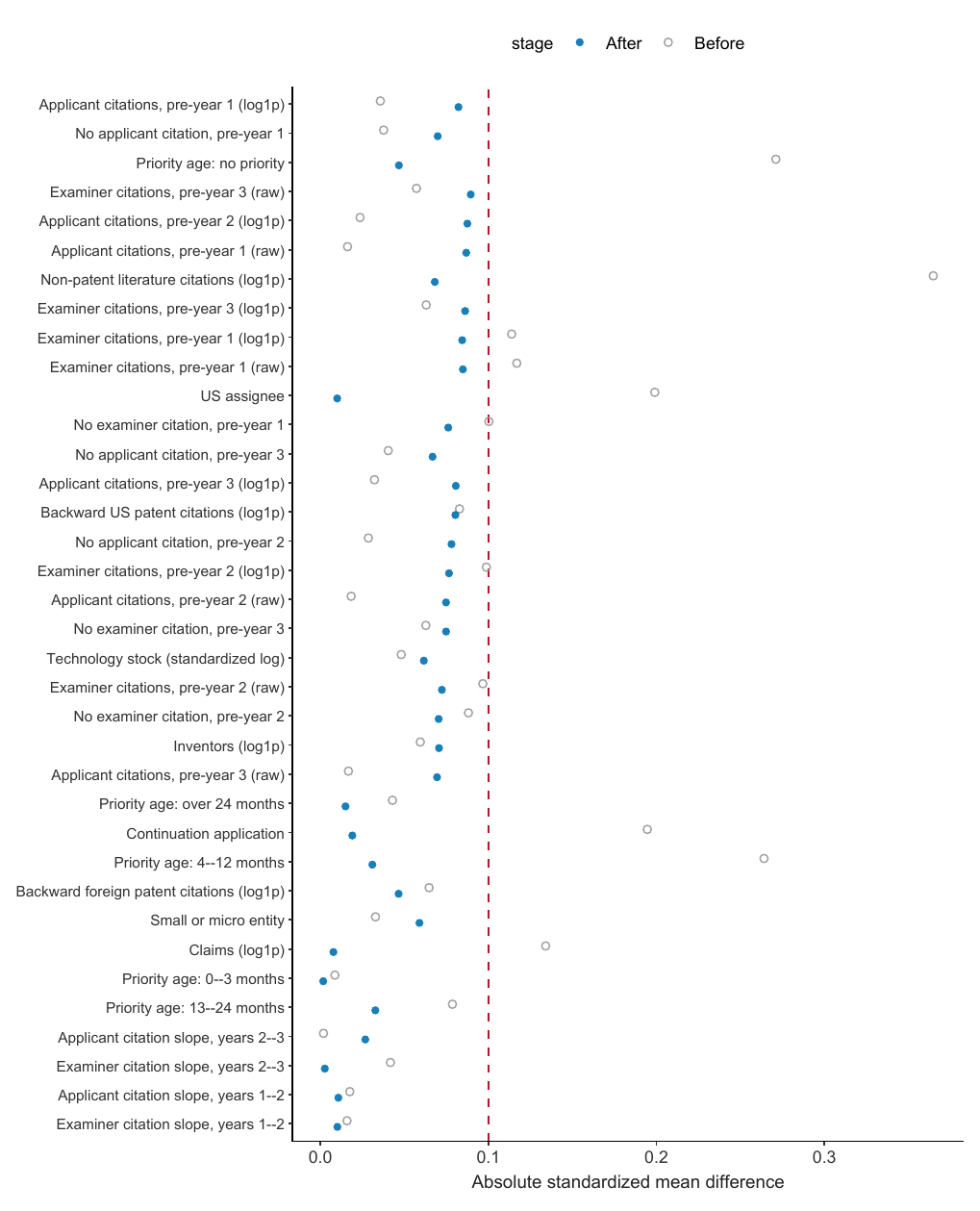}
  {the 2013 category-creation treatment}
  {fig:smd-creation}

\balancefigure
  {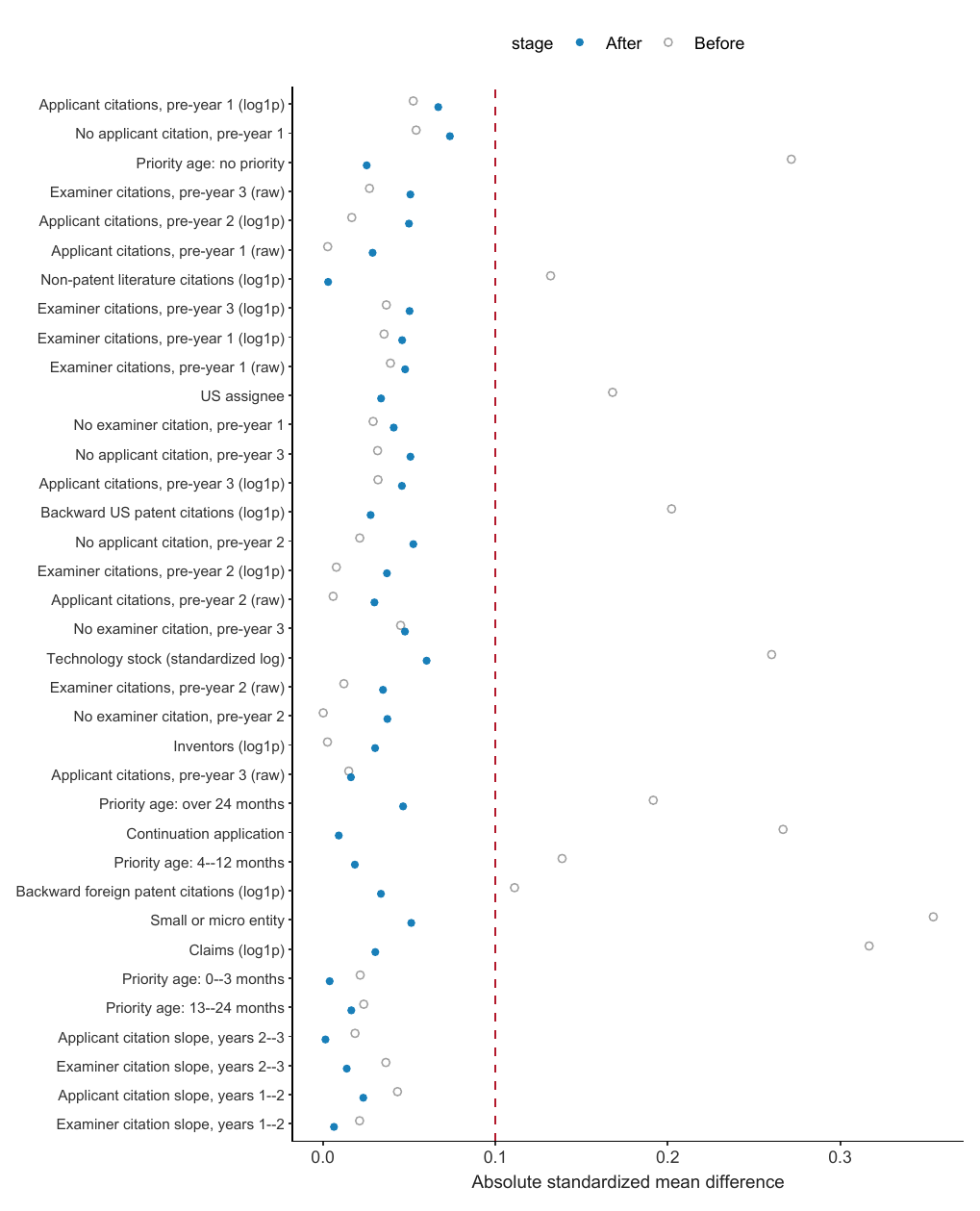}
  {the 2020 addition treatment}
  {fig:smd-addition}

\balancefigure
  {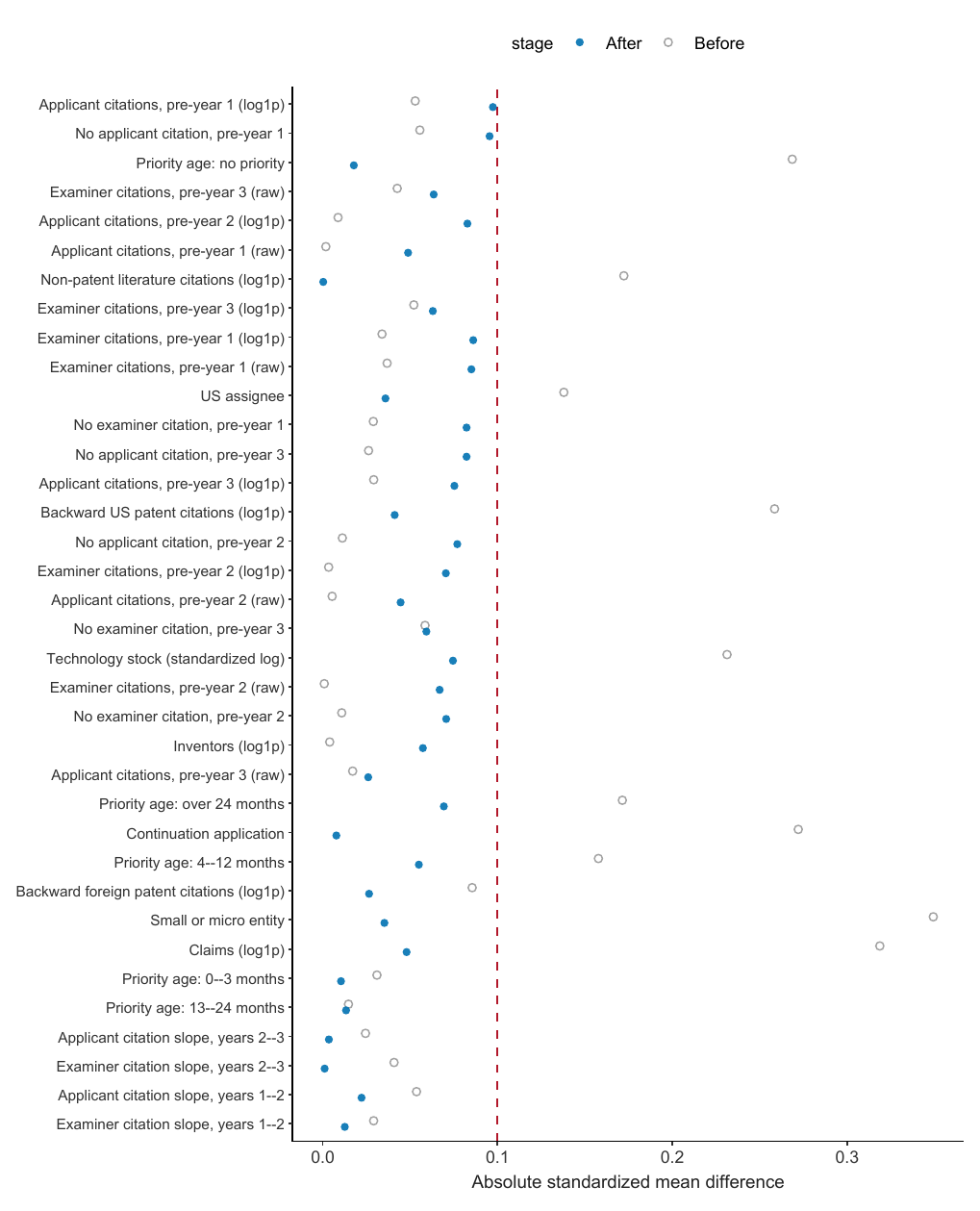}
  {the 2020 strict-addition treatment}
  {fig:smd-strict-addition}

\balancefigure
  {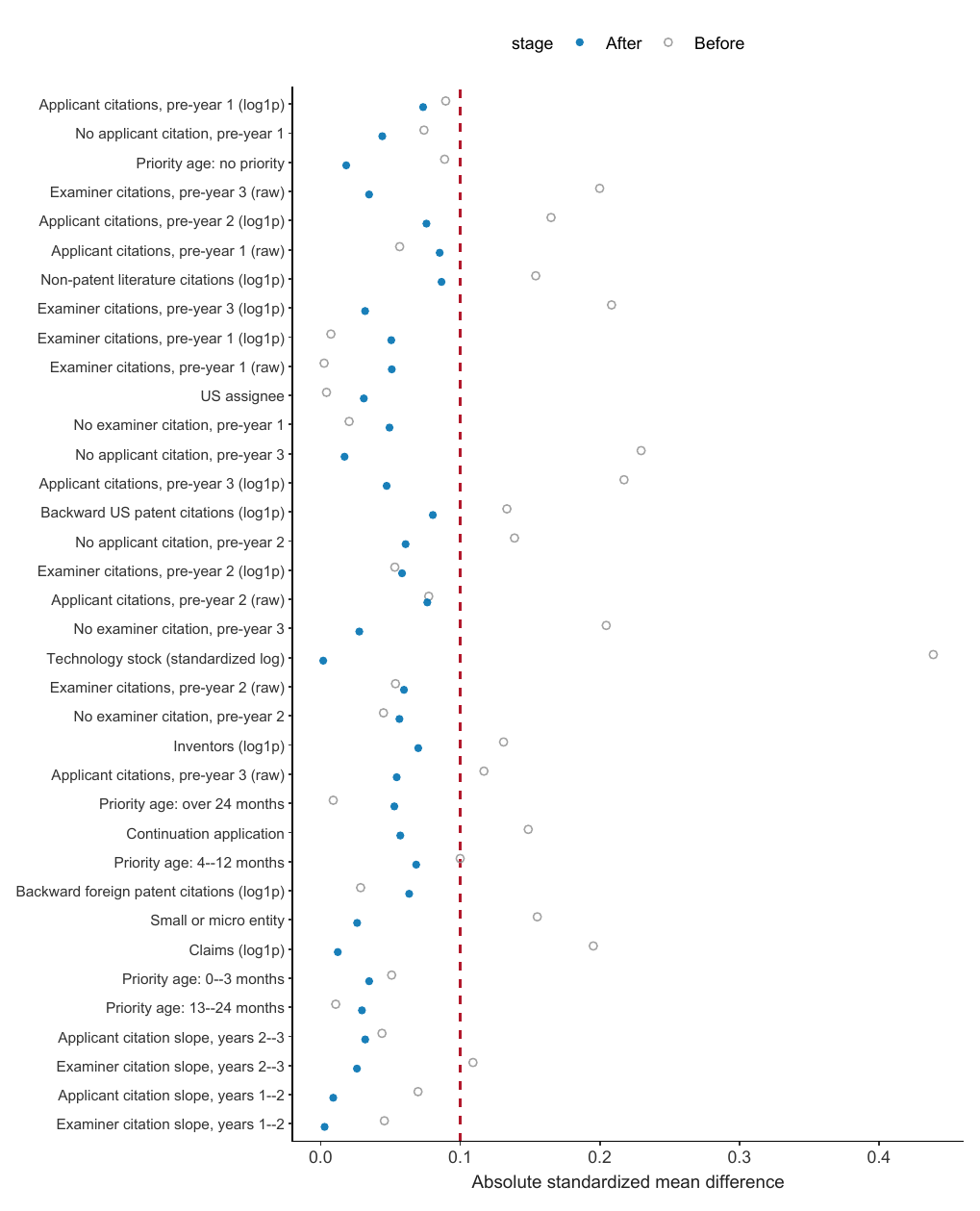}
  {the 2020 replacement treatment}
  {fig:smd-replacement}

\balancefigure
  {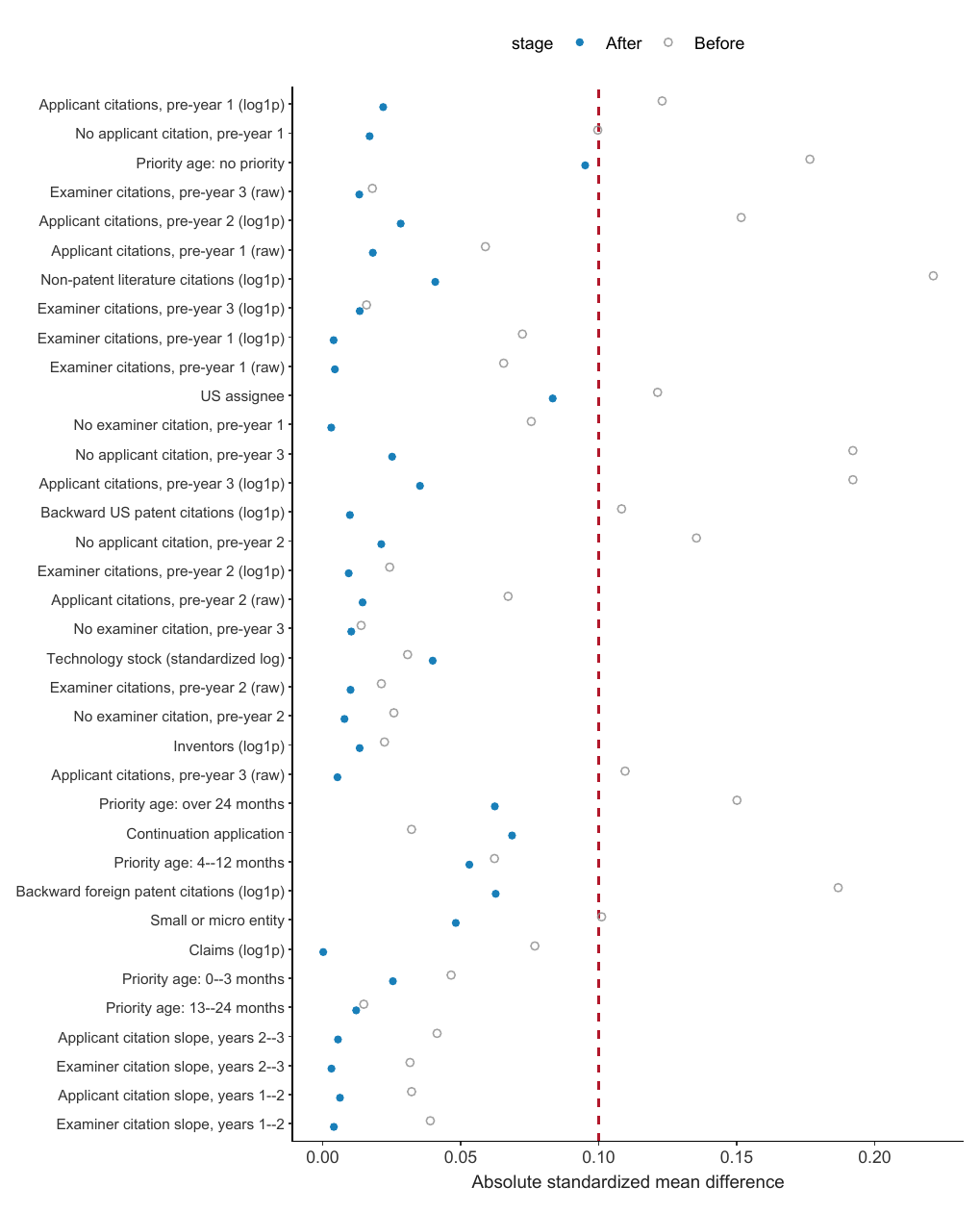}
  {the 2020 cleanup treatment}
  {fig:smd-cleanup}

\subsection{Matched-sample descriptives}
\label{app:outcome-descriptives}

Table~\ref{tab:outcome-descriptives} reports means, standard deviations, and zero shares for examiner- and applicant-added citation counts in the pre- and post-event windows of the final matched samples. These are unadjusted descriptive statistics; the adjusted citation-volume effects are estimated by the models described in Section~\ref{subsec:citation-volume}. 
Because fixed-effect Poisson estimation drops matched-pair strata with no positive counts for the analyzed audience and window, the reported number of patent-period observations can be smaller than twice the matched-patent count.

\IfFileExists{generated/tables/table_outcome_descriptives.tex}{
  \begin{table}[htbp]
\centering
\caption{Matched-sample citation-count descriptives}
\label{tab:outcome-descriptives}
\small
\begin{adjustbox}{max width=\linewidth,max totalheight=0.82\textheight,keepaspectratio,center}
\begin{threeparttable}
\begin{tabular}{llrrrrrr}
\toprule
Audience & Group & \shortstack{Pre\\mean} & \shortstack{Pre\\SD} & \shortstack{Post\\mean} & \shortstack{Post\\SD} & \shortstack{Zero\\pre} & \shortstack{Zero\\post} \\
\midrule
\multicolumn{8}{l}{\textit{Creation; matched pairs: 6,419}} \\[2pt]
Examiner & Control & 0.371 & 0.850 & 0.341 & 0.795 & 75.8\% & 76.7\% \\
 & Treated & 0.463 & 1.012 & 0.446 & 0.943 & 72.1\% & 71.9\% \\
Applicant & Control & 0.764 & 2.406 & 1.271 & 4.279 & 74.2\% & 66.8\% \\
 & Treated & 1.095 & 4.439 & 1.525 & 5.602 & 70.8\% & 63.2\% \\
\addlinespace
\multicolumn{8}{l}{\textit{Addition; matched pairs: 9,556}} \\[2pt]
Examiner & Control & 0.195 & 0.557 & 0.135 & 0.458 & 85.1\% & 89.3\% \\
 & Treated & 0.167 & 0.502 & 0.104 & 0.370 & 87.0\% & 91.2\% \\
Applicant & Control & 1.309 & 8.107 & 0.587 & 4.106 & 70.9\% & 83.1\% \\
 & Treated & 1.077 & 7.124 & 0.474 & 3.830 & 74.5\% & 85.8\% \\
\addlinespace
\multicolumn{8}{l}{\textit{Strict addition; matched pairs: 7,478}} \\[2pt]
Examiner & Control & 0.216 & 0.602 & 0.144 & 0.468 & 84.1\% & 88.7\% \\
 & Treated & 0.162 & 0.498 & 0.108 & 0.385 & 87.4\% & 90.9\% \\
Applicant & Control & 1.466 & 8.714 & 0.669 & 4.578 & 70.5\% & 82.3\% \\
 & Treated & 1.088 & 6.684 & 0.486 & 3.499 & 75.4\% & 85.8\% \\
\addlinespace
\multicolumn{8}{l}{\textit{Replacement; matched pairs: 2,336}} \\[2pt]
Examiner & Control & 0.134 & 0.411 & 0.133 & 0.471 & 88.6\% & 89.6\% \\
 & Treated & 0.167 & 0.473 & 0.130 & 0.446 & 86.1\% & 89.8\% \\
Applicant & Control & 0.362 & 1.238 & 0.242 & 1.170 & 82.8\% & 89.1\% \\
 & Treated & 0.539 & 2.328 & 0.323 & 2.298 & 80.2\% & 87.4\% \\
\addlinespace
\multicolumn{8}{l}{\textit{Cleanup; matched pairs: 46,420}} \\[2pt]
Examiner & Control & 0.223 & 0.590 & 0.152 & 0.471 & 83.4\% & 88.0\% \\
 & Treated & 0.228 & 0.596 & 0.175 & 0.508 & 83.0\% & 86.3\% \\
Applicant & Control & 0.779 & 12.428 & 0.466 & 8.073 & 80.5\% & 88.3\% \\
 & Treated & 0.619 & 5.187 & 0.347 & 3.406 & 79.5\% & 87.5\% \\
\addlinespace
\bottomrule
\end{tabular}
\begin{tablenotes}[flushleft]\footnotesize
\item[]
Statistics describe the matched samples used in the 24m count analyses. Control denotes matched control patents. Means and SDs summarize pair-deduplicated citations in the corresponding 24m windows; zero columns report the percentage without a citation. Each treatment heading reports the number of matched pairs.
\end{tablenotes}
\end{threeparttable}
\end{adjustbox}
\end{table}

}{\placeholder{Run the R script to generate the matched-outcome descriptive table.}}

\FloatBarrier

\FloatBarrier
\section{Estimator details}
\label{app:estimator-details}

\FloatBarrier
\subsection{Outcome-model control set}
\label{app:outcome-controls}

\setcounter{table}{0}
\renewcommand{\thetable}{D\arabic{table}}
\setcounter{figure}{0}
\renewcommand{\thefigure}{D\arabic{figure}}

The outcome models use the patent-level controls for claims, inventors, backward citations, entity and assignee status, continuation status, priority age, and technological maturity, together with the three annual logged examiner and applicant citation levels. Zero-count indicators and adjacent-year citation changes are used in matching but are not entered separately in the outcome models; the changes are deterministic linear combinations of the annual logged levels. Matched-pair strata provide the pair-specific baseline hazard in Equation~\ref{eq:cox}, and matched-pair fixed effects enter Equation~\ref{eq:did}. Because headline matching is one-to-one without replacement, every retained patent has unit matching weight.

\FloatBarrier
\subsection{Survival-model diagnostics}
\label{app:survival-diagnostics}

Section~\ref{subsec:search-efficiency} defines the headline Cox model. We assess the proportional-hazards assumption using treatment-specific and global Schoenfeld-residual tests \citep{grambschtherneau1994}. A small $p$-value provides evidence against a constant hazard ratio over the 24-month follow-up period; where proportional hazards are rejected, we treat the reported HR as a summary of an association that may vary over follow-up.

As a complementary measure that does not require proportional hazards, we report treated-minus-control differences in RMST, which is in our context the citation-free time through 12 and 24 months \citep{roystonparmar2013}. We obtain confidence intervals through 499 bootstrap replications that resample matched pairs. 

Table~\ref{tab:headline} reports the Schoenfeld tests and RMST contrasts for all headline models. A small $p$-value indicates evidence against a constant HR over the 24-month follow-up period. Where proportional hazards are rejected, we interpret the reported HR as an average contrast over follow-up.
Table~\ref{tab:headline} also reports treated-minus-control differences in RMST. 
A negative RMST difference indicates that treated patents spend fewer days citation-free and receive their first recorded citation sooner. A positive difference indicates later first citation.

\FloatBarrier

\subsection{Construction of the pre-event falsification tests}
\label{app:pretrend-construction}

The pre-event diagnostics use the same matched samples as used in the main analysis but reassign the analysis window to periods before the actual classification event. They are falsification tests: a rejection indicates evidence that treated and matched control patents differed before  treatment. 

\paragraph{Citation-count diagnostics.}
Let $T$ denote the classification-event date and let $Y_{i,-3}^{a}$, $Y_{i,-2}^{a}$, and $Y_{i,-1}^{a}$ denote the numbers of citations to patent $i$ by audience $a$ in the three mutually exclusive annual intervals
\[
[T-36,T-24),\qquad
[T-24,T-12),\qquad
[T-12,T),
\]
respectively, where endpoints are measured in months.

We first estimate two adjacent-year pseudo-policy DiDs. For each transition $r\rightarrow s\in\{(-3,-2),(-2,-1)\}$, we stack the two corresponding annual counts and estimate 
\[
E[Y_{it}^{a}\mid D_i,t]
=
\exp\!\left(
\alpha_i+\lambda_t+
\delta_{rs}^{a}
D_i\mathbf{1}\{t=s\}
\right),
\]
where $D_i$ indicates treatment, $\alpha_i$ is a patent fixed effect, and $\lambda_t$ is a common period effect. The null $H_0:\delta_{rs}^{a}=0$ asks whether the treated--control difference changed between the two adjacent pre-event years. We estimate the same specification linearly in raw citation counts as a functional-form check.

We separately stack all three pre-event years and use year $-3$ as the reference period:
\[
E[Y_{iy}^{a}\mid D_i,y]
=
\exp\!\left(
\alpha_i+\lambda_y+
\beta_{-2}^{a}D_i\mathbf{1}\{y=-2\}
+
\beta_{-1}^{a}D_i\mathbf{1}\{y=-1\}
\right).
\]
The coefficients $\beta_{-2}^{a}$ and $\beta_{-1}^{a}$ measure the treated--control differences in years $-2$ and $-1$ relative to year $-3$. A joint Wald test evaluates 
\[
H_0:\beta_{-2}^{a}=\beta_{-1}^{a}=0.
\]
Again, we estimate both Poisson and linear versions, obtaining ten count diagnostics with five Poisson tests---two adjacent-year contrasts, two annual coefficients, and one joint test---and the corresponding five linear tests.
Note that an adjacent-year contrast and the corresponding annual coefficient can represent the same underlying difference, and the Poisson and linear specifications use the same matched observations under different functional forms. 
We therefore consider which temporal contrast rejects, whether it is reproduced across estimators, and the joint test.

\paragraph{First-citation timing diagnostics.}
The Cox falsification exercises use four pre-event risk windows:
\[
\begin{aligned}
W_{36} &= [T-36,T),\\
W_{24} &= [T-24,T),\\
W_{12} &= [T-12,T),\\
W_{-4} &= [T-48,T-36).
\end{aligned}
\]
The first three windows are overlapping windows that all end immediately before the classification event. The fourth-year window is a separate, nonoverlapping interval covering months $-48$ through $-36$.

For each window, a patent enters the placebo risk set only if it was publicly observable by the beginning of the window. The follow-up citation-timing measurement starts at the window boundary, the event is the first eligible citation recorded before the end of the placebo window, and patents without a citation are censored at that endpoint. We estimate the same adjusted, matched-pair-stratified Cox specification used for the headline timing analysis, with the null hypothesis that the treated--control HR equals one.

The separate fourth-year test warrants a somewhat different interpretation. Headline matching conditions on citation histories during the 36 months immediately preceding treatment, whereas the fourth-year interval lies outside that matching-history window.
It is therefore the least mechanically conditioned of the timing placebos and can also have a smaller eligible sample because patents must already have been public 48 months before the event. A rejection confined to this remote interval is less direct evidence of differential citation dynamics immediately approaching treatment than a rejection in the 12- or 24-month windows. It nevertheless indicates imperfect pre-event comparability among older patents.

\subsection{Robustness specifications}
\label{app:robustness-specifications}

To assess whether the headline results depend on particular design, measurement, or estimator choices, we implement the following sensitivity analyses.

\begin{enumerate}[leftmargin=*]

\item \textit{Control groups, exact-cell definitions, and exclusion windows:}
We re-estimate the models across six combinations formed by crossing green versus broad-patent controls with year--AU4, year--TC2, and quarter--TC2 exact matching cells. We additionally use a stricter 12-month control-exclusion window, requiring controls to remain unaffected by any CPC revision during the 12 months before and after the focal event rather than the six-month window used in the headline design.

\item \textit{Outcome horizons and implementation lags:}
We estimate Cox models over 12-, 24-, and 36-month horizons and citation-count models over 12- and 24-month windows for both events, with an additional 36-month count window for 2013. To allow for delayed implementation of system-level CPC changes, we shift the beginning of post-event measurement by 3, 6, and 12 months for both events and by 24 months for 2013, without redefining the institutional event date. The Cox lag checks use a nominal 36-month risk horizon after the lag and the count checks use 24-month windows. Because PatEx ends in early 2023, longer and delayed 2020 Cox specifications are administratively censored at the data endpoint.

\item \textit{Citation-date measurement:}
We compare the baseline citation-date proxy with the alternative earliest and latest eligible prosecution-document proxies defined in Appendix~\ref{app:citation-timing}. The alternative-date Cox models use 36-month follow-up and the corresponding citation-count models use 24-month windows.

\item \textit{Count functional form:}
We compare the matched-pair fixed-effect Poisson DiD with linear DiD in raw citation counts, linear DiD in $\log(1+\text{citations})$, and a linear probability model indicating whether the patent receives at least one citation in the respective window. These specifications distinguish proportional changes in citation rates from additive changes, transformed citation intensity, and changes on the extensive margin.

\item \textit{Cited-record and CPC-source definitions:}
We vary whether treatment status is constructed from grant-level or pre-grant-publication-level CPC assignments and whether subsequent citations are observed as citations to the focal grant or to its pre-grant publication. Each combination is rematched before estimation. These checks use a common year--AU4 exact-cell design for both events and therefore evaluate measurement sensitivity rather than changing only one component of the headline 2020 design.

\item \textit{Matching implementation:}
We re-estimate the headline green-control designs under nine matching specifications. The fixed-caliper variants use propensity-score calipers of 0.05, 0.10, and 0.20 on the probability scale and a 0.20 caliper on the logit linear predictor. Additional specifications vary the maximum number of candidate controls retained per treated patent within an exact cell and the random seed used to sample controls when that cap binds. All specifications retain one-to-one matching without replacement and the same matching covariates.

\end{enumerate}

Across these exercises, the event-specific treatment logic and matching covariates are retained except where the purpose of the check is explicitly to vary their measurement or matching implementation. Specifications are rematched whenever the comparison group, exact-cell definition, or CPC source changes.

\subsection{Exploratory supplementary analyses}
\label{app:exploratory-methods}

\subsubsection{Breadth of examiner search}
\label{app:breadth-methods}

Using examiner-added citations in symmetric 24-month pre- and post-event windows, we measure absolute search reach by the numbers of distinct citing examiners, citing four-digit art units (AU4s), and citing two-digit technology centers (TC2s). We estimate these count outcomes using the same matched-pair fixed-effect Poisson DiD framework as for citation volume.

We additionally measure compositional reach as the share of examiner citations originating outside the focal patent's AU4 or TC2, among citations for which the corresponding unit identifiers are observed. For technology level $g\in\{\mathrm{AU4},\mathrm{TC2}\}$,
\[
S_{it}^{g}
=
\frac{\sum_{c\in C_{it}}\mathbf{1}\{g_c\neq g_i\}}
{\sum_{c\in C_{it}}\mathbf{1}\{g_c\ \text{and}\ g_i\ \text{observed}\}},
\]
where $C_{it}$ is the set of examiner citations to patent $i$ in period $t$. We estimate these shares using matched-pair fixed-effect linear DiD models, so coefficients are percentage-point changes. The share is observed only when at least one examiner citation in the relevant period has valid unit identifiers for both the citing and focal patents; requiring observations in both periods therefore produces smaller effective samples than the absolute-reach analyses. All breadth models use the headline matched samples, the headline outcome-model controls, and matched-pair-clustered standard errors.

\subsubsection{Moderator analysis}
\label{app:moderator-methods}

We explore heterogeneity in treatment effects using characteristics that capture the technological and classificatory search environment, organizational context, application lineage, and prior exposure to the green-classification infrastructure. The moderators are small- or micro-entity status, U.S.-assignee status, above-average technological maturity, broad CPC scope, child-application status, continuation status, and, for 2020, persistent membership in the 2013 green-code family. Above-average maturity indicates that the standardized log pre-event USPC-subclass patent stock exceeds zero, and broad CPC scope indicates assignment to at least three distinct CPC subclasses at issue. Persistent green membership requires historical membership in the 2013 launch family and continued assignment to a launch-family code in current CPC records.

The moderator analyses use the headline matched samples rather than rematching within subgroups. In the Cox models, treatment is interacted with the moderator and the exponentiated interaction is reported as a ratio of hazard ratios. In the citation-count models, the treatment-by-post-by-moderator interaction is exponentiated to obtain a ratio of incidence-rate ratios. These quantities test whether the treatment contrast differs between moderator groups.

Because these analyses involve large families of interaction tests, we report both conventional $p$-values and Benjamini--Hochberg false-discovery-rate-adjusted $q$-values \citep{benjaminihochberg1995}. Adjustments are performed separately by event and model family. For 2013, the timing family contains 12 estimable interactions; the corresponding count family contains 11 because the examiner broad-CPC-scope interaction is excluded as numerically unstable. For 2020, each model family contains 56 interactions.

\section{Supplementary results}
\label{app:supp-results}

\setcounter{table}{0}
\renewcommand{\thetable}{E\arabic{table}}
\setcounter{figure}{0}
\renewcommand{\thefigure}{E\arabic{figure}}

\subsection{Placebo tests and post-event estimates}
\label{app:placebos}

\begin{figure}[p]
  \centering
  \includegraphics[
    width=\linewidth,
    trim=1mm 1mm 1mm 1mm,
    clip
  ]{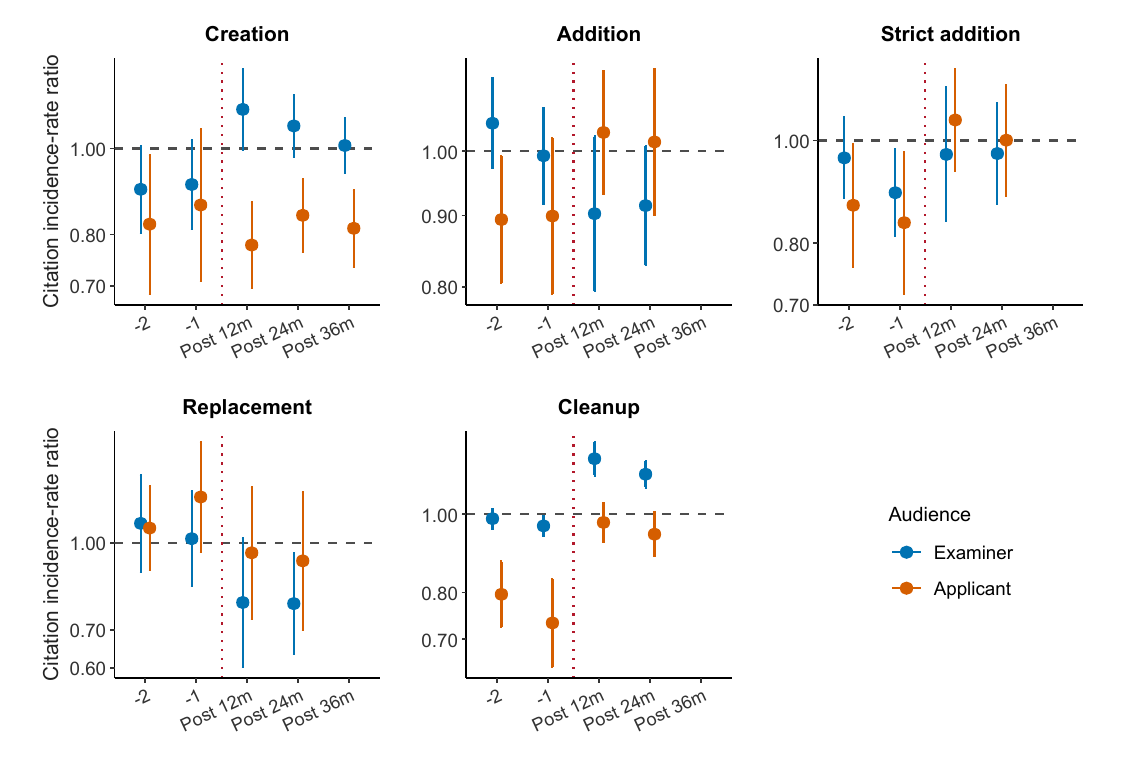}
  \caption{Estimated pre- and post-event effects on citation counts. The panels report exponentiated treatment-interaction coefficients from the matched-pair Poisson models. Pre-event coefficients compare annual periods with the omitted earliest pre-event period; post-event coefficients report effects over the available 12-, 24-, and 36-month windows. Points are estimates and whiskers are 95\% confidence intervals. The dashed line denotes an incidence-rate ratio of one, and the vertical line separates pre- and post-event estimates.}
  \label{fig:count-placebos-appendix}
\end{figure}

\begin{figure}[p]
  \centering
  \includegraphics[
    width=\linewidth,
    trim=1mm 1mm 1mm 1mm,
    clip
  ]{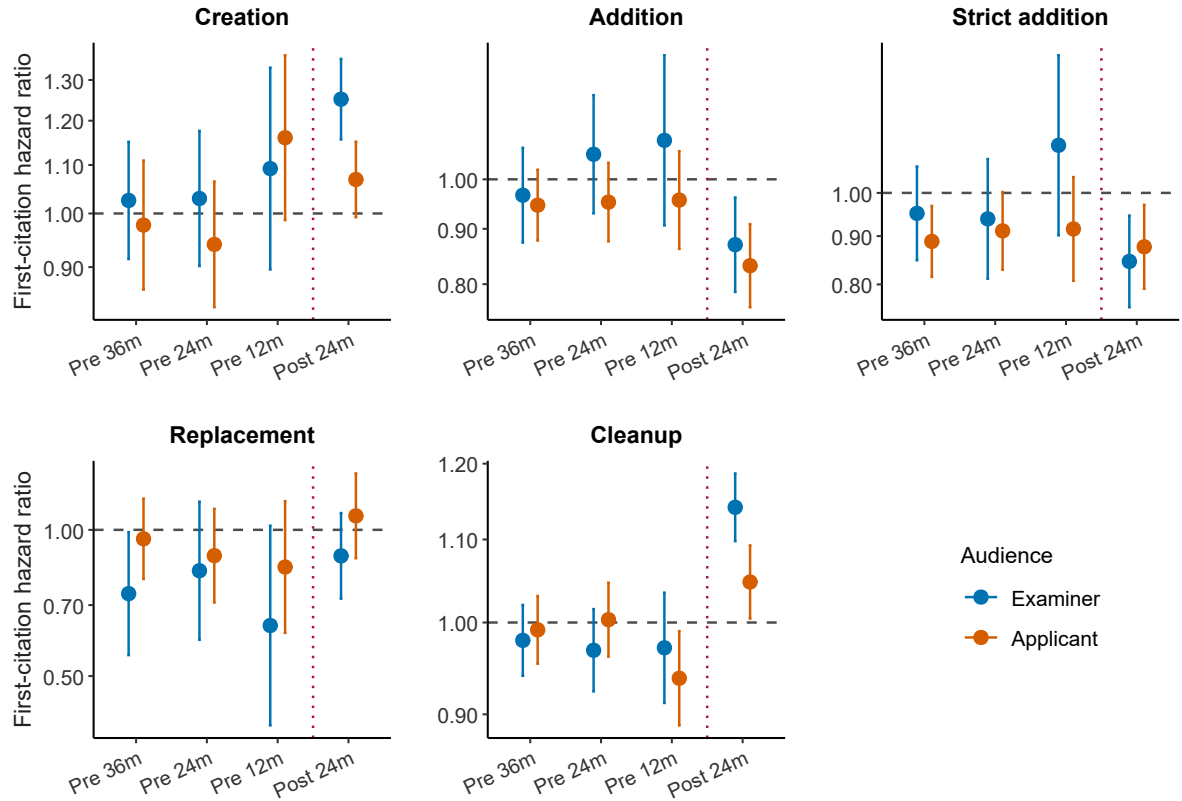}
  \caption{Pre-event timing-placebo estimates and the headline post-event estimate. Each treatment panel reports Cox HRs for the overlapping 36-, 24-, and 12-month pre-event risk windows and the 24-month post-event window. Points are estimates and whiskers are 95\% confidence intervals. The dashed horizontal line denotes one and the vertical line separates pre- and post-event estimates.}
  \label{fig:timing-placebos-appendix}
\end{figure}

Section~\ref{subsec:pretrend-results} summarizes the pre-event falsification evidence using minimum $p$-values and rejection counts. The tables below detail the underlying estimates. The first table shows the nearest pre-event count and timing tests alongside the corresponding 24-month post-event estimates. Tables~\ref{tab:count-pretrend-details} and~\ref{tab:timing-pretrend-details} report individual count and Cox falsification tests. Figures~\ref{fig:count-placebos-appendix} and~\ref{fig:timing-placebos-appendix} provide the corresponding visual summaries. 

\IfFileExists{generated/tables/table_headline_placebo_post24.tex}{
  \begin{table}[htbp]
\centering
\caption{Pre-event diagnostics and post-event estimates}
\label{tab:headline-placebo-post24}
\small
\begin{adjustbox}{max width=\linewidth,max totalheight=0.82\textheight,keepaspectratio,center}
\begin{threeparttable}
\begin{tabular}{lrrrrrrrr}
\toprule
\multicolumn{1}{c}{} & \multicolumn{4}{c}{Citation counts} & \multicolumn{4}{c}{Time to first citation} \\
\cmidrule(lr){2-5}\cmidrule(lr){6-9}
Audience & \shortstack{Placebo\\IRR} & $p$ & \shortstack{Post\\IRR} & $p$ & \shortstack{Pre 24m\\HR} & $p$ & \shortstack{Post 24m\\HR} & $p$ \\
\midrule
\multicolumn{9}{l}{\textit{Creation}} \\[2pt]
Examiner & $1.01$ & $.805$ & $1.05$ & $.233$ & $1.03$ & $.663$ & $1.25^{***}$ & $<.001$ \\
 & $(0.05)$ &  & $(0.04)$ &  & $(0.07)$ &  & $(0.05)$ &  \\
Applicant & $1.05$ & $.442$ & $0.84^{***}$ & $<.001$ & $0.94$ & $.335$ & $1.07^{*}$ & $.078$ \\
 & $(0.07)$ &  & $(0.04)$ &  & $(0.06)$ &  & $(0.04)$ &  \\
\addlinespace
\multicolumn{9}{l}{\textit{Addition}} \\[2pt]
Examiner & $0.95$ & $.179$ & $0.89^{**}$ & $.026$ & $1.05$ & $.406$ & $0.87^{***}$ & $.007$ \\
 & $(0.04)$ &  & $(0.04)$ &  & $(0.07)$ &  & $(0.04)$ &  \\
Applicant & $1.01$ & $.900$ & $0.98$ & $.751$ & $0.95$ & $.255$ & $0.83^{***}$ & $<.001$ \\
 & $(0.05)$ &  & $(0.06)$ &  & $(0.04)$ &  & $(0.04)$ &  \\
\addlinespace
\multicolumn{9}{l}{\textit{Strict addition}} \\[2pt]
Examiner & $0.93$ & $.133$ & $1.00$ & $.991$ & $0.94$ & $.395$ & $0.85^{***}$ & $.003$ \\
 & $(0.05)$ &  & $(0.06)$ &  & $(0.07)$ &  & $(0.05)$ &  \\
Applicant & $0.96$ & $.464$ & $0.98$ & $.723$ & $0.91^{*}$ & $.055$ & $0.88^{**}$ & $.012$ \\
 & $(0.05)$ &  & $(0.06)$ &  & $(0.04)$ &  & $(0.05)$ &  \\
\addlinespace
\multicolumn{9}{l}{\textit{Replacement}} \\[2pt]
Examiner & $0.94$ & $.524$ & $0.78^{**}$ & $.022$ & $0.82$ & $.246$ & $0.88$ & $.233$ \\
 & $(0.09)$ &  & $(0.08)$ &  & $(0.14)$ &  & $(0.09)$ &  \\
Applicant & $1.14$ & $.221$ & $0.90$ & $.420$ & $0.88$ & $.281$ & $1.07$ & $.518$ \\
 & $(0.12)$ &  & $(0.12)$ &  & $(0.10)$ &  & $(0.11)$ &  \\
\addlinespace
\multicolumn{9}{l}{\textit{Cleanup}} \\[2pt]
Examiner & $0.98$ & $.119$ & $1.12^{***}$ & $<.001$ & $0.97$ & $.186$ & $1.14^{***}$ & $<.001$ \\
 & $(0.01)$ &  & $(0.02)$ &  & $(0.02)$ &  & $(0.02)$ &  \\
Applicant & $0.92^{***}$ & $.009$ & $0.94^{**}$ & $.049$ & $1.00$ & $.879$ & $1.05^{**}$ & $.030$ \\
 & $(0.03)$ &  & $(0.03)$ &  & $(0.02)$ &  & $(0.02)$ &  \\
\addlinespace
\bottomrule
\end{tabular}
\begin{tablenotes}[flushleft]\footnotesize
\item[]
The first row for each audience reports the incidence-rate ratio or hazard ratio and its unadjusted two-sided $p$-value. The second row reports the corresponding delta-method standard error on the ratio scale in parentheses. The count placebo covers the transition from the middle to the nearest annual pre-event window and is not a 24-month placebo. The count post estimate compares symmetric 24-month pre- and post-event windows. The Cox estimates come from separate 24-month pre-event and post-event first-citation models. All estimates use the baseline matched samples and citations recorded to granted patents.
\end{tablenotes}
\end{threeparttable}
\end{adjustbox}
\end{table}

}{\placeholder{Run the R script to generate the headline placebo and post-event table.}}

\IfFileExists{generated/tables/table_count_pretrend_details.tex}{
  \begin{landscape}
\scriptsize
\setlength{\tabcolsep}{3pt}
\begin{ThreePartTable}
\begin{TableNotes}[flushleft]\footnotesize
\item[]
The Poisson coefficient rows report incidence-rate ratios (IRRs); linear rows report additive citation-count coefficients. The two pseudo-policy DiDs compare years -3 to -2 and -2 to -1. The annual coefficients use year -3 as the omitted pre-event period. The joint Wald row tests that both annual treatment coefficients equal zero. These are pre-event falsification tests rather than post-event treatment effects. Standard errors are clustered by matched pair. $^{***}p<.01$, $^{**}p<.05$, $^{*}p<.10$.
\end{TableNotes}
\begin{longtable}{lllclrrr}
\caption{Pre-event falsification tests for citation counts}\label{tab:count-pretrend-details}\\
\toprule
Audience & Estimator & Falsification test & Unit & Effect & 95\% CI & $p$ & $N$ \\
\midrule
\endfirsthead
\multicolumn{8}{c}{\tablename\ \thetable\ -- continued} \\
\toprule
Audience & Estimator & Falsification test & Unit & Effect & 95\% CI & $p$ & $N$ \\
\midrule
\endhead
\midrule
\multicolumn{8}{r}{\footnotesize Continued on next page} \\
\endfoot
\bottomrule
\insertTableNotes
\endlastfoot
\addlinespace\multicolumn{8}{l}{\textit{Creation}} \\[2pt]
Examiner & Poisson & Pseudo-policy DiD: year -3 to -2 & IRR & 0.900$^{*}$ & [0.804, 1.007] & $.065$ & 5,356 \\
Examiner & Poisson & Pseudo-policy DiD: year -2 to -1 & IRR & 1.013 & [0.917, 1.118] & $.805$ & 6,686 \\
Examiner & Poisson & Annual coefficient: year -2 vs. -3 & IRR & 0.900$^{*}$ & [0.804, 1.007] & $.065$ & 11,916 \\
Examiner & Poisson & Annual coefficient: year -1 vs. -3 & IRR & 0.911 & [0.811, 1.023] & $.115$ & 11,916 \\
Examiner & Poisson & Joint Wald test & Wald & --- & --- & $.158$ & 11,916 \\
Examiner & Linear & Pseudo-policy DiD: year -3 to -2 & Count & -0.002 & [-0.020, 0.016] & $.815$ & 25,676 \\
Examiner & Linear & Pseudo-policy DiD: year -2 to -1 & Count & 0.009 & [-0.011, 0.029] & $.383$ & 25,676 \\
Examiner & Linear & Annual coefficient: year -2 vs. -3 & Count & -0.002 & [-0.020, 0.016] & $.815$ & 38,514 \\
Examiner & Linear & Annual coefficient: year -1 vs. -3 & Count & 0.007 & [-0.013, 0.027] & $.504$ & 38,514 \\
Examiner & Linear & Joint Wald test & Wald & --- & --- & $.672$ & 38,514 \\
Applicant & Poisson & Pseudo-policy DiD: year -3 to -2 & IRR & 0.822$^{**}$ & [0.686, 0.984] & $.033$ & 5,144 \\
Applicant & Poisson & Pseudo-policy DiD: year -2 to -1 & IRR & 1.051 & [0.926, 1.193] & $.442$ & 7,058 \\
Applicant & Poisson & Annual coefficient: year -2 vs. -3 & IRR & 0.822$^{**}$ & [0.686, 0.984] & $.033$ & 11,598 \\
Applicant & Poisson & Annual coefficient: year -1 vs. -3 & IRR & 0.864 & [0.710, 1.051] & $.144$ & 11,598 \\
Applicant & Poisson & Joint Wald test & Wald & --- & --- & $.098$ & 11,598 \\
Applicant & Linear & Pseudo-policy DiD: year -3 to -2 & Count & 0.005 & [-0.044, 0.054] & $.848$ & 25,676 \\
Applicant & Linear & Pseudo-policy DiD: year -2 to -1 & Count & 0.093$^{***}$ & [0.027, 0.159] & $.006$ & 25,676 \\
Applicant & Linear & Annual coefficient: year -2 vs. -3 & Count & 0.005 & [-0.044, 0.054] & $.848$ & 38,514 \\
Applicant & Linear & Annual coefficient: year -1 vs. -3 & Count & 0.098$^{***}$ & [0.028, 0.168] & $.006$ & 38,514 \\
Applicant & Linear & Joint Wald test & Wald & --- & --- & $.013$ & 38,514 \\
\addlinespace\multicolumn{8}{l}{\textit{Addition}} \\[2pt]
Examiner & Poisson & Pseudo-policy DiD: year -3 to -2 & IRR & 1.047 & [0.973, 1.127] & $.219$ & 5,780 \\
Examiner & Poisson & Pseudo-policy DiD: year -2 to -1 & IRR & 0.948 & [0.877, 1.025] & $.179$ & 5,326 \\
Examiner & Poisson & Annual coefficient: year -2 vs. -3 & IRR & 1.047 & [0.973, 1.127] & $.219$ & 11,463 \\
Examiner & Poisson & Annual coefficient: year -1 vs. -3 & IRR & 0.993 & [0.917, 1.074] & $.856$ & 11,463 \\
Examiner & Poisson & Joint Wald test & Wald & --- & --- & $.320$ & 11,463 \\
Examiner & Linear & Pseudo-policy DiD: year -3 to -2 & Count & 0.008$^{**}$ & [0.000, 0.015] & $.046$ & 38,224 \\
Examiner & Linear & Pseudo-policy DiD: year -2 to -1 & Count & -0.004 & [-0.011, 0.003] & $.219$ & 38,224 \\
Examiner & Linear & Annual coefficient: year -2 vs. -3 & Count & 0.008$^{**}$ & [0.000, 0.015] & $.046$ & 57,336 \\
Examiner & Linear & Annual coefficient: year -1 vs. -3 & Count & 0.003 & [-0.005, 0.011] & $.409$ & 57,336 \\
Examiner & Linear & Joint Wald test & Wald & --- & --- & $.123$ & 57,336 \\
Applicant & Poisson & Pseudo-policy DiD: year -3 to -2 & IRR & 0.894$^{**}$ & [0.806, 0.992] & $.035$ & 10,680 \\
Applicant & Poisson & Pseudo-policy DiD: year -2 to -1 & IRR & 1.006 & [0.919, 1.100] & $.900$ & 10,422 \\
Applicant & Poisson & Annual coefficient: year -2 vs. -3 & IRR & 0.894$^{**}$ & [0.806, 0.992] & $.035$ & 18,996 \\
Applicant & Poisson & Annual coefficient: year -1 vs. -3 & IRR & 0.899 & [0.792, 1.022] & $.103$ & 18,996 \\
Applicant & Poisson & Joint Wald test & Wald & --- & --- & $.107$ & 18,996 \\
Applicant & Linear & Pseudo-policy DiD: year -3 to -2 & Count & -0.059$^{*}$ & [-0.120, 0.002] & $.059$ & 38,224 \\
Applicant & Linear & Pseudo-policy DiD: year -2 to -1 & Count & -0.004 & [-0.060, 0.053] & $.901$ & 38,224 \\
Applicant & Linear & Annual coefficient: year -2 vs. -3 & Count & -0.059$^{*}$ & [-0.120, 0.002] & $.059$ & 57,336 \\
Applicant & Linear & Annual coefficient: year -1 vs. -3 & Count & -0.062 & [-0.141, 0.016] & $.121$ & 57,336 \\
Applicant & Linear & Joint Wald test & Wald & --- & --- & $.160$ & 57,336 \\
\addlinespace\multicolumn{8}{l}{\textit{Strict addition}} \\[2pt]
Examiner & Poisson & Pseudo-policy DiD: year -3 to -2 & IRR & 0.963 & [0.882, 1.051] & $.399$ & 4,752 \\
Examiner & Poisson & Pseudo-policy DiD: year -2 to -1 & IRR & 0.927 & [0.840, 1.023] & $.133$ & 4,270 \\
Examiner & Poisson & Annual coefficient: year -2 vs. -3 & IRR & 0.963 & [0.882, 1.051] & $.399$ & 9,315 \\
Examiner & Poisson & Annual coefficient: year -1 vs. -3 & IRR & 0.893$^{**}$ & [0.813, 0.981] & $.018$ & 9,315 \\
Examiner & Poisson & Joint Wald test & Wald & --- & --- & $.061$ & 9,315 \\
Examiner & Linear & Pseudo-policy DiD: year -3 to -2 & Count & 0.003 & [-0.007, 0.012] & $.564$ & 29,912 \\
Examiner & Linear & Pseudo-policy DiD: year -2 to -1 & Count & -0.007 & [-0.016, 0.002] & $.120$ & 29,912 \\
Examiner & Linear & Annual coefficient: year -2 vs. -3 & Count & 0.003 & [-0.007, 0.012] & $.564$ & 44,868 \\
Examiner & Linear & Annual coefficient: year -1 vs. -3 & Count & -0.004 & [-0.014, 0.006] & $.390$ & 44,868 \\
Examiner & Linear & Joint Wald test & Wald & --- & --- & $.298$ & 44,868 \\
Applicant & Poisson & Pseudo-policy DiD: year -3 to -2 & IRR & 0.869$^{**}$ & [0.760, 0.993] & $.039$ & 8,350 \\
Applicant & Poisson & Pseudo-policy DiD: year -2 to -1 & IRR & 0.963 & [0.870, 1.066] & $.464$ & 8,098 \\
Applicant & Poisson & Annual coefficient: year -2 vs. -3 & IRR & 0.869$^{**}$ & [0.760, 0.993] & $.039$ & 14,757 \\
Applicant & Poisson & Annual coefficient: year -1 vs. -3 & IRR & 0.837$^{**}$ & [0.717, 0.976] & $.023$ & 14,757 \\
Applicant & Poisson & Joint Wald test & Wald & --- & --- & $.067$ & 14,757 \\
Applicant & Linear & Pseudo-policy DiD: year -3 to -2 & Count & -0.076$^{*}$ & [-0.161, 0.010] & $.083$ & 29,912 \\
Applicant & Linear & Pseudo-policy DiD: year -2 to -1 & Count & -0.035 & [-0.103, 0.033] & $.314$ & 29,912 \\
Applicant & Linear & Annual coefficient: year -2 vs. -3 & Count & -0.076$^{*}$ & [-0.161, 0.010] & $.083$ & 44,868 \\
Applicant & Linear & Annual coefficient: year -1 vs. -3 & Count & -0.110$^{**}$ & [-0.215, -0.006] & $.038$ & 44,868 \\
Applicant & Linear & Joint Wald test & Wald & --- & --- & $.112$ & 44,868 \\
\addlinespace\multicolumn{8}{l}{\textit{Replacement}} \\[2pt]
Examiner & Poisson & Pseudo-policy DiD: year -3 to -2 & IRR & 1.083 & [0.888, 1.321] & $.431$ & 920 \\
Examiner & Poisson & Pseudo-policy DiD: year -2 to -1 & IRR & 0.939 & [0.775, 1.139] & $.524$ & 1,182 \\
Examiner & Poisson & Annual coefficient: year -2 vs. -3 & IRR & 1.083 & [0.888, 1.320] & $.431$ & 2,208 \\
Examiner & Poisson & Annual coefficient: year -1 vs. -3 & IRR & 1.017 & [0.837, 1.236] & $.863$ & 2,208 \\
Examiner & Poisson & Joint Wald test & Wald & --- & --- & $.705$ & 2,208 \\
Examiner & Linear & Pseudo-policy DiD: year -3 to -2 & Count & 0.009 & [-0.002, 0.021] & $.111$ & 9,344 \\
Examiner & Linear & Pseudo-policy DiD: year -2 to -1 & Count & -0.002 & [-0.017, 0.013] & $.775$ & 9,344 \\
Examiner & Linear & Annual coefficient: year -2 vs. -3 & Count & 0.009 & [-0.002, 0.021] & $.111$ & 14,016 \\
Examiner & Linear & Annual coefficient: year -1 vs. -3 & Count & 0.007 & [-0.005, 0.020] & $.255$ & 14,016 \\
Examiner & Linear & Joint Wald test & Wald & --- & --- & $.212$ & 14,016 \\
Applicant & Poisson & Pseudo-policy DiD: year -3 to -2 & IRR & 1.062 & [0.896, 1.260] & $.486$ & 1,396 \\
Applicant & Poisson & Pseudo-policy DiD: year -2 to -1 & IRR & 1.135 & [0.927, 1.391] & $.221$ & 1,728 \\
Applicant & Poisson & Annual coefficient: year -2 vs. -3 & IRR & 1.062 & [0.896, 1.260] & $.486$ & 3,006 \\
Applicant & Poisson & Annual coefficient: year -1 vs. -3 & IRR & 1.206 & [0.963, 1.510] & $.102$ & 3,006 \\
Applicant & Poisson & Joint Wald test & Wald & --- & --- & $.261$ & 3,006 \\
Applicant & Linear & Pseudo-policy DiD: year -3 to -2 & Count & 0.019 & [-0.012, 0.050] & $.230$ & 9,344 \\
Applicant & Linear & Pseudo-policy DiD: year -2 to -1 & Count & 0.057$^{*}$ & [-0.002, 0.117] & $.059$ & 9,344 \\
Applicant & Linear & Annual coefficient: year -2 vs. -3 & Count & 0.019 & [-0.012, 0.050] & $.230$ & 14,016 \\
Applicant & Linear & Annual coefficient: year -1 vs. -3 & Count & 0.076$^{**}$ & [0.014, 0.138] & $.016$ & 14,016 \\
Applicant & Linear & Joint Wald test & Wald & --- & --- & $.051$ & 14,016 \\
\addlinespace\multicolumn{8}{l}{\textit{Cleanup}} \\[2pt]
Examiner & Poisson & Pseudo-policy DiD: year -3 to -2 & IRR & 0.988 & [0.960, 1.016] & $.388$ & 29,702 \\
Examiner & Poisson & Pseudo-policy DiD: year -2 to -1 & IRR & 0.980 & [0.955, 1.005] & $.119$ & 31,186 \\
Examiner & Poisson & Annual coefficient: year -2 vs. -3 & IRR & 0.988 & [0.960, 1.016] & $.388$ & 62,577 \\
Examiner & Poisson & Annual coefficient: year -1 vs. -3 & IRR & 0.967$^{**}$ & [0.941, 0.995] & $.019$ & 62,577 \\
Examiner & Poisson & Joint Wald test & Wald & --- & --- & $.055$ & 62,577 \\
Examiner & Linear & Pseudo-policy DiD: year -3 to -2 & Count & -0.001 & [-0.004, 0.002] & $.382$ & 185,680 \\
Examiner & Linear & Pseudo-policy DiD: year -2 to -1 & Count & -0.002 & [-0.005, 0.001] & $.169$ & 185,680 \\
Examiner & Linear & Annual coefficient: year -2 vs. -3 & Count & -0.001 & [-0.004, 0.002] & $.382$ & 278,520 \\
Examiner & Linear & Annual coefficient: year -1 vs. -3 & Count & -0.003$^{**}$ & [-0.007, -0.000] & $.032$ & 278,520 \\
Examiner & Linear & Joint Wald test & Wald & --- & --- & $.093$ & 278,520 \\
Applicant & Poisson & Pseudo-policy DiD: year -3 to -2 & IRR & 0.796$^{***}$ & [0.725, 0.874] & $<.001$ & 34,624 \\
Applicant & Poisson & Pseudo-policy DiD: year -2 to -1 & IRR & 0.922$^{***}$ & [0.867, 0.980] & $.009$ & 37,144 \\
Applicant & Poisson & Annual coefficient: year -2 vs. -3 & IRR & 0.796$^{***}$ & [0.725, 0.874] & $<.001$ & 67,998 \\
Applicant & Poisson & Annual coefficient: year -1 vs. -3 & IRR & 0.734$^{***}$ & [0.647, 0.831] & $<.001$ & 67,998 \\
Applicant & Poisson & Joint Wald test & Wald & --- & --- & $<.001$ & 67,998 \\
Applicant & Linear & Pseudo-policy DiD: year -3 to -2 & Count & -0.073$^{***}$ & [-0.110, -0.036] & $<.001$ & 185,680 \\
Applicant & Linear & Pseudo-policy DiD: year -2 to -1 & Count & -0.035$^{***}$ & [-0.060, -0.011] & $.005$ & 185,680 \\
Applicant & Linear & Annual coefficient: year -2 vs. -3 & Count & -0.073$^{***}$ & [-0.110, -0.036] & $<.001$ & 278,520 \\
Applicant & Linear & Annual coefficient: year -1 vs. -3 & Count & -0.108$^{***}$ & [-0.162, -0.055] & $<.001$ & 278,520 \\
Applicant & Linear & Joint Wald test & Wald & --- & --- & $<.001$ & 278,520 \\
\end{longtable}
\end{ThreePartTable}
\normalsize
\end{landscape}

}{\placeholder{Run the R script to generate the detailed count falsification table.}}

\IfFileExists{generated/tables/table_timing_pretrend_details.tex}{
  \scriptsize
\setlength{\tabcolsep}{3pt}
\begin{ThreePartTable}
\begin{TableNotes}[flushleft]\footnotesize
\item[]
Each row estimates the treated-versus-control first-citation hazard during a pre-event placebo risk window. The null is HR = 1. A rejection therefore indicates evidence of a treated--control difference in citation timing before the actual classification event. The 12-, 24-, and 36-month windows overlap; the separate year -4 test is nonoverlapping. Standard errors are clustered by matched pair.
\end{TableNotes}
\begin{longtable}{llrrrr}
\caption{Detailed pre-event Cox falsification tests}\label{tab:timing-pretrend-details}\\
\toprule
Audience & Pre-event window & HR & 95\% CI & $p$ & $N$ \\
\midrule
\endfirsthead
\multicolumn{6}{c}{\tablename\ \thetable\ -- continued} \\
\toprule
Audience & Pre-event window & HR & 95\% CI & $p$ & $N$ \\
\midrule
\endhead
\midrule
\multicolumn{6}{r}{\footnotesize Continued on next page} \\
\endfoot
\bottomrule
\insertTableNotes
\endlastfoot
\addlinespace\multicolumn{6}{l}{\textit{Creation}} \\[2pt]
Examiner & Separate year -4 & 0.976 & [0.845, 1.126] & $.736$ & 9,549 \\
Examiner & Prior 36 months & 1.026 & [0.914, 1.151] & $.662$ & 12,838 \\
Examiner & Prior 24 months & 1.030 & [0.902, 1.176] & $.663$ & 12,838 \\
Examiner & Prior 12 months & 1.092 & [0.896, 1.332] & $.383$ & 12,838 \\
Applicant & Separate year -4 & 1.128 & [0.913, 1.393] & $.264$ & 9,549 \\
Applicant & Prior 36 months & 0.977 & [0.861, 1.110] & $.725$ & 12,838 \\
Applicant & Prior 24 months & 0.941 & [0.832, 1.065] & $.335$ & 12,838 \\
Applicant & Prior 12 months & 1.161 & [0.987, 1.365] & $.071$ & 12,838 \\
\addlinespace\multicolumn{6}{l}{\textit{Addition}} \\[2pt]
Examiner & Separate year -4 & 0.997 & [0.905, 1.099] & $.954$ & 17,597 \\
Examiner & Prior 36 months & 0.967 & [0.874, 1.069] & $.511$ & 19,112 \\
Examiner & Prior 24 months & 1.055 & [0.930, 1.196] & $.406$ & 19,112 \\
Examiner & Prior 12 months & 1.086 & [0.907, 1.302] & $.369$ & 19,112 \\
Applicant & Separate year -4 & 1.062 & [0.980, 1.150] & $.142$ & 17,597 \\
Applicant & Prior 36 months & 0.946 & [0.878, 1.021] & $.153$ & 19,112 \\
Applicant & Prior 24 months & 0.953 & [0.876, 1.036] & $.255$ & 19,112 \\
Applicant & Prior 12 months & 0.957 & [0.862, 1.062] & $.406$ & 19,112 \\
\addlinespace\multicolumn{6}{l}{\textit{Strict addition}} \\[2pt]
Examiner & Separate year -4 & 0.995 & [0.889, 1.113] & $.929$ & 13,661 \\
Examiner & Prior 36 months & 0.951 & [0.848, 1.067] & $.393$ & 14,956 \\
Examiner & Prior 24 months & 0.939 & [0.811, 1.086] & $.395$ & 14,956 \\
Examiner & Prior 12 months & 1.123 & [0.902, 1.399] & $.300$ & 14,956 \\
Applicant & Separate year -4 & 0.954 & [0.868, 1.048] & $.329$ & 13,661 \\
Applicant & Prior 36 months & 0.888 & [0.815, 0.969] & $.007$ & 14,956 \\
Applicant & Prior 24 months & 0.912 & [0.829, 1.002] & $.055$ & 14,956 \\
Applicant & Prior 12 months & 0.916 & [0.807, 1.039] & $.173$ & 14,956 \\
\addlinespace\multicolumn{6}{l}{\textit{Replacement}} \\[2pt]
Examiner & Separate year -4 & 0.591 & [0.369, 0.944] & $.028$ & 3,278 \\
Examiner & Prior 36 months & 0.739 & [0.552, 0.989] & $.042$ & 4,672 \\
Examiner & Prior 24 months & 0.824 & [0.594, 1.143] & $.246$ & 4,672 \\
Examiner & Prior 12 months & 0.635 & [0.396, 1.020] & $.060$ & 4,672 \\
Applicant & Separate year -4 & 0.939 & [0.655, 1.348] & $.734$ & 3,278 \\
Applicant & Prior 36 months & 0.958 & [0.792, 1.159] & $.661$ & 4,672 \\
Applicant & Prior 24 months & 0.885 & [0.709, 1.105] & $.281$ & 4,672 \\
Applicant & Prior 12 months & 0.838 & [0.613, 1.145] & $.268$ & 4,672 \\
\addlinespace\multicolumn{6}{l}{\textit{Cleanup}} \\[2pt]
Examiner & Separate year -4 & 0.916 & [0.869, 0.966] & $.001$ & 74,884 \\
Examiner & Prior 36 months & 0.980 & [0.941, 1.020] & $.323$ & 92,840 \\
Examiner & Prior 24 months & 0.969 & [0.924, 1.016] & $.186$ & 92,840 \\
Examiner & Prior 12 months & 0.971 & [0.912, 1.035] & $.369$ & 92,840 \\
Applicant & Separate year -4 & 0.829 & [0.787, 0.872] & $<.001$ & 74,884 \\
Applicant & Prior 36 months & 0.992 & [0.954, 1.031] & $.672$ & 92,840 \\
Applicant & Prior 24 months & 1.003 & [0.962, 1.047] & $.879$ & 92,840 \\
Applicant & Prior 12 months & 0.938 & [0.889, 0.990] & $.020$ & 92,840 \\
\end{longtable}
\end{ThreePartTable}
\normalsize

}{\placeholder{Run the R script to generate the detailed timing falsification table.}}

\FloatBarrier

\begin{figure}[p]
  \centering
  \includegraphics[
    width=\linewidth,
    trim=1mm 1mm 1mm 1mm,
    clip
  ]{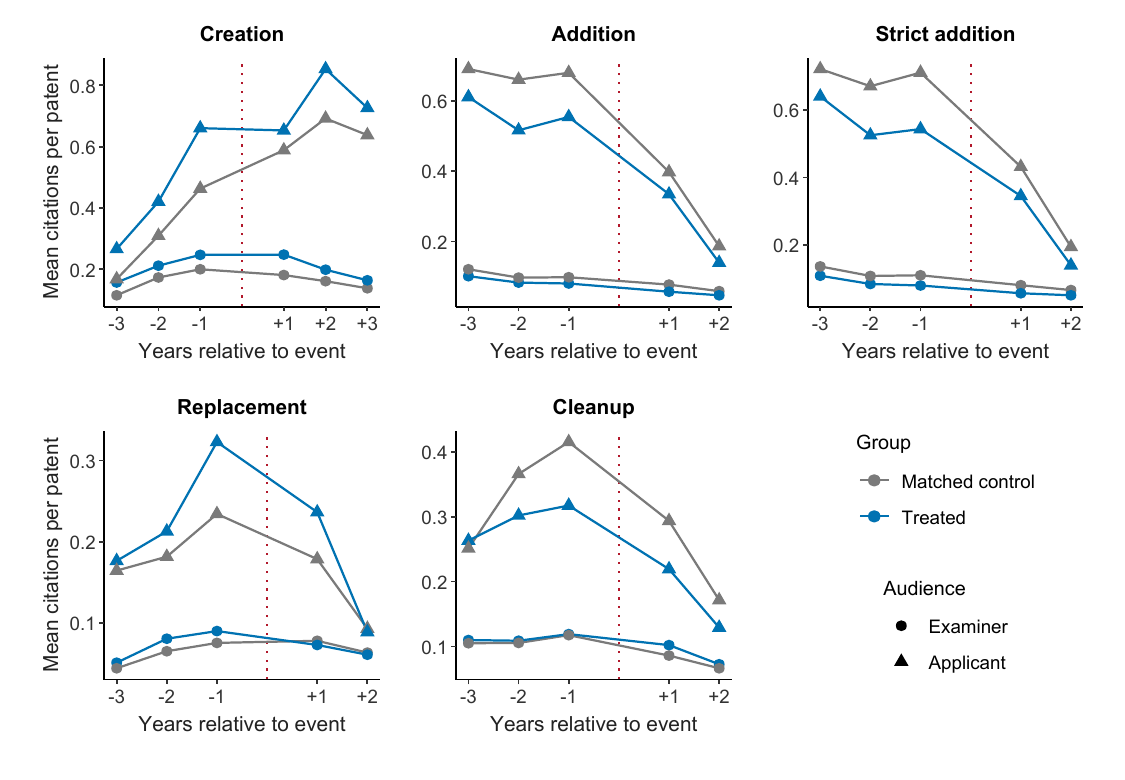}
  \caption{Annual citation counts before and after classification changes. The panels report unadjusted means for treated and matched control patents by audience in successive 12-month periods. The vertical line marks the classification event. These plots describe the matched data and are not adjusted treatment-effect estimates. Panel-specific vertical scales are used.}
  \label{fig:count-prepost-appendix}
\end{figure}

\FloatBarrier

\subsection{Full baseline-model coefficient estimates}
\label{app:full-coefficients}

Table~\ref{tab:headline} in Section~\ref{subsec:headline-results} reports the treatment effects and diagnostics presented as our main results. The following tables report the complete coefficient vectors from those same Cox and Poisson models, including the post-matching adjustment covariates. 

\IfFileExists{generated/tables/table_full_cox_coefficients.tex}{
  \begin{table}[p]
\centering
\caption{Full coefficient estimates: time to first citation}\label{tab:full-cox}
\scriptsize
\setlength{\tabcolsep}{1.8pt}
\renewcommand{\arraystretch}{0.94}
\begin{adjustbox}{max width=\linewidth,max totalheight=0.80\textheight,keepaspectratio,center}
\begin{threeparttable}
\begin{tabular}{l*{10}{c}}
\toprule
 & \multicolumn{2}{c}{\shortstack{2013\\Creation}} & \multicolumn{2}{c}{\shortstack{2020\\Addition}} & \multicolumn{2}{c}{\shortstack{2020\\Strict add.}} & \multicolumn{2}{c}{\shortstack{2020\\Replacement}} & \multicolumn{2}{c}{\shortstack{2020\\Cleanup}} \\
\cmidrule(lr){2-3}\cmidrule(lr){4-5}\cmidrule(lr){6-7}\cmidrule(lr){8-9}\cmidrule(lr){10-11}
Variable & Examiner & Applicant & Examiner & Applicant & Examiner & Applicant & Examiner & Applicant & Examiner & Applicant \\
\midrule
Treated & \shortstack{$0.225^{***}$\\{\scriptsize $(0.040)$}} & \shortstack{$0.067^{*}$\\{\scriptsize $(0.038)$}} & \shortstack{$-0.139^{***}$\\{\scriptsize $(0.051)$}} & \shortstack{$-0.184^{***}$\\{\scriptsize $(0.045)$}} & \shortstack{$-0.167^{***}$\\{\scriptsize $(0.057)$}} & \shortstack{$-0.131^{**}$\\{\scriptsize $(0.052)$}} & \shortstack{$-0.123$\\{\scriptsize $(0.104)$}} & \shortstack{$0.066$\\{\scriptsize $(0.103)$}} & \shortstack{$0.132^{***}$\\{\scriptsize $(0.020)$}} & \shortstack{$0.047^{**}$\\{\scriptsize $(0.021)$}} \\
Exam. cit., pre y1 & \shortstack{$0.375^{**}$\\{\scriptsize $(0.151)$}} & \shortstack{$0.419^{**}$\\{\scriptsize $(0.167)$}} & \shortstack{$1.102^{***}$\\{\scriptsize $(0.305)$}} & \shortstack{$0.637^{***}$\\{\scriptsize $(0.242)$}} & \shortstack{$0.361$\\{\scriptsize $(0.252)$}} & \shortstack{$1.008^{***}$\\{\scriptsize $(0.241)$}} & \shortstack{$1.643^{***}$\\{\scriptsize $(0.539)$}} & \shortstack{$1.476^{**}$\\{\scriptsize $(0.604)$}} & \shortstack{$1.205^{***}$\\{\scriptsize $(0.153)$}} & \shortstack{$0.740^{***}$\\{\scriptsize $(0.162)$}} \\
Exam. cit., pre y2 & \shortstack{$0.339^{**}$\\{\scriptsize $(0.137)$}} & \shortstack{$-0.057$\\{\scriptsize $(0.150)$}} & \shortstack{$0.864^{***}$\\{\scriptsize $(0.285)$}} & \shortstack{$-0.033$\\{\scriptsize $(0.238)$}} & \shortstack{$0.645^{**}$\\{\scriptsize $(0.263)$}} & \shortstack{$0.411$\\{\scriptsize $(0.256)$}} & \shortstack{$1.178^{*}$\\{\scriptsize $(0.637)$}} & \shortstack{$0.698$\\{\scriptsize $(0.588)$}} & \shortstack{$0.881^{***}$\\{\scriptsize $(0.127)$}} & \shortstack{$0.445^{***}$\\{\scriptsize $(0.132)$}} \\
Exam. cit., pre y3 & \shortstack{$0.263^{**}$\\{\scriptsize $(0.133)$}} & \shortstack{$0.038$\\{\scriptsize $(0.142)$}} & \shortstack{$0.792^{***}$\\{\scriptsize $(0.228)$}} & \shortstack{$0.193$\\{\scriptsize $(0.218)$}} & \shortstack{$0.635^{**}$\\{\scriptsize $(0.249)$}} & \shortstack{$0.238$\\{\scriptsize $(0.246)$}} & \shortstack{$-0.418$\\{\scriptsize $(1.053)$}} & \shortstack{$-1.984^{*}$\\{\scriptsize $(1.059)$}} & \shortstack{$0.508^{***}$\\{\scriptsize $(0.116)$}} & \shortstack{$0.072$\\{\scriptsize $(0.116)$}} \\
Appl. cit., pre y1 & \shortstack{$0.125$\\{\scriptsize $(0.094)$}} & \shortstack{$0.614^{***}$\\{\scriptsize $(0.106)$}} & \shortstack{$0.094$\\{\scriptsize $(0.147)$}} & \shortstack{$0.888^{***}$\\{\scriptsize $(0.129)$}} & \shortstack{$0.369^{**}$\\{\scriptsize $(0.160)$}} & \shortstack{$0.728^{***}$\\{\scriptsize $(0.138)$}} & \shortstack{$0.598^{*}$\\{\scriptsize $(0.338)$}} & \shortstack{$1.368^{***}$\\{\scriptsize $(0.346)$}} & \shortstack{$0.074$\\{\scriptsize $(0.075)$}} & \shortstack{$0.868^{***}$\\{\scriptsize $(0.074)$}} \\
Appl. cit., pre y2 & \shortstack{$0.149$\\{\scriptsize $(0.094)$}} & \shortstack{$0.621^{***}$\\{\scriptsize $(0.102)$}} & \shortstack{$-0.299^{*}$\\{\scriptsize $(0.156)$}} & \shortstack{$0.737^{***}$\\{\scriptsize $(0.124)$}} & \shortstack{$-0.267$\\{\scriptsize $(0.168)$}} & \shortstack{$0.812^{***}$\\{\scriptsize $(0.139)$}} & \shortstack{$-0.093$\\{\scriptsize $(0.427)$}} & \shortstack{$0.313$\\{\scriptsize $(0.434)$}} & \shortstack{$-0.153^{*}$\\{\scriptsize $(0.086)$}} & \shortstack{$0.395^{***}$\\{\scriptsize $(0.083)$}} \\
Appl. cit., pre y3 & \shortstack{$-0.258^{**}$\\{\scriptsize $(0.106)$}} & \shortstack{$0.258^{*}$\\{\scriptsize $(0.134)$}} & \shortstack{$0.443^{***}$\\{\scriptsize $(0.148)$}} & \shortstack{$-0.003$\\{\scriptsize $(0.114)$}} & \shortstack{$0.154$\\{\scriptsize $(0.156)$}} & \shortstack{$0.076$\\{\scriptsize $(0.122)$}} & \shortstack{$-0.276$\\{\scriptsize $(0.455)$}} & \shortstack{$-0.731$\\{\scriptsize $(0.541)$}} & \shortstack{$-0.353^{***}$\\{\scriptsize $(0.125)$}} & \shortstack{$0.024$\\{\scriptsize $(0.115)$}} \\
Claims & \shortstack{$0.219^{***}$\\{\scriptsize $(0.074)$}} & \shortstack{$0.373^{***}$\\{\scriptsize $(0.082)$}} & \shortstack{$0.491^{*}$\\{\scriptsize $(0.288)$}} & \shortstack{$0.154$\\{\scriptsize $(0.246)$}} & \shortstack{$0.924^{***}$\\{\scriptsize $(0.329)$}} & \shortstack{$-0.086$\\{\scriptsize $(0.303)$}} & \shortstack{$0.381$\\{\scriptsize $(0.345)$}} & \shortstack{$0.345$\\{\scriptsize $(0.328)$}} & \shortstack{$0.168^{***}$\\{\scriptsize $(0.028)$}} & \shortstack{$0.199^{***}$\\{\scriptsize $(0.031)$}} \\
Inventors & \shortstack{$0.023$\\{\scriptsize $(0.071)$}} & \shortstack{$0.088$\\{\scriptsize $(0.066)$}} & \shortstack{$-0.050$\\{\scriptsize $(0.139)$}} & \shortstack{$0.019$\\{\scriptsize $(0.119)$}} & \shortstack{$-0.224$\\{\scriptsize $(0.156)$}} & \shortstack{$0.116$\\{\scriptsize $(0.147)$}} & \shortstack{$0.242$\\{\scriptsize $(0.197)$}} & \shortstack{$-0.138$\\{\scriptsize $(0.196)$}} & \shortstack{$-0.042$\\{\scriptsize $(0.034)$}} & \shortstack{$-0.044$\\{\scriptsize $(0.037)$}} \\
Bwd. U.S. cit. & \shortstack{$-0.022$\\{\scriptsize $(0.054)$}} & \shortstack{$-0.049$\\{\scriptsize $(0.061)$}} & \shortstack{$-0.135$\\{\scriptsize $(0.203)$}} & \shortstack{$0.030$\\{\scriptsize $(0.174)$}} & \shortstack{$-0.515^{*}$\\{\scriptsize $(0.266)$}} & \shortstack{$0.166$\\{\scriptsize $(0.253)$}} & \shortstack{$-0.043$\\{\scriptsize $(0.176)$}} & \shortstack{$0.026$\\{\scriptsize $(0.176)$}} & \shortstack{$0.033$\\{\scriptsize $(0.025)$}} & \shortstack{$-0.033$\\{\scriptsize $(0.026)$}} \\
Bwd. foreign cit. & \shortstack{$-0.195^{**}$\\{\scriptsize $(0.093)$}} & \shortstack{$-0.219^{**}$\\{\scriptsize $(0.107)$}} & \shortstack{$0.046$\\{\scriptsize $(0.042)$}} & \shortstack{$-0.009$\\{\scriptsize $(0.037)$}} & \shortstack{$-0.003$\\{\scriptsize $(0.045)$}} & \shortstack{$-0.011$\\{\scriptsize $(0.041)$}} & \shortstack{$-0.085$\\{\scriptsize $(0.087)$}} & \shortstack{$-0.040$\\{\scriptsize $(0.093)$}} & \shortstack{$0.315^{***}$\\{\scriptsize $(0.091)$}} & \shortstack{$0.301^{***}$\\{\scriptsize $(0.090)$}} \\
NPL refs. & \shortstack{$0.371^{**}$\\{\scriptsize $(0.172)$}} & \shortstack{$0.422^{**}$\\{\scriptsize $(0.202)$}} & \shortstack{$0.120$\\{\scriptsize $(0.092)$}} & \shortstack{$0.025$\\{\scriptsize $(0.078)$}} & \shortstack{$0.375^{***}$\\{\scriptsize $(0.132)$}} & \shortstack{$-0.058$\\{\scriptsize $(0.125)$}} & \shortstack{$0.174$\\{\scriptsize $(0.144)$}} & \shortstack{$0.085$\\{\scriptsize $(0.138)$}} & \shortstack{$-0.228^{***}$\\{\scriptsize $(0.071)$}} & \shortstack{$-0.183^{***}$\\{\scriptsize $(0.069)$}} \\
Small/micro entity & \shortstack{$-0.195^{*}$\\{\scriptsize $(0.105)$}} & \shortstack{$-0.046$\\{\scriptsize $(0.112)$}} & \shortstack{$-0.556$\\{\scriptsize $(0.646)$}} & \shortstack{$0.020$\\{\scriptsize $(0.555)$}} & \shortstack{$-1.432^{**}$\\{\scriptsize $(0.695)$}} & \shortstack{$0.762$\\{\scriptsize $(0.663)$}} & \shortstack{$-0.286$\\{\scriptsize $(0.354)$}} & \shortstack{$-0.172$\\{\scriptsize $(0.346)$}} & \shortstack{$-0.038$\\{\scriptsize $(0.039)$}} & \shortstack{$0.089^{**}$\\{\scriptsize $(0.040)$}} \\
U.S. assignee & \shortstack{$-0.032$\\{\scriptsize $(0.076)$}} & \shortstack{$0.057$\\{\scriptsize $(0.071)$}} & \shortstack{$0.030$\\{\scriptsize $(0.123)$}} & \shortstack{$0.200^{*}$\\{\scriptsize $(0.109)$}} & \shortstack{$-0.220$\\{\scriptsize $(0.163)$}} & \shortstack{$0.299^{*}$\\{\scriptsize $(0.153)$}} & \shortstack{$-0.628^{*}$\\{\scriptsize $(0.327)$}} & \shortstack{$-0.224$\\{\scriptsize $(0.298)$}} & \shortstack{$0.020$\\{\scriptsize $(0.067)$}} & \shortstack{$0.329^{***}$\\{\scriptsize $(0.069)$}} \\
Continuation & \shortstack{$0.471^{*}$\\{\scriptsize $(0.269)$}} & \shortstack{$0.804^{**}$\\{\scriptsize $(0.314)$}} & \shortstack{$-0.096$\\{\scriptsize $(0.432)$}} & \shortstack{$-0.000$\\{\scriptsize $(0.369)$}} & \shortstack{$0.538$\\{\scriptsize $(0.481)$}} & \shortstack{$-0.120$\\{\scriptsize $(0.453)$}} & \shortstack{$-0.231$\\{\scriptsize $(0.337)$}} & \shortstack{$0.423$\\{\scriptsize $(0.326)$}} & \shortstack{$-0.313^{***}$\\{\scriptsize $(0.043)$}} & \shortstack{$0.102^{**}$\\{\scriptsize $(0.042)$}} \\
Tech. maturity & \shortstack{$-0.006$\\{\scriptsize $(0.034)$}} & \shortstack{$0.003$\\{\scriptsize $(0.034)$}} & \shortstack{$0.185$\\{\scriptsize $(0.126)$}} & \shortstack{$-0.089$\\{\scriptsize $(0.108)$}} & \shortstack{$0.229^{**}$\\{\scriptsize $(0.105)$}} & \shortstack{$-0.107$\\{\scriptsize $(0.097)$}} & \shortstack{$0.469$\\{\scriptsize $(0.440)$}} & \shortstack{$0.406$\\{\scriptsize $(0.451)$}} & \shortstack{$0.050^{***}$\\{\scriptsize $(0.016)$}} & \shortstack{$0.031^{*}$\\{\scriptsize $(0.018)$}} \\
Priority: none & \shortstack{$0.396^{***}$\\{\scriptsize $(0.130)$}} & \shortstack{$0.453^{***}$\\{\scriptsize $(0.137)$}} & \shortstack{$0.278$\\{\scriptsize $(0.460)$}} & \shortstack{$0.119$\\{\scriptsize $(0.408)$}} & \shortstack{$1.070^{**}$\\{\scriptsize $(0.524)$}} & \shortstack{$-0.192$\\{\scriptsize $(0.496)$}} & --- & --- & \shortstack{$0.067$\\{\scriptsize $(0.057)$}} & \shortstack{$0.048$\\{\scriptsize $(0.061)$}} \\
Priority: 0--3m & \shortstack{$0.767^{**}$\\{\scriptsize $(0.343)$}} & \shortstack{$0.295$\\{\scriptsize $(0.326)$}} & \shortstack{$0.360$\\{\scriptsize $(0.447)$}} & \shortstack{$0.151$\\{\scriptsize $(0.409)$}} & \shortstack{$0.301$\\{\scriptsize $(0.435)$}} & \shortstack{$-0.684$\\{\scriptsize $(0.421)$}} & \shortstack{$-2.449^{*}$\\{\scriptsize $(1.294)$}} & \shortstack{$-0.828$\\{\scriptsize $(1.217)$}} & \shortstack{$-0.748^{***}$\\{\scriptsize $(0.242)$}} & \shortstack{$-0.557^{**}$\\{\scriptsize $(0.251)$}} \\
Priority: 4--12m & \shortstack{$-0.459^{**}$\\{\scriptsize $(0.217)$}} & \shortstack{$-0.609^{**}$\\{\scriptsize $(0.246)$}} & \shortstack{$-0.135$\\{\scriptsize $(0.170)$}} & \shortstack{$-0.071$\\{\scriptsize $(0.163)$}} & \shortstack{$-0.001$\\{\scriptsize $(0.167)$}} & \shortstack{$-0.017$\\{\scriptsize $(0.162)$}} & \shortstack{$-0.766$\\{\scriptsize $(0.538)$}} & \shortstack{$-0.666$\\{\scriptsize $(0.538)$}} & \shortstack{$-0.031$\\{\scriptsize $(0.052)$}} & \shortstack{$-0.145^{**}$\\{\scriptsize $(0.059)$}} \\
Priority: 13--24m & \shortstack{$-0.523^{*}$\\{\scriptsize $(0.269)$}} & \shortstack{$-0.708^{**}$\\{\scriptsize $(0.305)$}} & \shortstack{$0.314$\\{\scriptsize $(0.305)$}} & \shortstack{$0.181$\\{\scriptsize $(0.324)$}} & \shortstack{$0.348$\\{\scriptsize $(0.325)$}} & \shortstack{$-0.127$\\{\scriptsize $(0.361)$}} & \shortstack{$-0.537$\\{\scriptsize $(0.605)$}} & \shortstack{$-0.671$\\{\scriptsize $(0.708)$}} & \shortstack{$-0.080$\\{\scriptsize $(0.115)$}} & \shortstack{$-0.287^{**}$\\{\scriptsize $(0.130)$}} \\
Priority: $>24$m & \shortstack{---\\{\scriptsize $(0.000)$}} & \shortstack{---\\{\scriptsize $(0.000)$}} & \shortstack{---\\{\scriptsize $(0.000)$}} & \shortstack{---\\{\scriptsize $(0.000)$}} & \shortstack{---\\{\scriptsize $(0.000)$}} & \shortstack{---\\{\scriptsize $(0.000)$}} & \shortstack{$-0.750^{*}$\\{\scriptsize $(0.402)$}} & \shortstack{$-0.333$\\{\scriptsize $(0.388)$}} & \shortstack{---\\{\scriptsize $(0.000)$}} & \shortstack{---\\{\scriptsize $(0.000)$}} \\
\addlinespace[2pt]\multicolumn{11}{l}{\emph{Model statistics}} \\[1pt]
$N$ & 12,838 & 12,838 & 19,112 & 19,112 & 14,956 & 14,956 & 4,672 & 4,672 & 92,840 & 92,840 \\
Events & 3,302 & 4,495 & 1,864 & 2,973 & 1,525 & 2,390 & 483 & 548 & 11,915 & 11,223 \\
Pairs & 6,419 & 6,419 & 9,556 & 9,556 & 7,478 & 7,478 & 2,336 & 2,336 & 46,420 & 46,420 \\
Concordance & 0.622 & 0.693 & 0.614 & 0.679 & 0.611 & 0.718 & 0.645 & 0.664 & 0.582 & 0.626 \\
\bottomrule
\end{tabular}
\begin{tablenotes}[flushleft]\footnotesize
\item[]
Cells report log-hazard coefficients, with matched-pair-clustered standard errors in parentheses. Models are stratified by matched pair and use the narrow citation-date proxy with 24 months of follow-up. Claims, inventors, backward citations, and annual pre-event citation counts enter as $\log(1+x)$. Citation slopes are adjacent-year changes in logged citation counts; technological maturity is a standardized logged technology-stock measure. Exam. denotes examiner; Appl. denotes applicant; Bwd. denotes backward; NPL denotes non-patent literature. A dash denotes a coefficient that is not separately estimable because of collinearity or insufficient within-pair variation. $^{***}p<.01$, $^{**}p<.05$, $^{*}p<.10$.
\end{tablenotes}
\end{threeparttable}
\end{adjustbox}
\renewcommand{\arraystretch}{1}
\end{table}

}{\placeholder{Run the R script to generate the full Cox coefficient table.}}

\IfFileExists{generated/tables/table_full_count_coefficients.tex}{
  \begin{table}[p]
\centering
\caption{Full coefficient estimates: 24-month citation counts}\label{tab:full-count}
\scriptsize
\setlength{\tabcolsep}{1.8pt}
\renewcommand{\arraystretch}{0.94}
\begin{adjustbox}{max width=\linewidth,max totalheight=0.80\textheight,keepaspectratio,center}
\begin{threeparttable}
\begin{tabular}{l*{10}{c}}
\toprule
 & \multicolumn{2}{c}{\shortstack{2013\\Creation}} & \multicolumn{2}{c}{\shortstack{2020\\Addition}} & \multicolumn{2}{c}{\shortstack{2020\\Strict add.}} & \multicolumn{2}{c}{\shortstack{2020\\Replacement}} & \multicolumn{2}{c}{\shortstack{2020\\Cleanup}} \\
\cmidrule(lr){2-3}\cmidrule(lr){4-5}\cmidrule(lr){6-7}\cmidrule(lr){8-9}\cmidrule(lr){10-11}
Variable & Examiner & Applicant & Examiner & Applicant & Examiner & Applicant & Examiner & Applicant & Examiner & Applicant \\
\midrule
Treated $\times$ Post & \shortstack{$0.049$\\{\scriptsize $(0.041)$}} & \shortstack{$-0.178^{***}$\\{\scriptsize $(0.048)$}} & \shortstack{$-0.111^{**}$\\{\scriptsize $(0.050)$}} & \shortstack{$-0.018$\\{\scriptsize $(0.058)$}} & \shortstack{$0.001$\\{\scriptsize $(0.057)$}} & \shortstack{$-0.022$\\{\scriptsize $(0.062)$}} & \shortstack{$-0.243^{**}$\\{\scriptsize $(0.106)$}} & \shortstack{$-0.108$\\{\scriptsize $(0.134)$}} & \shortstack{$0.117^{***}$\\{\scriptsize $(0.019)$}} & \shortstack{$-0.063^{**}$\\{\scriptsize $(0.032)$}} \\
Treated & \shortstack{$0.121^{***}$\\{\scriptsize $(0.017)$}} & \shortstack{$0.100^{***}$\\{\scriptsize $(0.031)$}} & \shortstack{$-0.072^{***}$\\{\scriptsize $(0.013)$}} & \shortstack{$-0.056^{***}$\\{\scriptsize $(0.018)$}} & \shortstack{$-0.121^{***}$\\{\scriptsize $(0.018)$}} & \shortstack{$-0.063^{***}$\\{\scriptsize $(0.022)$}} & \shortstack{$0.053$\\{\scriptsize $(0.040)$}} & \shortstack{$0.031$\\{\scriptsize $(0.045)$}} & \shortstack{$0.020^{***}$\\{\scriptsize $(0.004)$}} & \shortstack{$0.094^{***}$\\{\scriptsize $(0.011)$}} \\
Post & \shortstack{$-0.085^{***}$\\{\scriptsize $(0.032)$}} & \shortstack{$0.509^{***}$\\{\scriptsize $(0.032)$}} & \shortstack{$-0.365^{***}$\\{\scriptsize $(0.039)$}} & \shortstack{$-0.802^{***}$\\{\scriptsize $(0.045)$}} & \shortstack{$-0.405^{***}$\\{\scriptsize $(0.043)$}} & \shortstack{$-0.784^{***}$\\{\scriptsize $(0.044)$}} & \shortstack{$-0.006$\\{\scriptsize $(0.090)$}} & \shortstack{$-0.404^{***}$\\{\scriptsize $(0.097)$}} & \shortstack{$-0.382^{***}$\\{\scriptsize $(0.016)$}} & \shortstack{$-0.514^{***}$\\{\scriptsize $(0.021)$}} \\
Exam. cit., pre y1 & \shortstack{$1.195^{***}$\\{\scriptsize $(0.079)$}} & \shortstack{$-0.275^{***}$\\{\scriptsize $(0.097)$}} & \shortstack{$1.725^{***}$\\{\scriptsize $(0.105)$}} & \shortstack{$0.263^{**}$\\{\scriptsize $(0.113)$}} & \shortstack{$1.582^{***}$\\{\scriptsize $(0.106)$}} & \shortstack{$0.394^{***}$\\{\scriptsize $(0.102)$}} & \shortstack{$2.165^{***}$\\{\scriptsize $(0.224)$}} & \shortstack{$1.224^{***}$\\{\scriptsize $(0.233)$}} & \shortstack{$1.298^{***}$\\{\scriptsize $(0.055)$}} & \shortstack{$0.361^{***}$\\{\scriptsize $(0.073)$}} \\
Exam. cit., pre y2 & \shortstack{$1.202^{***}$\\{\scriptsize $(0.065)$}} & \shortstack{$-0.429^{***}$\\{\scriptsize $(0.109)$}} & \shortstack{$1.617^{***}$\\{\scriptsize $(0.108)$}} & \shortstack{$-0.176$\\{\scriptsize $(0.141)$}} & \shortstack{$1.604^{***}$\\{\scriptsize $(0.105)$}} & \shortstack{$-0.042$\\{\scriptsize $(0.063)$}} & \shortstack{$2.199^{***}$\\{\scriptsize $(0.254)$}} & \shortstack{$0.228$\\{\scriptsize $(0.217)$}} & \shortstack{$1.169^{***}$\\{\scriptsize $(0.047)$}} & \shortstack{$0.251^{***}$\\{\scriptsize $(0.052)$}} \\
Exam. cit., pre y3 & \shortstack{$0.124^{*}$\\{\scriptsize $(0.064)$}} & \shortstack{$-0.131$\\{\scriptsize $(0.084)$}} & \shortstack{$0.073$\\{\scriptsize $(0.087)$}} & \shortstack{$0.086$\\{\scriptsize $(0.074)$}} & \shortstack{$0.033$\\{\scriptsize $(0.111)$}} & \shortstack{$-0.206^{***}$\\{\scriptsize $(0.078)$}} & \shortstack{$0.253$\\{\scriptsize $(0.424)$}} & \shortstack{$-1.346^{***}$\\{\scriptsize $(0.410)$}} & \shortstack{$0.096^{**}$\\{\scriptsize $(0.037)$}} & \shortstack{$0.078$\\{\scriptsize $(0.055)$}} \\
Appl. cit., pre y1 & \shortstack{$0.093^{**}$\\{\scriptsize $(0.046)$}} & \shortstack{$0.769^{***}$\\{\scriptsize $(0.050)$}} & \shortstack{$-0.029$\\{\scriptsize $(0.060)$}} & \shortstack{$0.723^{***}$\\{\scriptsize $(0.049)$}} & \shortstack{$-0.003$\\{\scriptsize $(0.063)$}} & \shortstack{$0.647^{***}$\\{\scriptsize $(0.053)$}} & \shortstack{$0.267^{**}$\\{\scriptsize $(0.128)$}} & \shortstack{$1.529^{***}$\\{\scriptsize $(0.109)$}} & \shortstack{$0.052^{**}$\\{\scriptsize $(0.026)$}} & \shortstack{$0.761^{***}$\\{\scriptsize $(0.033)$}} \\
Appl. cit., pre y2 & \shortstack{$0.108^{**}$\\{\scriptsize $(0.050)$}} & \shortstack{$0.704^{***}$\\{\scriptsize $(0.058)$}} & \shortstack{$0.020$\\{\scriptsize $(0.068)$}} & \shortstack{$0.753^{***}$\\{\scriptsize $(0.047)$}} & \shortstack{$-0.003$\\{\scriptsize $(0.074)$}} & \shortstack{$0.845^{***}$\\{\scriptsize $(0.060)$}} & \shortstack{$0.115$\\{\scriptsize $(0.168)$}} & \shortstack{$0.394^{***}$\\{\scriptsize $(0.138)$}} & \shortstack{$0.025$\\{\scriptsize $(0.035)$}} & \shortstack{$0.430^{***}$\\{\scriptsize $(0.049)$}} \\
Appl. cit., pre y3 & \shortstack{$-0.158^{**}$\\{\scriptsize $(0.063)$}} & \shortstack{$-0.119^{*}$\\{\scriptsize $(0.063)$}} & \shortstack{$0.072$\\{\scriptsize $(0.065)$}} & \shortstack{$-0.046$\\{\scriptsize $(0.044)$}} & \shortstack{$0.082$\\{\scriptsize $(0.072)$}} & \shortstack{$-0.019$\\{\scriptsize $(0.048)$}} & \shortstack{$0.025$\\{\scriptsize $(0.220)$}} & \shortstack{$-0.565^{***}$\\{\scriptsize $(0.164)$}} & \shortstack{$0.014$\\{\scriptsize $(0.051)$}} & \shortstack{$-0.345^{***}$\\{\scriptsize $(0.074)$}} \\
Claims & \shortstack{$0.079^{**}$\\{\scriptsize $(0.040)$}} & \shortstack{$0.431^{***}$\\{\scriptsize $(0.045)$}} & \shortstack{$-0.275^{**}$\\{\scriptsize $(0.123)$}} & \shortstack{$-0.141$\\{\scriptsize $(0.108)$}} & \shortstack{$-0.042$\\{\scriptsize $(0.161)$}} & \shortstack{$-0.365^{***}$\\{\scriptsize $(0.136)$}} & \shortstack{$-0.058$\\{\scriptsize $(0.145)$}} & \shortstack{$0.417^{***}$\\{\scriptsize $(0.113)$}} & \shortstack{$0.075^{***}$\\{\scriptsize $(0.013)$}} & \shortstack{$0.110^{***}$\\{\scriptsize $(0.018)$}} \\
Inventors & \shortstack{$0.052$\\{\scriptsize $(0.038)$}} & \shortstack{$0.133^{***}$\\{\scriptsize $(0.050)$}} & \shortstack{$0.181^{***}$\\{\scriptsize $(0.062)$}} & \shortstack{$0.077$\\{\scriptsize $(0.049)$}} & \shortstack{$0.078$\\{\scriptsize $(0.074)$}} & \shortstack{$0.179^{***}$\\{\scriptsize $(0.056)$}} & \shortstack{$-0.007$\\{\scriptsize $(0.101)$}} & \shortstack{$0.035$\\{\scriptsize $(0.081)$}} & \shortstack{$0.013$\\{\scriptsize $(0.015)$}} & \shortstack{$-0.040^{*}$\\{\scriptsize $(0.024)$}} \\
Bwd. U.S. cit. & \shortstack{$0.030$\\{\scriptsize $(0.027)$}} & \shortstack{$-0.159^{***}$\\{\scriptsize $(0.034)$}} & \shortstack{$0.310^{***}$\\{\scriptsize $(0.087)$}} & \shortstack{$0.153^{**}$\\{\scriptsize $(0.077)$}} & \shortstack{$0.152$\\{\scriptsize $(0.125)$}} & \shortstack{$0.370^{***}$\\{\scriptsize $(0.110)$}} & \shortstack{$0.085$\\{\scriptsize $(0.075)$}} & \shortstack{$-0.136^{**}$\\{\scriptsize $(0.068)$}} & \shortstack{$0.050^{***}$\\{\scriptsize $(0.010)$}} & \shortstack{$-0.080^{***}$\\{\scriptsize $(0.015)$}} \\
Bwd. foreign cit. & \shortstack{$-0.024$\\{\scriptsize $(0.047)$}} & \shortstack{$-0.395^{***}$\\{\scriptsize $(0.062)$}} & \shortstack{$0.019$\\{\scriptsize $(0.018)$}} & \shortstack{$-0.031^{*}$\\{\scriptsize $(0.016)$}} & \shortstack{$-0.011$\\{\scriptsize $(0.021)$}} & \shortstack{$-0.058^{***}$\\{\scriptsize $(0.017)$}} & \shortstack{$-0.114^{**}$\\{\scriptsize $(0.046)$}} & \shortstack{$-0.018$\\{\scriptsize $(0.039)$}} & \shortstack{$-0.023$\\{\scriptsize $(0.036)$}} & \shortstack{$0.373^{***}$\\{\scriptsize $(0.052)$}} \\
NPL refs. & \shortstack{$0.066$\\{\scriptsize $(0.082)$}} & \shortstack{$0.803^{***}$\\{\scriptsize $(0.105)$}} & \shortstack{$-0.102^{***}$\\{\scriptsize $(0.039)$}} & \shortstack{$-0.053$\\{\scriptsize $(0.033)$}} & \shortstack{$-0.021$\\{\scriptsize $(0.063)$}} & \shortstack{$-0.139^{***}$\\{\scriptsize $(0.051)$}} & \shortstack{$0.013$\\{\scriptsize $(0.061)$}} & \shortstack{$0.221^{***}$\\{\scriptsize $(0.050)$}} & \shortstack{$0.025$\\{\scriptsize $(0.028)$}} & \shortstack{$-0.276^{***}$\\{\scriptsize $(0.041)$}} \\
Small/micro entity & \shortstack{$-0.003$\\{\scriptsize $(0.053)$}} & \shortstack{$-0.293^{***}$\\{\scriptsize $(0.070)$}} & \shortstack{$0.949^{***}$\\{\scriptsize $(0.281)$}} & \shortstack{$0.458^{*}$\\{\scriptsize $(0.246)$}} & \shortstack{$0.290$\\{\scriptsize $(0.318)$}} & \shortstack{$1.041^{***}$\\{\scriptsize $(0.292)$}} & \shortstack{$-0.034$\\{\scriptsize $(0.158)$}} & \shortstack{$-0.249^{**}$\\{\scriptsize $(0.120)$}} & \shortstack{$-0.014$\\{\scriptsize $(0.017)$}} & \shortstack{$0.013$\\{\scriptsize $(0.021)$}} \\
U.S. assignee & \shortstack{$-0.047$\\{\scriptsize $(0.042)$}} & \shortstack{$0.061$\\{\scriptsize $(0.049)$}} & \shortstack{$0.162^{***}$\\{\scriptsize $(0.055)$}} & \shortstack{$0.130^{***}$\\{\scriptsize $(0.045)$}} & \shortstack{$0.085$\\{\scriptsize $(0.073)$}} & \shortstack{$0.289^{***}$\\{\scriptsize $(0.066)$}} & \shortstack{$-0.136$\\{\scriptsize $(0.151)$}} & \shortstack{$-0.315^{***}$\\{\scriptsize $(0.111)$}} & \shortstack{$-0.054^{**}$\\{\scriptsize $(0.027)$}} & \shortstack{$0.320^{***}$\\{\scriptsize $(0.043)$}} \\
Continuation & \shortstack{$0.007$\\{\scriptsize $(0.132)$}} & \shortstack{$1.180^{***}$\\{\scriptsize $(0.166)$}} & \shortstack{$-0.836^{***}$\\{\scriptsize $(0.187)$}} & \shortstack{$-0.195$\\{\scriptsize $(0.164)$}} & \shortstack{$-0.394^{*}$\\{\scriptsize $(0.223)$}} & \shortstack{$-0.596^{***}$\\{\scriptsize $(0.196)$}} & \shortstack{$-0.251$\\{\scriptsize $(0.156)$}} & \shortstack{$0.464^{***}$\\{\scriptsize $(0.119)$}} & \shortstack{$-0.128^{***}$\\{\scriptsize $(0.019)$}} & \shortstack{$0.067^{***}$\\{\scriptsize $(0.023)$}} \\
Tech. maturity & \shortstack{$-0.017$\\{\scriptsize $(0.018)$}} & \shortstack{$0.031$\\{\scriptsize $(0.026)$}} & \shortstack{$-0.143^{***}$\\{\scriptsize $(0.055)$}} & \shortstack{$-0.047$\\{\scriptsize $(0.049)$}} & \shortstack{$-0.053$\\{\scriptsize $(0.051)$}} & \shortstack{$-0.120^{***}$\\{\scriptsize $(0.042)$}} & \shortstack{$-0.067$\\{\scriptsize $(0.184)$}} & \shortstack{$0.520^{***}$\\{\scriptsize $(0.148)$}} & \shortstack{$0.006$\\{\scriptsize $(0.007)$}} & \shortstack{$0.017$\\{\scriptsize $(0.012)$}} \\
Priority: none & \shortstack{$0.212^{***}$\\{\scriptsize $(0.067)$}} & \shortstack{$0.593^{***}$\\{\scriptsize $(0.087)$}} & \shortstack{$-0.715^{***}$\\{\scriptsize $(0.202)$}} & \shortstack{$-0.300^{*}$\\{\scriptsize $(0.180)$}} & \shortstack{$-0.274$\\{\scriptsize $(0.257)$}} & \shortstack{$-0.702^{***}$\\{\scriptsize $(0.216)$}} & --- & --- & \shortstack{$0.075^{***}$\\{\scriptsize $(0.024)$}} & \shortstack{$-0.088^{**}$\\{\scriptsize $(0.035)$}} \\
Priority: 0--3m & \shortstack{$0.307^{**}$\\{\scriptsize $(0.143)$}} & \shortstack{$0.356^{**}$\\{\scriptsize $(0.163)$}} & \shortstack{$-0.208$\\{\scriptsize $(0.180)$}} & \shortstack{$-0.291^{*}$\\{\scriptsize $(0.151)$}} & \shortstack{$-0.184$\\{\scriptsize $(0.218)$}} & \shortstack{$-0.301^{*}$\\{\scriptsize $(0.178)$}} & \shortstack{$-2.057^{**}$\\{\scriptsize $(0.844)$}} & \shortstack{$-1.073^{**}$\\{\scriptsize $(0.534)$}} & \shortstack{$-0.018$\\{\scriptsize $(0.104)$}} & \shortstack{$-0.817^{***}$\\{\scriptsize $(0.128)$}} \\
Priority: 4--12m & \shortstack{$-0.105$\\{\scriptsize $(0.108)$}} & \shortstack{$-1.133^{***}$\\{\scriptsize $(0.142)$}} & \shortstack{$-0.172^{**}$\\{\scriptsize $(0.076)$}} & \shortstack{$-0.096$\\{\scriptsize $(0.084)$}} & \shortstack{$-0.082$\\{\scriptsize $(0.083)$}} & \shortstack{$-0.070$\\{\scriptsize $(0.074)$}} & \shortstack{$-0.028$\\{\scriptsize $(0.245)$}} & \shortstack{$-0.745^{***}$\\{\scriptsize $(0.222)$}} & \shortstack{$0.050^{**}$\\{\scriptsize $(0.022)$}} & \shortstack{$-0.174^{***}$\\{\scriptsize $(0.049)$}} \\
Priority: 13--24m & \shortstack{$-0.230^{*}$\\{\scriptsize $(0.137)$}} & \shortstack{$-1.272^{***}$\\{\scriptsize $(0.183)$}} & \shortstack{$0.051$\\{\scriptsize $(0.126)$}} & \shortstack{$-0.103$\\{\scriptsize $(0.128)$}} & \shortstack{$0.115$\\{\scriptsize $(0.190)$}} & \shortstack{$-0.164$\\{\scriptsize $(0.129)$}} & \shortstack{$-0.041$\\{\scriptsize $(0.309)$}} & \shortstack{$-0.680^{*}$\\{\scriptsize $(0.389)$}} & \shortstack{$0.091^{*}$\\{\scriptsize $(0.048)$}} & \shortstack{$-0.278^{***}$\\{\scriptsize $(0.069)$}} \\
Priority: $>24$m & --- & --- & --- & --- & --- & --- & \shortstack{$-0.077$\\{\scriptsize $(0.187)$}} & \shortstack{$-0.483^{***}$\\{\scriptsize $(0.165)$}} & --- & --- \\
\addlinespace[2pt]\multicolumn{11}{l}{\emph{Model statistics}} \\[1pt]
$N$ & 15,068 & 16,524 & 11,004 & 15,140 & 8,948 & 12,004 & 2,784 & 3,136 & 61,728 & 58,960 \\
Pairs & 3,767 & 4,131 & 2,751 & 3,785 & 2,237 & 3,001 & 696 & 784 & 15,432 & 14,740 \\
Pseudo $R^2$ & 0.255 & 0.606 & 0.178 & 0.706 & 0.185 & 0.721 & 0.162 & 0.494 & 0.164 & 0.733 \\
\bottomrule
\end{tabular}
\begin{tablenotes}[flushleft]\footnotesize
\item[]
Cells report log-incidence-rate coefficients, with matched-pair-clustered standard errors in parentheses. Models include matched-pair fixed effects and compare symmetric 24-month pre- and post-event count windows. Claims, inventors, backward citations, and annual pre-event citation counts enter as $\log(1+x)$. Citation slopes are adjacent-year changes in logged citation counts; technological maturity is a standardized logged technology-stock measure. Exam. denotes examiner; Appl. denotes applicant; Bwd. denotes backward; NPL denotes non-patent literature. A dash denotes a coefficient that is not separately estimable because of collinearity or insufficient within-pair variation. $^{***}p<.01$, $^{**}p<.05$, $^{*}p<.10$.
\end{tablenotes}
\end{threeparttable}
\end{adjustbox}
\renewcommand{\arraystretch}{1}
\end{table}

}{\placeholder{Run the R script to generate the full count-model coefficient table.}}

\FloatBarrier
\subsection{Robustness}
\label{app:robustness}

\subsubsection{Control-pool and exact-cell robustness}
\label{app:control-pool}

The six-design exercise varies both the comparison group and the exact technology--cohort cell; Table~\ref{tab:robustness-summary} summarizes the results.

\IfFileExists{generated/tables/table_robustness_summary.tex}{
\begin{table}[htbp]
\centering
\caption{Stability across six matching designs}
\label{tab:robustness-summary}
\small
\begin{adjustbox}{max width=\linewidth,max totalheight=0.82\textheight,keepaspectratio,center}
\begin{threeparttable}
\begin{tabular}{lcrrrrrrrr}
\toprule
\multicolumn{2}{c}{} & \multicolumn{4}{c}{Cox timing} & \multicolumn{4}{c}{Citation counts} \\
\cmidrule(lr){3-6}\cmidrule(lr){7-10}
Audience & \shortstack{Balanced\\designs} & \shortstack{Median\\HR} & \shortstack{HR\\range} & \shortstack{Sig.\\$>1$} & \shortstack{Sig.\\$<1$} & \shortstack{Median\\effect} & \shortstack{Effect\\range} & \shortstack{Sig.\\$>0$} & \shortstack{Sig.\\$<0$} \\
\midrule
\multicolumn{10}{l}{\textit{Creation}} \\[2pt]
Applicant & 1/6 & 1.12 & 1.02--1.21 & 3/6 & 0/6 & -10.6\% & -16.3---3.2\% & 0/6 & 4/6 \\
Examiner & 1/6 & 1.24 & 1.19--1.52 & 6/6 & 0/6 & 24.0\% & 5.0--38.9\% & 5/6 & 0/6 \\
\addlinespace
\multicolumn{10}{l}{\textit{Addition}} \\[2pt]
Applicant & 5/6 & 0.83 & 0.78--0.97 & 0/6 & 5/6 & -5.3\% & -15.8---1.8\% & 0/6 & 2/6 \\
Examiner & 5/6 & 0.82 & 0.76--0.92 & 0/6 & 6/6 & -21.8\% & -30.2---10.5\% & 0/6 & 6/6 \\
\addlinespace
\multicolumn{10}{l}{\textit{Strict addition}} \\[2pt]
Applicant & 4/6 & 0.88 & 0.84--0.99 & 0/6 & 4/6 & -7.7\% & -16.1---2.2\% & 0/6 & 2/6 \\
Examiner & 4/6 & 0.79 & 0.73--0.90 & 0/6 & 6/6 & -20.3\% & -30.0--0.1\% & 0/6 & 5/6 \\
\addlinespace
\multicolumn{10}{l}{\textit{Replacement}} \\[2pt]
Applicant & 4/6 & 1.07 & 0.93--1.33 & 0/6 & 0/6 & 0.1\% & -20.9--4.1\% & 0/6 & 0/6 \\
Examiner & 4/6 & 1.03 & 0.77--1.14 & 0/6 & 0/6 & -6.9\% & -46.0--5.5\% & 0/6 & 3/6 \\
\addlinespace
\multicolumn{10}{l}{\textit{Cleanup}} \\[2pt]
Applicant & 4/6 & 1.06 & 0.99--1.16 & 4/6 & 0/6 & 2.1\% & -6.1--9.6\% & 0/6 & 1/6 \\
Examiner & 4/6 & 1.18 & 1.13--1.20 & 6/6 & 0/6 & 12.1\% & -2.1--14.8\% & 5/6 & 0/6 \\
\addlinespace
\bottomrule
\end{tabular}
\begin{tablenotes}[flushleft]\footnotesize
\item[]
Each row summarizes six separately matched designs formed by crossing current-green versus universe controls with publication-year-by-AU4, publication-year-by-TC2, and publication-quarter-by-TC2 exact cells. The Cox columns report hazard ratios. Count effects and their ranges are percentage changes implied by the Poisson DiD coefficients. Balanced reports the number of matching designs with Max SMD below .10 divided by the number of available designs. The positive and negative columns report the number of specifications with an effect in the indicated direction and a two-sided $p$-value below .05, divided by the number of available specifications. These are descriptive stability counts, not multiple-testing-adjusted tests or meta-analytic estimates. Standard errors in the underlying models are clustered by matched pair.
\end{tablenotes}
\end{threeparttable}
\end{adjustbox}
\end{table}

}{\placeholder{Run the R script to generate the six-design robustness summary.}}

Further, Table~\ref{tab:control-pool-headline} compares green and broad-patent controls while holding the baseline exact-cell definition fixed (year--AU4 in 2013 and year--TC2 in 2020). 
Every row is based on a separately matched sample and differences in coefficients should be read together with differences in common support and post-match balance.

\IfFileExists{generated/tables/table_control_pool_headline_cells.tex}{
  \begin{landscape}
\begin{table}[htbp]
\centering
\caption{Comparison of green and broad-patent controls}
\label{tab:control-pool-headline}
\small
\begin{adjustbox}{max width=\linewidth,max totalheight=0.82\textheight,keepaspectratio,center}
\begin{threeparttable}
\begin{tabular}{llrrrrrr}
\toprule
Control pool & Audience & \shortstack{Matched\\pairs} & \shortstack{Max\\SMD} & HR [95\% CI] & $p$ & \shortstack{Count effect\\{[95\% CI]}} & $p$ \\
\midrule
\multicolumn{8}{l}{\textit{Creation}} \\[2pt]
Unchanged green & Examiner & 6,419 & 0.089 & 1.252$^{***}$ [1.157, 1.355] & $<.001$ & 5.0\% [-3.1, 13.7] & $.233$ \\
Unchanged green & Applicant & 6,419 & 0.089 & 1.069$^{*}$ [0.993, 1.151] & $.078$ & -16.3\%$^{***}$ [-23.8, -8.0] & $<.001$ \\
Broad patent & Examiner & 34,170 & 0.144 & 1.235$^{***}$ [1.196, 1.275] & $<.001$ & 13.3\%$^{***}$ [10.2, 16.5] & $<.001$ \\
Broad patent & Applicant & 34,170 & 0.144 & 1.208$^{***}$ [1.170, 1.247] & $<.001$ & -9.1\%$^{***}$ [-12.6, -5.4] & $<.001$ \\
\addlinespace
\multicolumn{8}{l}{\textit{Addition}} \\[2pt]
Unchanged green & Examiner & 9,556 & 0.074 & 0.870$^{***}$ [0.787, 0.962] & $.007$ & -10.5\%$^{**}$ [-18.9, -1.3] & $.026$ \\
Unchanged green & Applicant & 9,556 & 0.074 & 0.832$^{***}$ [0.762, 0.909] & $<.001$ & -1.8\% [-12.3, 10.0] & $.751$ \\
Broad patent & Examiner & 17,052 & 0.081 & 0.816$^{***}$ [0.758, 0.878] & $<.001$ & -24.5\%$^{***}$ [-29.3, -19.2] & $<.001$ \\
Broad patent & Applicant & 17,052 & 0.081 & 0.931$^{**}$ [0.871, 0.994] & $.033$ & -13.9\%$^{***}$ [-20.7, -6.6] & $<.001$ \\
\addlinespace
\multicolumn{8}{l}{\textit{Strict addition}} \\[2pt]
Unchanged green & Examiner & 7,478 & 0.097 & 0.846$^{***}$ [0.757, 0.946] & $.003$ & 0.1\% [-10.5, 11.8] & $.991$ \\
Unchanged green & Applicant & 7,478 & 0.097 & 0.877$^{**}$ [0.792, 0.971] & $.012$ & -2.2\% [-13.4, 10.5] & $.723$ \\
Broad patent & Examiner & 14,249 & 0.083 & 0.832$^{***}$ [0.769, 0.900] & $<.001$ & -24.0\%$^{***}$ [-29.4, -18.2] & $<.001$ \\
Broad patent & Applicant & 14,249 & 0.083 & 0.909$^{**}$ [0.845, 0.978] & $.010$ & -12.9\%$^{***}$ [-20.0, -5.1] & $.002$ \\
\addlinespace
\multicolumn{8}{l}{\textit{Replacement}} \\[2pt]
Unchanged green & Examiner & 2,336 & 0.087 & 0.884 [0.722, 1.083] & $.233$ & -21.6\%$^{**}$ [-36.2, -3.5] & $.022$ \\
Unchanged green & Applicant & 2,336 & 0.087 & 1.069 [0.874, 1.306] & $.518$ & -10.3\% [-31.0, 16.8] & $.420$ \\
Broad patent & Examiner & 2,957 & 0.076 & 1.129 [0.946, 1.348] & $.179$ & 5.3\% [-10.3, 23.5] & $.527$ \\
Broad patent & Applicant & 2,957 & 0.076 & 1.093 [0.914, 1.306] & $.330$ & -0.9\% [-23.2, 27.9] & $.948$ \\
\addlinespace
\multicolumn{8}{l}{\textit{Cleanup}} \\[2pt]
Unchanged green & Examiner & 46,420 & 0.095 & 1.141$^{***}$ [1.098, 1.187] & $<.001$ & 12.4\%$^{***}$ [8.2, 16.7] & $<.001$ \\
Unchanged green & Applicant & 46,420 & 0.095 & 1.048$^{**}$ [1.005, 1.092] & $.030$ & -6.1\%$^{**}$ [-11.9, -0.0] & $.049$ \\
Broad patent & Examiner & 55,549 & 0.135 & 1.188$^{***}$ [1.147, 1.230] & $<.001$ & 8.1\%$^{***}$ [4.6, 11.7] & $<.001$ \\
Broad patent & Applicant & 55,549 & 0.135 & 1.062$^{***}$ [1.022, 1.104] & $.002$ & 3.9\% [-2.0, 10.1] & $.198$ \\
\addlinespace
\bottomrule
\end{tabular}
\begin{tablenotes}[flushleft]\footnotesize
\item[]
The 2013 rows use publication-year--AU4 exact cells; the 2020 rows use publication-year--TC2 exact cells. Each control-pool specification is rematched. The current-green specification is the paper-facing design. Broad-patent controls provide a larger candidate pool but can leave greater residual imbalance. Count effects are percentage changes implied by Poisson DiD coefficients. $^{***}p<.01$, $^{**}p<.05$, $^{*}p<.10$.
\end{tablenotes}
\end{threeparttable}
\end{adjustbox}
\end{table}
\end{landscape}

}{\placeholder{Run the R script to generate the direct control-pool comparison.}}

\FloatBarrier

\subsubsection{Matching-implementation sensitivity}
\label{app:matching-sensitivity}

Table~\ref{tab:matching-sensitivity} reports all nine matching implementation checks underlying the ranges summarized in Section~\ref{subsec:results_robustness}.
The specifications vary the propensity-score caliper, its scale, the pre-matching control-pool cap, or the random subset of controls admitted when an exact cell exceeds that cap. Treatment definitions, the baseline green-control group, and the event-specific headline exact-cell definition are retained.

\IfFileExists{generated/tables/table_matching_sensitivity.tex}{
  \begin{landscape}
\scriptsize
\setlength{\tabcolsep}{3pt}
\begin{ThreePartTable}
\begin{TableNotes}[flushleft]\footnotesize
\item[]
All rows retain the paper-facing unchanged-green control group and exact-cell definition. Only the matching implementation changes. The reference uses the adaptive matching procedure. Fixed-caliper rows prevent adaptive caliper switching. The logit-scale row imposes a 0.20 caliper on the log-odds linear predictor rather than on the propensity-score probability. Control-pool-cap and seed specifications change only the pre-MatchIt candidate-control subsampling. Final matching remains one-to-one and without replacement. $^{***}p<.01$, $^{**}p<.05$, $^{*}p<.10$.
\end{TableNotes}
\begin{longtable}{llrrrrrr}
\caption{Sensitivity to the matching procedure}\label{tab:matching-sensitivity}\\
\toprule
Matching specification & Audience & \shortstack{Matched\\pairs} & \shortstack{Max\\SMD} & HR [95\% CI] & $p$ & \shortstack{Count effect\\{[95\% CI]}} & $p$ \\
\midrule
\endfirsthead
\multicolumn{8}{c}{\tablename\ \thetable\ -- continued} \\
\toprule
Matching specification & Audience & \shortstack{Matched\\pairs} & \shortstack{Max\\SMD} & HR [95\% CI] & $p$ & \shortstack{Count effect\\{[95\% CI]}} & $p$ \\
\midrule
\endhead
\midrule
\multicolumn{8}{r}{\footnotesize Continued on next page} \\
\endfoot
\bottomrule
\insertTableNotes
\endlastfoot
\addlinespace\multicolumn{8}{l}{\textit{Creation}} \\[2pt]
Adaptive reference & Examiner & 6,419 & 0.089 & 1.252$^{***}$ [1.157, 1.355] & $<.001$ & 5.0\% [-3.1, 13.7] & $.233$ \\
Adaptive reference & Applicant & 6,419 & 0.089 & 1.069$^{*}$ [0.993, 1.151] & $.078$ & -16.3\%$^{***}$ [-23.8, -8.0] & $<.001$ \\
Fixed caliper 0.05 & Examiner & 5,203 & 0.102 & 1.284$^{***}$ [1.174, 1.403] & $<.001$ & 2.4\% [-6.5, 12.3] & $.604$ \\
Fixed caliper 0.05 & Applicant & 5,203 & 0.102 & 1.071 [0.984, 1.166] & $.113$ & -12.3\%$^{**}$ [-21.0, -2.6] & $.014$ \\
Fixed caliper 0.10 & Examiner & 6,419 & 0.089 & 1.252$^{***}$ [1.157, 1.355] & $<.001$ & 5.0\% [-3.1, 13.7] & $.233$ \\
Fixed caliper 0.10 & Applicant & 6,419 & 0.089 & 1.069$^{*}$ [0.993, 1.151] & $.078$ & -16.3\%$^{***}$ [-23.8, -8.0] & $<.001$ \\
Fixed caliper 0.20 & Examiner & 7,365 & 0.082 & 1.231$^{***}$ [1.143, 1.326] & $<.001$ & 8.8\%$^{**}$ [1.0, 17.1] & $.026$ \\
Fixed caliper 0.20 & Applicant & 7,365 & 0.082 & 1.001 [0.936, 1.071] & $.969$ & -17.3\%$^{***}$ [-24.2, -9.8] & $<.001$ \\
Logit-scale caliper 0.20 & Examiner & 7,361 & 0.087 & 1.245$^{***}$ [1.156, 1.341] & $<.001$ & 6.2\% [-1.4, 14.4] & $.113$ \\
Logit-scale caliper 0.20 & Applicant & 7,361 & 0.087 & 1.002 [0.936, 1.072] & $.963$ & -16.5\%$^{***}$ [-23.5, -8.9] & $<.001$ \\
Lower control-pool cap & Examiner & 6,419 & 0.096 & 1.264$^{***}$ [1.167, 1.370] & $<.001$ & 5.5\% [-2.6, 14.1] & $.188$ \\
Lower control-pool cap & Applicant & 6,419 & 0.096 & 1.051 [0.976, 1.132] & $.187$ & -16.7\%$^{***}$ [-24.2, -8.6] & $<.001$ \\
Higher control-pool cap & Examiner & 6,424 & 0.088 & 1.255$^{***}$ [1.159, 1.358] & $<.001$ & 7.6\%$^{*}$ [-0.7, 16.6] & $.075$ \\
Higher control-pool cap & Applicant & 6,424 & 0.088 & 1.067$^{*}$ [0.991, 1.149] & $.086$ & -14.7\%$^{***}$ [-22.3, -6.3] & $<.001$ \\
Alternative pool seed 1 & Examiner & 6,402 & 0.092 & 1.262$^{***}$ [1.166, 1.366] & $<.001$ & 5.2\% [-2.9, 14.1] & $.214$ \\
Alternative pool seed 1 & Applicant & 6,402 & 0.092 & 1.061 [0.986, 1.142] & $.115$ & -14.3\%$^{***}$ [-22.0, -5.8] & $.001$ \\
Alternative pool seed 2 & Examiner & 6,424 & 0.086 & 1.261$^{***}$ [1.165, 1.364] & $<.001$ & 6.3\% [-1.9, 15.2] & $.136$ \\
Alternative pool seed 2 & Applicant & 6,424 & 0.086 & 1.055 [0.980, 1.136] & $.156$ & -15.1\%$^{***}$ [-22.9, -6.6] & $<.001$ \\
\addlinespace\multicolumn{8}{l}{\textit{Addition}} \\[2pt]
Adaptive reference & Examiner & 9,562 & 0.071 & 0.885$^{**}$ [0.801, 0.978] & $.016$ & -8.3\%$^{*}$ [-16.9, 1.2] & $.084$ \\
Adaptive reference & Applicant & 9,562 & 0.071 & 0.834$^{***}$ [0.763, 0.912] & $<.001$ & 5.3\% [-7.3, 19.5] & $.428$ \\
Fixed caliper 0.05 & Examiner & 9,414 & 0.079 & 0.834$^{***}$ [0.754, 0.923] & $<.001$ & -3.4\% [-12.8, 7.0] & $.507$ \\
Fixed caliper 0.05 & Applicant & 9,414 & 0.079 & 0.873$^{***}$ [0.795, 0.958] & $.004$ & -1.9\% [-13.5, 11.3] & $.768$ \\
Fixed caliper 0.10 & Examiner & 9,562 & 0.071 & 0.885$^{**}$ [0.801, 0.978] & $.016$ & -8.3\%$^{*}$ [-16.9, 1.2] & $.084$ \\
Fixed caliper 0.10 & Applicant & 9,562 & 0.071 & 0.834$^{***}$ [0.763, 0.912] & $<.001$ & 5.3\% [-7.3, 19.5] & $.428$ \\
Fixed caliper 0.20 & Examiner & 9,652 & 0.054 & 0.915$^{*}$ [0.829, 1.009] & $.075$ & -8.1\%$^{*}$ [-16.5, 1.1] & $.084$ \\
Fixed caliper 0.20 & Applicant & 9,652 & 0.054 & 0.886$^{***}$ [0.814, 0.965] & $.005$ & 4.0\% [-8.0, 17.7] & $.527$ \\
Logit-scale caliper 0.20 & Examiner & 9,586 & 0.056 & 0.917$^{*}$ [0.831, 1.012] & $.086$ & -4.5\% [-13.2, 5.0] & $.339$ \\
Logit-scale caliper 0.20 & Applicant & 9,586 & 0.056 & 0.900$^{**}$ [0.826, 0.980] & $.015$ & 3.4\% [-8.7, 17.2] & $.596$ \\
Lower control-pool cap & Examiner & 9,539 & 0.061 & 0.847$^{***}$ [0.767, 0.934] & $<.001$ & -6.9\% [-15.6, 2.6] & $.150$ \\
Lower control-pool cap & Applicant & 9,539 & 0.061 & 0.842$^{***}$ [0.771, 0.921] & $<.001$ & -3.9\% [-13.7, 7.1] & $.473$ \\
Higher control-pool cap & Examiner & 9,571 & 0.069 & 0.881$^{**}$ [0.797, 0.974] & $.013$ & -7.9\%$^{*}$ [-16.5, 1.5] & $.099$ \\
Higher control-pool cap & Applicant & 9,571 & 0.069 & 0.857$^{***}$ [0.785, 0.935] & $<.001$ & 1.4\% [-11.0, 15.5] & $.834$ \\
Alternative pool seed 1 & Examiner & 9,553 & 0.071 & 0.893$^{**}$ [0.809, 0.986] & $.026$ & -8.5\%$^{*}$ [-17.0, 0.9] & $.075$ \\
Alternative pool seed 1 & Applicant & 9,553 & 0.071 & 0.833$^{***}$ [0.763, 0.909] & $<.001$ & 1.2\% [-10.7, 14.8] & $.849$ \\
Alternative pool seed 2 & Examiner & 9,556 & 0.075 & 0.885$^{**}$ [0.800, 0.978] & $.017$ & -8.5\%$^{*}$ [-17.1, 0.9] & $.075$ \\
Alternative pool seed 2 & Applicant & 9,556 & 0.075 & 0.824$^{***}$ [0.754, 0.901] & $<.001$ & 1.9\% [-9.5, 14.9] & $.751$ \\
\addlinespace\multicolumn{8}{l}{\textit{Strict addition}} \\[2pt]
Adaptive reference & Examiner & 7,477 & 0.100 & 0.828$^{***}$ [0.741, 0.925] & $<.001$ & -2.8\% [-13.0, 8.5] & $.613$ \\
Adaptive reference & Applicant & 7,477 & 0.100 & 0.891$^{**}$ [0.805, 0.987] & $.027$ & -2.1\% [-13.4, 10.8] & $.741$ \\
Fixed caliper 0.05 & Examiner & 7,311 & 0.119 & 0.834$^{***}$ [0.743, 0.936] & $.002$ & 7.3\% [-4.5, 20.6] & $.233$ \\
Fixed caliper 0.05 & Applicant & 7,311 & 0.119 & 0.847$^{***}$ [0.758, 0.946] & $.003$ & -5.6\% [-17.2, 7.6] & $.388$ \\
Fixed caliper 0.10 & Examiner & 7,477 & 0.100 & 0.828$^{***}$ [0.741, 0.925] & $<.001$ & -2.8\% [-13.0, 8.5] & $.613$ \\
Fixed caliper 0.10 & Applicant & 7,477 & 0.100 & 0.891$^{**}$ [0.805, 0.987] & $.027$ & -2.1\% [-13.4, 10.8] & $.741$ \\
Fixed caliper 0.20 & Examiner & 7,588 & 0.081 & 0.879$^{**}$ [0.789, 0.980] & $.020$ & -4.4\% [-14.1, 6.3] & $.407$ \\
Fixed caliper 0.20 & Applicant & 7,588 & 0.081 & 0.902$^{**}$ [0.817, 0.996] & $.041$ & -1.1\% [-12.3, 11.6] & $.856$ \\
Logit-scale caliper 0.20 & Examiner & 7,510 & 0.083 & 0.851$^{***}$ [0.762, 0.949] & $.004$ & -3.0\% [-12.8, 7.8] & $.568$ \\
Logit-scale caliper 0.20 & Applicant & 7,510 & 0.083 & 0.946 [0.857, 1.043] & $.265$ & -3.6\% [-14.7, 9.1] & $.562$ \\
Lower control-pool cap & Examiner & 7,431 & 0.084 & 0.830$^{***}$ [0.742, 0.927] & $<.001$ & -2.1\% [-12.3, 9.3] & $.707$ \\
Lower control-pool cap & Applicant & 7,431 & 0.084 & 0.851$^{***}$ [0.768, 0.944] & $.002$ & -1.2\% [-13.3, 12.6] & $.858$ \\
Higher control-pool cap & Examiner & 7,480 & 0.097 & 0.815$^{***}$ [0.730, 0.911] & $<.001$ & -3.7\% [-13.6, 7.4] & $.501$ \\
Higher control-pool cap & Applicant & 7,480 & 0.097 & 0.880$^{**}$ [0.795, 0.975] & $.014$ & 0.2\% [-11.2, 13.1] & $.973$ \\
Alternative pool seed 1 & Examiner & 7,475 & 0.102 & 0.867$^{**}$ [0.775, 0.969] & $.012$ & -2.5\% [-12.7, 9.0] & $.660$ \\
Alternative pool seed 1 & Applicant & 7,475 & 0.102 & 0.870$^{***}$ [0.787, 0.962] & $.007$ & -3.1\% [-15.1, 10.6] & $.644$ \\
Alternative pool seed 2 & Examiner & 7,483 & 0.099 & 0.854$^{***}$ [0.764, 0.956] & $.006$ & -1.6\% [-11.8, 9.9] & $.777$ \\
Alternative pool seed 2 & Applicant & 7,483 & 0.099 & 0.895$^{**}$ [0.808, 0.991] & $.033$ & -0.3\% [-11.5, 12.3] & $.962$ \\
\addlinespace\multicolumn{8}{l}{\textit{Replacement}} \\[2pt]
Adaptive reference & Examiner & 2,381 & 0.090 & 0.879 [0.725, 1.067] & $.194$ & -16.4\%$^{*}$ [-31.2, 1.6] & $.072$ \\
Adaptive reference & Applicant & 2,381 & 0.090 & 0.951 [0.787, 1.151] & $.608$ & -9.4\% [-28.9, 15.6] & $.429$ \\
Fixed caliper 0.05 & Examiner & 2,345 & 0.107 & 0.929 [0.763, 1.132] & $.467$ & -23.7\%$^{**}$ [-38.0, -6.0] & $.011$ \\
Fixed caliper 0.05 & Applicant & 2,345 & 0.107 & 0.978 [0.804, 1.190] & $.826$ & -5.7\% [-27.3, 22.5] & $.662$ \\
Fixed caliper 0.10 & Examiner & 2,381 & 0.090 & 0.879 [0.725, 1.067] & $.194$ & -16.4\%$^{*}$ [-31.2, 1.6] & $.072$ \\
Fixed caliper 0.10 & Applicant & 2,381 & 0.090 & 0.951 [0.787, 1.151] & $.608$ & -9.4\% [-28.9, 15.6] & $.429$ \\
Fixed caliper 0.20 & Examiner & 2,404 & 0.096 & 0.946 [0.781, 1.147] & $.575$ & -16.1\%$^{*}$ [-30.3, 1.0] & $.064$ \\
Fixed caliper 0.20 & Applicant & 2,404 & 0.096 & 0.937 [0.774, 1.135] & $.508$ & -14.1\% [-33.4, 10.9] & $.244$ \\
Logit-scale caliper 0.20 & Examiner & 2,423 & 0.102 & 0.908 [0.749, 1.100] & $.324$ & -17.8\%$^{**}$ [-31.8, -0.9] & $.040$ \\
Logit-scale caliper 0.20 & Applicant & 2,423 & 0.102 & 0.965 [0.805, 1.156] & $.697$ & -12.1\% [-31.3, 12.4] & $.304$ \\
Lower control-pool cap & Examiner & 2,378 & 0.084 & 0.928 [0.765, 1.125] & $.445$ & -14.7\% [-29.8, 3.8] & $.112$ \\
Lower control-pool cap & Applicant & 2,378 & 0.084 & 1.027 [0.845, 1.248] & $.787$ & -14.7\% [-33.7, 9.7] & $.214$ \\
Higher control-pool cap & Examiner & 2,392 & 0.124 & 0.872 [0.717, 1.059] & $.167$ & -13.2\% [-28.1, 4.7] & $.138$ \\
Higher control-pool cap & Applicant & 2,392 & 0.124 & 0.983 [0.817, 1.184] & $.860$ & -20.4\%$^{*}$ [-38.0, 2.0] & $.072$ \\
Alternative pool seed 1 & Examiner & 2,378 & 0.116 & 0.977 [0.801, 1.192] & $.821$ & -9.7\% [-25.8, 9.8] & $.307$ \\
Alternative pool seed 1 & Applicant & 2,378 & 0.116 & 1.069 [0.881, 1.296] & $.498$ & -4.1\% [-25.2, 23.0] & $.744$ \\
Alternative pool seed 2 & Examiner & 2,337 & 0.107 & 0.901 [0.738, 1.099] & $.303$ & -15.5\% [-30.9, 3.3] & $.101$ \\
Alternative pool seed 2 & Applicant & 2,337 & 0.107 & 0.987 [0.812, 1.199] & $.895$ & -10.0\% [-30.4, 16.3] & $.421$ \\
\addlinespace\multicolumn{8}{l}{\textit{Cleanup}} \\[2pt]
Adaptive reference & Examiner & 46,419 & 0.095 & 1.136$^{***}$ [1.093, 1.181] & $<.001$ & 11.8\%$^{***}$ [7.7, 16.1] & $<.001$ \\
Adaptive reference & Applicant & 46,419 & 0.095 & 1.044$^{**}$ [1.001, 1.089] & $.044$ & -6.0\%$^{*}$ [-11.8, 0.2] & $.057$ \\
Fixed caliper 0.05 & Examiner & 46,052 & 0.087 & 1.141$^{***}$ [1.097, 1.187] & $<.001$ & 11.0\%$^{***}$ [6.8, 15.4] & $<.001$ \\
Fixed caliper 0.05 & Applicant & 46,052 & 0.087 & 1.075$^{***}$ [1.030, 1.122] & $<.001$ & -3.5\% [-9.5, 2.9] & $.283$ \\
Fixed caliper 0.10 & Examiner & 46,419 & 0.095 & 1.136$^{***}$ [1.093, 1.181] & $<.001$ & 11.8\%$^{***}$ [7.7, 16.1] & $<.001$ \\
Fixed caliper 0.10 & Applicant & 46,419 & 0.095 & 1.044$^{**}$ [1.001, 1.089] & $.044$ & -6.0\%$^{*}$ [-11.8, 0.2] & $.057$ \\
Fixed caliper 0.20 & Examiner & 46,842 & 0.108 & 1.121$^{***}$ [1.078, 1.165] & $<.001$ & 10.4\%$^{***}$ [6.4, 14.6] & $<.001$ \\
Fixed caliper 0.20 & Applicant & 46,842 & 0.108 & 1.041$^{*}$ [0.999, 1.084] & $.056$ & -5.3\%$^{*}$ [-11.1, 0.9] & $.092$ \\
Logit-scale caliper 0.20 & Examiner & 46,924 & 0.109 & 1.126$^{***}$ [1.083, 1.170] & $<.001$ & 11.5\%$^{***}$ [7.5, 15.7] & $<.001$ \\
Logit-scale caliper 0.20 & Applicant & 46,924 & 0.109 & 1.042$^{**}$ [1.000, 1.086] & $.049$ & -4.9\% [-10.5, 1.2] & $.113$ \\
Lower control-pool cap & Examiner & 46,375 & 0.091 & 1.129$^{***}$ [1.086, 1.174] & $<.001$ & 11.9\%$^{***}$ [7.7, 16.2] & $<.001$ \\
Lower control-pool cap & Applicant & 46,375 & 0.091 & 1.040$^{*}$ [0.997, 1.084] & $.068$ & -6.5\%$^{**}$ [-12.5, -0.2] & $.044$ \\
Higher control-pool cap & Examiner & 46,080 & 0.096 & 1.155$^{***}$ [1.110, 1.201] & $<.001$ & 9.7\%$^{***}$ [5.6, 14.1] & $<.001$ \\
Higher control-pool cap & Applicant & 46,080 & 0.096 & 1.065$^{***}$ [1.020, 1.112] & $.004$ & -4.5\% [-10.5, 2.0] & $.169$ \\
Alternative pool seed 1 & Examiner & 46,419 & 0.094 & 1.139$^{***}$ [1.095, 1.184] & $<.001$ & 12.1\%$^{***}$ [8.0, 16.4] & $<.001$ \\
Alternative pool seed 1 & Applicant & 46,419 & 0.094 & 1.051$^{**}$ [1.008, 1.097] & $.019$ & -5.9\%$^{*}$ [-11.6, 0.2] & $.058$ \\
Alternative pool seed 2 & Examiner & 46,421 & 0.095 & 1.136$^{***}$ [1.093, 1.181] & $<.001$ & 11.9\%$^{***}$ [7.8, 16.2] & $<.001$ \\
Alternative pool seed 2 & Applicant & 46,421 & 0.095 & 1.038$^{*}$ [0.995, 1.082] & $.084$ & -5.4\%$^{*}$ [-11.4, 0.9] & $.089$ \\
\end{longtable}
\end{ThreePartTable}
\normalsize
\end{landscape}

}{\placeholder{Run the R script to generate the matching-sensitivity table.}}

\FloatBarrier

\subsubsection{Outcome horizons, implementation lags, and citation-date measurement}
\label{app:outcome-robustness}

Table~\ref{tab:outcome-robustness-details} reports the underlying specification-level estimates summarized in Section~\ref{subsec:results_robustness}. The Cox rows vary the follow-up horizon, the assumed delay between the institutional revision and the beginning of the risk window, and the citation-date proxy. The Poisson rows provide the corresponding citation-volume checks. These estimates use the earlier constructed matched samples. 

\IfFileExists{generated/tables/table_outcome_robustness_details.tex}{
  \begin{landscape}
\scriptsize
\setlength{\tabcolsep}{3pt}
\begin{ThreePartTable}
\begin{TableNotes}[flushleft]\footnotesize
\item[]
Cox rows report hazard ratios. Poisson rows report percentage changes implied by the treatment-by-post coefficient. Implementation-lag specifications shift the start of the measurement window without changing the institutional event date. Cox lag and alternative-date specifications use the horizons shown explicitly in the Specification column. All estimates are read from the existing outcome-robustness workbooks; no models are re-estimated by the paper-output script. $^{***}p<.01$, $^{**}p<.05$, $^{*}p<.10$.
\end{TableNotes}
\begin{longtable}{lllclrrr}
\caption{Outcome-horizon, implementation-lag, and citation-date robustness}\label{tab:outcome-robustness-details}\\
\toprule
Family & Audience & Specification & Unit & Effect & 95\% CI & $p$ & $N$ \\
\midrule
\endfirsthead
\multicolumn{8}{c}{\tablename\ \thetable\ -- continued} \\
\toprule
Family & Audience & Specification & Unit & Effect & 95\% CI & $p$ & $N$ \\
\midrule
\endhead
\midrule
\multicolumn{8}{r}{\footnotesize Continued on next page} \\
\endfoot
\bottomrule
\insertTableNotes
\endlastfoot
\addlinespace\multicolumn{8}{l}{\textit{Creation}} \\[2pt]
Cox timing & Examiner & Horizon 12m & HR & 1.290$^{***}$ & [1.170, 1.421] & $<.001$ & 12,862 \\
Cox timing & Examiner & Horizon 24m & HR & 1.254$^{***}$ & [1.159, 1.357] & $<.001$ & 12,862 \\
Cox timing & Examiner & Horizon 36m & HR & 1.226$^{***}$ & [1.141, 1.318] & $<.001$ & 12,862 \\
Cox timing & Examiner & Lag 3m; 36m horizon & HR & 1.233$^{***}$ & [1.146, 1.327] & $<.001$ & 12,862 \\
Cox timing & Examiner & Lag 6m; 36m horizon & HR & 1.157$^{***}$ & [1.074, 1.246] & $<.001$ & 12,862 \\
Cox timing & Examiner & Lag 12m; 36m horizon & HR & 1.126$^{***}$ & [1.043, 1.216] & $.002$ & 12,862 \\
Cox timing & Examiner & Lag 24m; 36m horizon & HR & 1.131$^{***}$ & [1.044, 1.226] & $.003$ & 12,862 \\
Cox timing & Examiner & Earliest date; 36m horizon & HR & 1.228$^{***}$ & [1.142, 1.319] & $<.001$ & 12,862 \\
Cox timing & Examiner & Latest date; 36m horizon & HR & 1.173$^{***}$ & [1.090, 1.262] & $<.001$ & 12,862 \\
Cox timing & Applicant & Horizon 12m & HR & 1.006 & [0.920, 1.099] & $.897$ & 12,862 \\
Cox timing & Applicant & Horizon 24m & HR & 1.055 & [0.980, 1.137] & $.153$ & 12,862 \\
Cox timing & Applicant & Horizon 36m & HR & 1.033 & [0.965, 1.106] & $.350$ & 12,862 \\
Cox timing & Applicant & Lag 3m; 36m horizon & HR & 1.063$^{*}$ & [0.992, 1.138] & $.084$ & 12,862 \\
Cox timing & Applicant & Lag 6m; 36m horizon & HR & 1.037 & [0.968, 1.111] & $.299$ & 12,862 \\
Cox timing & Applicant & Lag 12m; 36m horizon & HR & 1.067$^{*}$ & [0.996, 1.143] & $.064$ & 12,862 \\
Cox timing & Applicant & Lag 24m; 36m horizon & HR & 1.004 & [0.938, 1.075] & $.909$ & 12,862 \\
Cox timing & Applicant & Earliest date; 36m horizon & HR & 1.034 & [0.966, 1.107] & $.339$ & 12,862 \\
Cox timing & Applicant & Latest date; 36m horizon & HR & 1.001 & [0.923, 1.086] & $.980$ & 12,862 \\
Poisson count & Examiner & Window 12m & \% & 10.7$^{*}$ & [-0.3, 23.0] & $.058$ & 11,096 \\
Poisson count & Examiner & Window 24m & \% & 6.1 & [-2.1, 14.9] & $.150$ & 15,092 \\
Poisson count & Examiner & Window 36m & \% & 0.8 & [-6.1, 8.2] & $.826$ & 17,160 \\
Poisson count & Examiner & Lag 3m; 24m window & \% & 5.0 & [-3.3, 13.9] & $.245$ & 14,896 \\
Poisson count & Examiner & Lag 6m; 24m window & \% & 2.7 & [-5.6, 11.6] & $.539$ & 14,808 \\
Poisson count & Examiner & Lag 12m; 24m window & \% & -1.4 & [-9.5, 7.4] & $.745$ & 14,492 \\
Poisson count & Examiner & Lag 24m; 24m window & \% & -0.1 & [-9.0, 9.7] & $.981$ & 14,008 \\
Poisson count & Examiner & Earliest date; 24m window & \% & 6.0 & [-2.2, 14.8] & $.158$ & 15,096 \\
Poisson count & Examiner & Latest date; 24m window & \% & 2.5 & [-5.8, 11.5] & $.564$ & 14,732 \\
Poisson count & Applicant & Window 12m & \% & -22.2$^{***}$ & [-30.3, -13.0] & $<.001$ & 13,012 \\
Poisson count & Applicant & Window 24m & \% & -15.9$^{***}$ & [-23.5, -7.6] & $<.001$ & 16,612 \\
Poisson count & Applicant & Window 36m & \% & -18.7$^{***}$ & [-26.5, -10.1] & $<.001$ & 18,420 \\
Poisson count & Applicant & Lag 3m; 24m window & \% & -14.7$^{***}$ & [-22.7, -5.9] & $.002$ & 16,660 \\
Poisson count & Applicant & Lag 6m; 24m window & \% & -13.3$^{***}$ & [-21.7, -3.9] & $.006$ & 16,576 \\
Poisson count & Applicant & Lag 12m; 24m window & \% & -15.1$^{***}$ & [-24.1, -5.0] & $.004$ & 16,736 \\
Poisson count & Applicant & Lag 24m; 24m window & \% & -18.8$^{***}$ & [-29.0, -7.1] & $.002$ & 16,600 \\
Poisson count & Applicant & Earliest date; 24m window & \% & -16.0$^{***}$ & [-23.5, -7.7] & $<.001$ & 16,632 \\
Poisson count & Applicant & Latest date; 24m window & \% & 1.3 & [-12.2, 16.8] & $.860$ & 12,388 \\
\addlinespace\multicolumn{8}{l}{\textit{Addition}} \\[2pt]
Cox timing & Examiner & Horizon 12m & HR & 0.809$^{***}$ & [0.714, 0.918] & $<.001$ & 19,110 \\
Cox timing & Examiner & Horizon 24m & HR & 0.865$^{***}$ & [0.784, 0.955] & $.004$ & 19,110 \\
Cox timing & Examiner & Horizon 36m & HR & 0.847$^{***}$ & [0.770, 0.933] & $<.001$ & 19,110 \\
Cox timing & Examiner & Lag 3m; 36m horizon & HR & 0.846$^{***}$ & [0.764, 0.937] & $.001$ & 19,110 \\
Cox timing & Examiner & Lag 6m; 36m horizon & HR & 0.841$^{***}$ & [0.755, 0.937] & $.002$ & 19,110 \\
Cox timing & Examiner & Lag 12m; 36m horizon & HR & 0.844$^{**}$ & [0.739, 0.964] & $.012$ & 19,110 \\
Cox timing & Examiner & Earliest date; 36m horizon & HR & 0.846$^{***}$ & [0.768, 0.931] & $<.001$ & 19,110 \\
Cox timing & Examiner & Latest date; 36m horizon & HR & 0.840$^{***}$ & [0.770, 0.917] & $<.001$ & 19,110 \\
Cox timing & Applicant & Horizon 12m & HR & 0.900$^{**}$ & [0.818, 0.990] & $.031$ & 19,110 \\
Cox timing & Applicant & Horizon 24m & HR & 0.847$^{***}$ & [0.774, 0.926] & $<.001$ & 19,110 \\
Cox timing & Applicant & Horizon 36m & HR & 0.842$^{***}$ & [0.771, 0.920] & $<.001$ & 19,110 \\
Cox timing & Applicant & Lag 3m; 36m horizon & HR & 0.819$^{***}$ & [0.743, 0.904] & $<.001$ & 19,110 \\
Cox timing & Applicant & Lag 6m; 36m horizon & HR & 0.770$^{***}$ & [0.692, 0.856] & $<.001$ & 19,110 \\
Cox timing & Applicant & Lag 12m; 36m horizon & HR & 0.677$^{***}$ & [0.591, 0.776] & $<.001$ & 19,110 \\
Cox timing & Applicant & Earliest date; 36m horizon & HR & 0.842$^{***}$ & [0.770, 0.920] & $<.001$ & 19,110 \\
Cox timing & Applicant & Latest date; 36m horizon & HR & 0.810$^{***}$ & [0.754, 0.870] & $<.001$ & 19,110 \\
Poisson count & Examiner & Window 12m & \% & -9.7 & [-20.5, 2.5] & $.113$ & 7,040 \\
Poisson count & Examiner & Window 24m & \% & -8.5$^{*}$ & [-17.0, 0.8] & $.073$ & 11,000 \\
Poisson count & Examiner & Lag 3m; 24m window & \% & -9.8$^{**}$ & [-18.5, -0.2] & $.045$ & 10,656 \\
Poisson count & Examiner & Lag 6m; 24m window & \% & -9.0$^{*}$ & [-18.2, 1.2] & $.081$ & 10,232 \\
Poisson count & Examiner & Earliest date; 24m window & \% & -9.0$^{*}$ & [-17.4, 0.4] & $.060$ & 11,000 \\
Poisson count & Examiner & Latest date; 24m window & \% & -7.0 & [-15.1, 1.9] & $.118$ & 12,988 \\
Poisson count & Applicant & Window 12m & \% & 3.1 & [-6.8, 14.1] & $.548$ & 12,320 \\
Poisson count & Applicant & Window 24m & \% & 1.6 & [-9.9, 14.5] & $.801$ & 15,092 \\
Poisson count & Applicant & Lag 3m; 24m window & \% & 2.9 & [-9.9, 17.7] & $.670$ & 14,532 \\
Poisson count & Applicant & Lag 6m; 24m window & \% & -7.3 & [-21.6, 9.6] & $.377$ & 14,148 \\
Poisson count & Applicant & Earliest date; 24m window & \% & 1.3 & [-10.1, 14.3] & $.828$ & 15,088 \\
Poisson count & Applicant & Latest date; 24m window & \% & -13.1$^{***}$ & [-20.5, -5.2] & $.002$ & 18,456 \\
\addlinespace\multicolumn{8}{l}{\textit{Strict addition}} \\[2pt]
Cox timing & Examiner & Horizon 12m & HR & 0.824$^{***}$ & [0.713, 0.953] & $.009$ & 14,950 \\
Cox timing & Examiner & Horizon 24m & HR & 0.831$^{***}$ & [0.743, 0.930] & $.001$ & 14,950 \\
Cox timing & Examiner & Horizon 36m & HR & 0.827$^{***}$ & [0.741, 0.922] & $<.001$ & 14,950 \\
Cox timing & Examiner & Lag 3m; 36m horizon & HR & 0.819$^{***}$ & [0.729, 0.920] & $<.001$ & 14,950 \\
Cox timing & Examiner & Lag 6m; 36m horizon & HR & 0.835$^{***}$ & [0.738, 0.945] & $.004$ & 14,950 \\
Cox timing & Examiner & Lag 12m; 36m horizon & HR & 0.820$^{***}$ & [0.708, 0.951] & $.008$ & 14,950 \\
Cox timing & Examiner & Earliest date; 36m horizon & HR & 0.824$^{***}$ & [0.738, 0.920] & $<.001$ & 14,950 \\
Cox timing & Examiner & Latest date; 36m horizon & HR & 0.816$^{***}$ & [0.737, 0.902] & $<.001$ & 14,950 \\
Cox timing & Applicant & Horizon 12m & HR & 0.941 & [0.842, 1.051] & $.282$ & 14,950 \\
Cox timing & Applicant & Horizon 24m & HR & 0.888$^{**}$ & [0.801, 0.985] & $.025$ & 14,950 \\
Cox timing & Applicant & Horizon 36m & HR & 0.881$^{**}$ & [0.795, 0.976] & $.015$ & 14,950 \\
Cox timing & Applicant & Lag 3m; 36m horizon & HR & 0.803$^{***}$ & [0.718, 0.897] & $<.001$ & 14,950 \\
Cox timing & Applicant & Lag 6m; 36m horizon & HR & 0.729$^{***}$ & [0.646, 0.822] & $<.001$ & 14,950 \\
Cox timing & Applicant & Lag 12m; 36m horizon & HR & 0.695$^{***}$ & [0.595, 0.812] & $<.001$ & 14,950 \\
Cox timing & Applicant & Earliest date; 36m horizon & HR & 0.879$^{**}$ & [0.793, 0.974] & $.014$ & 14,950 \\
Cox timing & Applicant & Latest date; 36m horizon & HR & 0.821$^{***}$ & [0.756, 0.892] & $<.001$ & 14,950 \\
Poisson count & Examiner & Window 12m & \% & -3.0 & [-16.1, 12.2] & $.684$ & 5,756 \\
Poisson count & Examiner & Window 24m & \% & -2.8 & [-12.9, 8.5] & $.614$ & 8,992 \\
Poisson count & Examiner & Lag 3m; 24m window & \% & -1.9 & [-12.5, 9.9] & $.737$ & 8,692 \\
Poisson count & Examiner & Lag 6m; 24m window & \% & -0.4 & [-11.5, 12.1] & $.951$ & 8,344 \\
Poisson count & Examiner & Earliest date; 24m window & \% & -3.2 & [-13.3, 8.0] & $.560$ & 8,992 \\
Poisson count & Examiner & Latest date; 24m window & \% & -6.7 & [-15.7, 3.4] & $.186$ & 10,552 \\
Poisson count & Applicant & Window 12m & \% & 4.6 & [-6.4, 16.8] & $.430$ & 9,980 \\
Poisson count & Applicant & Window 24m & \% & 0.1 & [-11.3, 12.9] & $.990$ & 11,992 \\
Poisson count & Applicant & Lag 3m; 24m window & \% & 0.9 & [-11.7, 15.3] & $.896$ & 11,576 \\
Poisson count & Applicant & Lag 6m; 24m window & \% & -8.3 & [-22.5, 8.5] & $.312$ & 11,300 \\
Poisson count & Applicant & Earliest date; 24m window & \% & -0.0 & [-11.4, 12.8] & $.996$ & 11,992 \\
Poisson count & Applicant & Latest date; 24m window & \% & -15.3$^{***}$ & [-22.7, -7.1] & $<.001$ & 14,532 \\
\addlinespace\multicolumn{8}{l}{\textit{Replacement}} \\[2pt]
Cox timing & Examiner & Horizon 12m & HR & 0.929 & [0.720, 1.200] & $.573$ & 4,678 \\
Cox timing & Examiner & Horizon 24m & HR & 0.908 & [0.746, 1.106] & $.337$ & 4,678 \\
Cox timing & Examiner & Horizon 36m & HR & 0.904 & [0.745, 1.096] & $.304$ & 4,678 \\
Cox timing & Examiner & Lag 3m; 36m horizon & HR & 0.848 & [0.685, 1.050] & $.130$ & 4,678 \\
Cox timing & Examiner & Lag 6m; 36m horizon & HR & 0.782$^{**}$ & [0.618, 0.991] & $.042$ & 4,678 \\
Cox timing & Examiner & Lag 12m; 36m horizon & HR & 0.885 & [0.676, 1.159] & $.376$ & 4,678 \\
Cox timing & Examiner & Earliest date; 36m horizon & HR & 0.899 & [0.741, 1.091] & $.282$ & 4,678 \\
Cox timing & Examiner & Latest date; 36m horizon & HR & 0.835$^{*}$ & [0.693, 1.007] & $.059$ & 4,678 \\
Cox timing & Applicant & Horizon 12m & HR & 0.970 & [0.774, 1.215] & $.788$ & 4,678 \\
Cox timing & Applicant & Horizon 24m & HR & 0.979 & [0.804, 1.191] & $.829$ & 4,678 \\
Cox timing & Applicant & Horizon 36m & HR & 0.977 & [0.805, 1.186] & $.816$ & 4,678 \\
Cox timing & Applicant & Lag 3m; 36m horizon & HR & 0.933 & [0.754, 1.155] & $.525$ & 4,678 \\
Cox timing & Applicant & Lag 6m; 36m horizon & HR & 0.889 & [0.708, 1.116] & $.310$ & 4,678 \\
Cox timing & Applicant & Lag 12m; 36m horizon & HR & 1.068 & [0.798, 1.428] & $.659$ & 4,678 \\
Cox timing & Applicant & Earliest date; 36m horizon & HR & 0.965 & [0.794, 1.172] & $.718$ & 4,678 \\
Cox timing & Applicant & Latest date; 36m horizon & HR & 1.020 & [0.872, 1.193] & $.804$ & 4,678 \\
Poisson count & Examiner & Window 12m & \% & -21.6$^{*}$ & [-39.7, 1.9] & $.069$ & 1,812 \\
Poisson count & Examiner & Window 24m & \% & -22.0$^{**}$ & [-36.4, -4.3] & $.017$ & 2,836 \\
Poisson count & Examiner & Lag 3m; 24m window & \% & -26.1$^{***}$ & [-40.4, -8.5] & $.006$ & 2,664 \\
Poisson count & Examiner & Lag 6m; 24m window & \% & -28.2$^{***}$ & [-42.7, -10.0] & $.004$ & 2,540 \\
Poisson count & Examiner & Earliest date; 24m window & \% & -23.1$^{**}$ & [-37.4, -5.7] & $.012$ & 2,832 \\
Poisson count & Examiner & Latest date; 24m window & \% & -16.6$^{*}$ & [-32.0, 2.2] & $.080$ & 3,004 \\
Poisson count & Applicant & Window 12m & \% & -4.0 & [-26.6, 25.7] & $.768$ & 2,460 \\
Poisson count & Applicant & Window 24m & \% & -7.1 & [-30.0, 23.3] & $.611$ & 3,172 \\
Poisson count & Applicant & Lag 3m; 24m window & \% & -9.1 & [-32.7, 22.8] & $.534$ & 3,000 \\
Poisson count & Applicant & Lag 6m; 24m window & \% & -21.8 & [-44.7, 10.6] & $.164$ & 2,880 \\
Poisson count & Applicant & Earliest date; 24m window & \% & -7.4 & [-30.3, 23.0] & $.595$ & 3,172 \\
Poisson count & Applicant & Latest date; 24m window & \% & 36.2$^{**}$ & [0.0, 85.4] & $.050$ & 3,280 \\
\addlinespace\multicolumn{8}{l}{\textit{Cleanup}} \\[2pt]
Cox timing & Examiner & Horizon 12m & HR & 1.175$^{***}$ & [1.120, 1.233] & $<.001$ & 92,844 \\
Cox timing & Examiner & Horizon 24m & HR & 1.142$^{***}$ & [1.099, 1.187] & $<.001$ & 92,844 \\
Cox timing & Examiner & Horizon 36m & HR & 1.141$^{***}$ & [1.098, 1.185] & $<.001$ & 92,844 \\
Cox timing & Examiner & Lag 3m; 36m horizon & HR & 1.117$^{***}$ & [1.073, 1.164] & $<.001$ & 92,844 \\
Cox timing & Examiner & Lag 6m; 36m horizon & HR & 1.108$^{***}$ & [1.061, 1.156] & $<.001$ & 92,844 \\
Cox timing & Examiner & Lag 12m; 36m horizon & HR & 1.088$^{***}$ & [1.033, 1.146] & $.002$ & 92,844 \\
Cox timing & Examiner & Earliest date; 36m horizon & HR & 1.140$^{***}$ & [1.097, 1.184] & $<.001$ & 92,844 \\
Cox timing & Examiner & Latest date; 36m horizon & HR & 1.118$^{***}$ & [1.080, 1.159] & $<.001$ & 92,844 \\
Cox timing & Applicant & Horizon 12m & HR & 1.058$^{**}$ & [1.010, 1.109] & $.018$ & 92,844 \\
Cox timing & Applicant & Horizon 24m & HR & 1.053$^{**}$ & [1.010, 1.098] & $.015$ & 92,844 \\
Cox timing & Applicant & Horizon 36m & HR & 1.052$^{**}$ & [1.009, 1.096] & $.017$ & 92,844 \\
Cox timing & Applicant & Lag 3m; 36m horizon & HR & 1.052$^{**}$ & [1.006, 1.100] & $.026$ & 92,844 \\
Cox timing & Applicant & Lag 6m; 36m horizon & HR & 1.036 & [0.987, 1.088] & $.149$ & 92,844 \\
Cox timing & Applicant & Lag 12m; 36m horizon & HR & 1.060$^{*}$ & [0.998, 1.126] & $.060$ & 92,844 \\
Cox timing & Applicant & Earliest date; 36m horizon & HR & 1.051$^{**}$ & [1.009, 1.096] & $.018$ & 92,844 \\
Cox timing & Applicant & Latest date; 36m horizon & HR & 1.059$^{***}$ & [1.024, 1.095] & $<.001$ & 92,844 \\
Poisson count & Examiner & Window 12m & \% & 17.2$^{***}$ & [11.7, 22.9] & $<.001$ & 41,732 \\
Poisson count & Examiner & Window 24m & \% & 12.1$^{***}$ & [8.0, 16.4] & $<.001$ & 61,832 \\
Poisson count & Examiner & Lag 3m; 24m window & \% & 10.3$^{***}$ & [6.0, 14.8] & $<.001$ & 58,864 \\
Poisson count & Examiner & Lag 6m; 24m window & \% & 9.4$^{***}$ & [4.9, 14.2] & $<.001$ & 56,476 \\
Poisson count & Examiner & Earliest date; 24m window & \% & 12.0$^{***}$ & [7.8, 16.3] & $<.001$ & 61,828 \\
Poisson count & Examiner & Latest date; 24m window & \% & 7.9$^{***}$ & [3.8, 12.0] & $<.001$ & 66,956 \\
Poisson count & Applicant & Window 12m & \% & -2.3 & [-7.6, 3.3] & $.413$ & 44,876 \\
Poisson count & Applicant & Window 24m & \% & -5.5$^{*}$ & [-11.3, 0.6] & $.076$ & 58,948 \\
Poisson count & Applicant & Lag 3m; 24m window & \% & -7.5$^{**}$ & [-13.8, -0.7] & $.031$ & 56,520 \\
Poisson count & Applicant & Lag 6m; 24m window & \% & -7.7$^{*}$ & [-14.9, 0.1] & $.054$ & 54,084 \\
Poisson count & Applicant & Earliest date; 24m window & \% & -6.0$^{*}$ & [-11.7, 0.1] & $.055$ & 59,032 \\
Poisson count & Applicant & Latest date; 24m window & \% & 6.6$^{**}$ & [1.1, 12.3] & $.017$ & 68,248 \\
\end{longtable}
\end{ThreePartTable}
\normalsize
\end{landscape}

}{\placeholder{Run the R script to generate the detailed outcome-robustness table.}}

\FloatBarrier

\subsubsection{Cited-record and CPC-definition checks}
\label{app:route-checks}

Table~\ref{tab:routes} crosses two choices in the constructed data: whether treatment status is defined from the focal grant's or pre-grant publication's CPC assignments, and whether subsequent citations are observed in the PatentsView cited-grant or cited-publication edge table. The citing unit and audience-specific PatEx timing proxy remain unchanged. Each combination is rematched before estimation. 
The checks use a common year--AU4 exact-cell design for both events. 

\IfFileExists{generated/tables/table_route_effects.tex}{
  \begin{table}[htbp]
\centering
\caption{Cited-record and CPC-definition robustness checks}
\label{tab:routes}
\small
\begin{adjustbox}{max width=\linewidth,max totalheight=0.82\textheight,keepaspectratio,center}
\begin{threeparttable}
\begin{tabular}{llrrrrrrrrrr}
\toprule
\multicolumn{4}{c}{} & \multicolumn{4}{c}{Cox timing} & \multicolumn{4}{c}{Citation counts} \\
\cmidrule(lr){5-8}\cmidrule(lr){9-12}
\multicolumn{4}{c}{} & \multicolumn{2}{c}{Examiner} & \multicolumn{2}{c}{Applicant} & \multicolumn{2}{c}{Examiner} & \multicolumn{2}{c}{Applicant} \\
\cmidrule(lr){5-6}\cmidrule(lr){7-8}\cmidrule(lr){9-10}\cmidrule(lr){11-12}
\shortstack{CPC\\basis} & \shortstack{Cited\\record} & $N$ & \shortstack{Max\\SMD} & HR & $p$ & HR & $p$ & \shortstack{Effect\\(\%)} & $p$ & \shortstack{Effect\\(\%)} & $p$ \\
\midrule
\multicolumn{12}{l}{\textit{Creation}} \\[2pt]
Grant & Grant & 6,419 & 0.089 & 1.25$^{***}$ & $<.001$ & 1.07$^{*}$ & $.078$ & 5.0\% & $.233$ & -16.3\%$^{***}$ & $<.001$ \\
 &  &  &  & $(0.05)$ &  & $(0.04)$ &  & $(4.3\%)$ &  & $(4.0\%)$ &  \\
 & Pre-grant & 6,431 & 0.087 & 1.26$^{***}$ & $<.001$ & 1.14$^{***}$ & $<.001$ & 0.9\% & $.805$ & -10.2\%$^{**}$ & $.015$ \\
 &  &  &  & $(0.05)$ &  & $(0.04)$ &  & $(3.6\%)$ &  & $(4.0\%)$ &  \\
Pre-grant & Grant & 6,339 & 0.088 & 1.27$^{***}$ & $<.001$ & 1.07$^{*}$ & $.074$ & 9.7\%$^{**}$ & $.024$ & -13.9\%$^{***}$ & $.002$ \\
 &  &  &  & $(0.05)$ &  & $(0.04)$ &  & $(4.5\%)$ &  & $(4.1\%)$ &  \\
 & Pre-grant & 6,330 & 0.091 & 1.29$^{***}$ & $<.001$ & 1.15$^{***}$ & $<.001$ & -2.3\% & $.524$ & -12.3\%$^{***}$ & $.003$ \\
 &  &  &  & $(0.05)$ &  & $(0.04)$ &  & $(3.6\%)$ &  & $(3.9\%)$ &  \\
\addlinespace
\multicolumn{12}{l}{\textit{Addition}} \\[2pt]
Grant & Grant & 6,944 & 0.129 & 0.82$^{***}$ & $.002$ & 0.79$^{***}$ & $<.001$ & -26.6\%$^{***}$ & $<.001$ & -3.4\% & $.696$ \\
 &  &  &  & $(0.05)$ &  & $(0.05)$ &  & $(4.6\%)$ &  & $(8.6\%)$ &  \\
 & Pre-grant &  & 0.129 & 0.99 & $.887$ & 1.08$^{*}$ & $.074$ & -6.7\%$^{**}$ & $.039$ & 2.7\% & $.646$ \\
 &  &  &  & $(0.04)$ &  & $(0.05)$ &  & $(3.2\%)$ &  & $(6.1\%)$ &  \\
Pre-grant & Grant & 8,623 & 0.138 & 0.81$^{***}$ & $<.001$ & 0.90$^{**}$ & $.030$ & -26.7\%$^{***}$ & $<.001$ & 5.1\% & $.574$ \\
 &  &  &  & $(0.04)$ &  & $(0.05)$ &  & $(4.0\%)$ &  & $(9.2\%)$ &  \\
 & Pre-grant &  & 0.134 & 0.98 & $.618$ & 1.14$^{***}$ & $<.001$ & -6.8\%$^{**}$ & $.023$ & 8.4\% & $.130$ \\
 &  &  &  & $(0.03)$ &  & $(0.04)$ &  & $(2.9\%)$ &  & $(5.8\%)$ &  \\
\addlinespace
\multicolumn{12}{l}{\textit{Replacement}} \\[2pt]
Grant & Grant & 1,969 & 0.189 & 0.72$^{***}$ & $.008$ & 1.10 & $.393$ & -33.1\%$^{***}$ & $<.001$ & 2.0\% & $.881$ \\
 &  &  &  & $(0.09)$ &  & $(0.13)$ &  & $(7.2\%)$ &  & $(13.8\%)$ &  \\
 & Pre-grant & 1,039 & 0.183 & 1.09 & $.387$ & 1.57$^{***}$ & $<.001$ & -9.7\% & $.309$ & 24.0\% & $.162$ \\
 &  &  &  & $(0.11)$ &  & $(0.20)$ &  & $(9.1\%)$ &  & $(19.1\%)$ &  \\
\addlinespace
\multicolumn{12}{l}{\textit{Cleanup}} \\[2pt]
Grant & Grant & 32,214 & 0.081 & 1.13$^{***}$ & $<.001$ & 1.09$^{***}$ & $.002$ & 13.8\%$^{***}$ & $<.001$ & 11.7\%$^{**}$ & $.020$ \\
 &  &  &  & $(0.03)$ &  & $(0.03)$ &  & $(3.0\%)$ &  & $(5.3\%)$ &  \\
 & Pre-grant & 32,179 & 0.081 & 1.17$^{***}$ & $<.001$ & 1.11$^{***}$ & $<.001$ & 0.8\% & $.646$ & -5.9\%$^{**}$ & $.025$ \\
 &  &  &  & $(0.02)$ &  & $(0.02)$ &  & $(1.7\%)$ &  & $(2.5\%)$ &  \\
Pre-grant & Grant & 27 & 0.400 & --- & --- & --- & --- & -44.4\% & $.711$ & -40.0\% & $.611$ \\
 &  &  &  &  &  &  &  & $(88.2\%)$ &  & $(60.2\%)$ &  \\
 & Pre-grant & 28 & 0.392 & --- & --- & --- & --- & -55.6\% & $.336$ & -42.5\%$^{*}$ & $.096$ \\
 &  &  &  &  &  &  &  & $(37.4\%)$ &  & $(19.1\%)$ &  \\
\addlinespace
\bottomrule
\end{tabular}
\begin{tablenotes}[flushleft]\footnotesize
\item[]
Examiner and applicant estimates are reported in adjacent columns. The first row for each specification reports the point estimate and unadjusted two-sided $p$-value; the following row reports the delta-method SE in parentheses. For the Cox estimates, $SE(\exp\widehat\beta)=\exp(\widehat\beta)SE(\widehat\beta)$. For count effects, $SE\{100[\exp(\widehat\beta)-1]\}=100\exp(\widehat\beta)SE(\widehat\beta)$. Cox estimates are hazard ratios; count effects are percentage changes implied by the Poisson DiD coefficient. Rows cross the record defining CPC treatment with the record to which the subsequent citation is recorded. All specifications use current-green controls and publication-year-by-AU4 exact cells and are matched separately. $N$ is the number of treated patents retained; Max SMD below .10 indicates acceptable post-match balance. All cited-record and CPC-basis comparisons use publication-year-by-AU4 exact cells and are separately rematched within each source/target combination. Underlying model SEs are clustered by matched pair. Because support and matches change across rows, estimates are not causal contrasts between record definitions. $^{***}p<.01$, $^{**}p<.05$, $^{*}p<.10$.
\end{tablenotes}
\end{threeparttable}
\end{adjustbox}
\end{table}

}{\placeholder{Run the R script to generate the combined cited-record and CPC-definition table.}}

\FloatBarrier

\subsubsection{Count functional-form robustness}
\label{app:count-functional-form}

Table~\ref{tab:count-functional-form} compares the primary Poisson citation-volume estimator with additive linear, log-one-plus-count, and any-citation specifications. The estimators answer related but distinct questions and their coefficients therefore have different units. We use the alternatives to assess whether qualitative conclusions depend on the distributional representation of the count outcome.

\IfFileExists{generated/tables/table_count_functional_form.tex}{
  \scriptsize
\setlength{\tabcolsep}{3pt}
\begin{ThreePartTable}
\begin{TableNotes}[flushleft]\footnotesize
\item[]
All rows use the same matched samples, narrow citation-date proxy, and symmetric 24-month pre- and post-event windows as used for the baseline results. Poisson effects are percentage changes implied by the incidence-rate coefficient. Linear-level effects are additive citation-count changes. The log(1+y) model is reported in transformed outcome units. The any-citation linear probability model reports percentage-point changes in the probability of receiving at least one citation in the corresponding period. Standard errors are clustered by matched pair. $^{***}p<.01$, $^{**}p<.05$, $^{*}p<.10$.
\end{TableNotes}
\begin{longtable}{lllrrrr}
\caption{Functional-form robustness of the count estimates}\label{tab:count-functional-form}\\
\toprule
Audience & Estimator & Unit & Effect & 95\% CI & $p$ & $N$ \\
\midrule
\endfirsthead
\multicolumn{7}{c}{\tablename\ \thetable\ -- continued} \\
\toprule
Audience & Estimator & Unit & Effect & 95\% CI & $p$ & $N$ \\
\midrule
\endhead
\midrule
\multicolumn{7}{r}{\footnotesize Continued on next page} \\
\endfoot
\bottomrule
\insertTableNotes
\endlastfoot
\addlinespace\multicolumn{7}{l}{\textit{Creation}} \\[2pt]
Examiner & Poisson & \% & $6.054$ & [-2.105, 14.894] & $.150$ & 15,092 \\
Examiner & Linear levels & citations & $0.019$ & [-0.014, 0.051] & $.260$ & 25,724 \\
Examiner & Linear log(1+y) & $\log(1+y)$ & $0.011$ & [-0.005, 0.026] & $.181$ & 25,724 \\
Examiner & Any-citation LPM & pp & $1.120$ & [-0.618, 2.857] & $.207$ & 25,724 \\
Applicant & Poisson & \% & $-15.919^{***}$ & [-23.473, -7.620] & $<.001$ & 16,612 \\
Applicant & Linear levels & citations & $-0.083$ & [-0.207, 0.042] & $.193$ & 25,724 \\
Applicant & Linear log(1+y) & $\log(1+y)$ & $-0.008$ & [-0.028, 0.012] & $.426$ & 25,724 \\
Applicant & Any-citation LPM & pp & $0.249$ & [-1.422, 1.919] & $.770$ & 25,724 \\
\addlinespace\multicolumn{7}{l}{\textit{Addition}} \\[2pt]
Examiner & Poisson & \% & $-8.525^{*}$ & [-17.014, 0.832] & $.073$ & 11,000 \\
Examiner & Linear levels & citations & $0.000$ & [-0.013, 0.013] & $.987$ & 38,220 \\
Examiner & Linear log(1+y) & $\log(1+y)$ & $0.000$ & [-0.007, 0.007] & $.988$ & 38,220 \\
Examiner & Any-citation LPM & pp & $0.126$ & [-0.781, 1.032] & $.786$ & 38,220 \\
Applicant & Poisson & \% & $1.555$ & [-9.915, 14.487] & $.801$ & 15,092 \\
Applicant & Linear levels & citations & $0.159^{**}$ & [0.022, 0.296] & $.023$ & 38,220 \\
Applicant & Linear log(1+y) & $\log(1+y)$ & $0.016^{***}$ & [0.006, 0.027] & $.003$ & 38,220 \\
Applicant & Any-citation LPM & pp & $0.827$ & [-0.163, 1.817] & $.102$ & 38,220 \\
\addlinespace\multicolumn{7}{l}{\textit{Strict addition}} \\[2pt]
Examiner & Poisson & \% & $-2.786$ & [-12.893, 8.494] & $.614$ & 8,992 \\
Examiner & Linear levels & citations & $0.015^{*}$ & [-0.001, 0.030] & $.064$ & 29,900 \\
Examiner & Linear log(1+y) & $\log(1+y)$ & $0.008^{*}$ & [-0.001, 0.016] & $.074$ & 29,900 \\
Examiner & Any-citation LPM & pp & $0.870$ & [-0.195, 1.935] & $.110$ & 29,900 \\
Applicant & Poisson & \% & $0.075$ & [-11.257, 12.854] & $.990$ & 11,992 \\
Applicant & Linear levels & citations & $0.172^{**}$ & [0.040, 0.304] & $.011$ & 29,900 \\
Applicant & Linear log(1+y) & $\log(1+y)$ & $0.026^{***}$ & [0.014, 0.038] & $<.001$ & 29,900 \\
Applicant & Any-citation LPM & pp & $1.431^{**}$ & [0.297, 2.566] & $.013$ & 29,900 \\
\addlinespace\multicolumn{7}{l}{\textit{Replacement}} \\[2pt]
Examiner & Poisson & \% & $-21.973^{**}$ & [-36.404, -4.267] & $.017$ & 2,836 \\
Examiner & Linear levels & citations & $-0.038^{**}$ & [-0.066, -0.009] & $.010$ & 9,356 \\
Examiner & Linear log(1+y) & $\log(1+y)$ & $-0.021^{***}$ & [-0.037, -0.005] & $.009$ & 9,356 \\
Examiner & Any-citation LPM & pp & $-2.608^{**}$ & [-4.593, -0.623] & $.010$ & 9,356 \\
Applicant & Poisson & \% & $-7.084$ & [-29.970, 23.281] & $.611$ & 3,172 \\
Applicant & Linear levels & citations & $-0.066$ & [-0.165, 0.033] & $.189$ & 9,356 \\
Applicant & Linear log(1+y) & $\log(1+y)$ & $-0.019^{*}$ & [-0.039, 0.001] & $.056$ & 9,356 \\
Applicant & Any-citation LPM & pp & $-0.556$ & [-2.564, 1.453] & $.588$ & 9,356 \\
\addlinespace\multicolumn{7}{l}{\textit{Cleanup}} \\[2pt]
Examiner & Poisson & \% & $12.112^{***}$ & [7.963, 16.420] & $<.001$ & 61,832 \\
Examiner & Linear levels & citations & $0.017^{***}$ & [0.011, 0.024] & $<.001$ & 185,688 \\
Examiner & Linear log(1+y) & $\log(1+y)$ & $0.010^{***}$ & [0.007, 0.014] & $<.001$ & 185,688 \\
Examiner & Any-citation LPM & pp & $1.267^{***}$ & [0.830, 1.704] & $<.001$ & 185,688 \\
Applicant & Poisson & \% & $-5.519^{*}$ & [-11.262, 0.597] & $.076$ & 58,948 \\
Applicant & Linear levels & citations & $0.045^{*}$ & [-0.003, 0.093] & $.065$ & 185,688 \\
Applicant & Linear log(1+y) & $\log(1+y)$ & $-0.007^{***}$ & [-0.011, -0.003] & $.001$ & 185,688 \\
Applicant & Any-citation LPM & pp & $-0.142$ & [-0.551, 0.267] & $.496$ & 185,688 \\
\end{longtable}
\end{ThreePartTable}
\normalsize

}{\placeholder{Run the R script to generate the count functional-form table.}}

\subsection{Exploratory moderator analyses}
\label{app:moderators}

Section~\ref{subsec:heterogeneity-results} summarizes the interaction patterns between treatment variables and moderators as explained and motivated in Appendix~\ref{app:moderator-methods}. Table~\ref{tab:moderators-bh} reports the complete exploratory family. Timing cells contain ratios of hazard ratios (RoHRs), and citation-volume cells contain ratios of incidence-rate ratios (RoIRRs).
Each cell reports the interaction estimate, 95\% confidence interval, unadjusted $p$-value, and Benjamini--Hochberg-adjusted $q$-value. The models use the baseline matched samples rather than rematching within moderator subgroups. The 2013 examiner broad-CPC-scope count interaction is reported as unstable because of numerical separation and is excluded from its BH family.

\IfFileExists{generated/tables/table_moderators_bh.tex}{
  \scriptsize
\setlength{\tabcolsep}{3pt}
\begin{ThreePartTable}
\begin{TableNotes}[flushleft]\footnotesize
\item[]
Each cell reports the exponentiated interaction estimate with its 95\% confidence interval in brackets; the second line reports the unadjusted two-sided p-value and Benjamini--Hochberg-adjusted q-value. RoHR denotes the ratio of hazard ratios obtained from the Cox treatment-by-moderator interaction. RoIRR denotes the ratio of incidence-rate ratios obtained from the Poisson treatment-by-post-by-moderator interaction. Values above one indicate a more positive treatment contrast when the binary moderator equals one; values below one indicate a less positive contrast. BH adjustments are performed separately by event and model family. The 2013 examiner broad-CPC-scope count interaction is marked unstable because of numerical separation and is excluded from the count multiplicity family. Persistent green membership is defined only for 2020. Standard errors in the underlying models are clustered by matched pair. $^{***}p<.01$, $^{**}p<.05$, $^{*}p<.10$.
\end{TableNotes}
\begin{longtable}{l@{\hspace{10pt}}c@{\hspace{14pt}}c@{\hspace{18pt}}c@{\hspace{14pt}}c}
\caption{Moderator estimates for category revisions}\label{tab:moderators-bh}\\
\toprule
\multicolumn{1}{l}{} & \multicolumn{2}{c}{Timing (RoHR)} & \multicolumn{2}{c}{Count (RoIRR)} \\\cmidrule(lr){2-3}\cmidrule(lr){4-5}
Moderator & Examiner & Applicant & Examiner & Applicant \\
\midrule
\endfirsthead
\multicolumn{5}{c}{\tablename\ \thetable\ -- continued} \\
\toprule
\multicolumn{1}{l}{} & \multicolumn{2}{c}{Timing (RoHR)} & \multicolumn{2}{c}{Count (RoIRR)} \\\cmidrule(lr){2-3}\cmidrule(lr){4-5}
Moderator & Examiner & Applicant & Examiner & Applicant \\
\midrule
\endhead
\midrule
\multicolumn{5}{r}{\footnotesize Continued on next page} \\
\endfoot
\bottomrule
\insertTableNotes
\endlastfoot
\addlinespace\multicolumn{5}{l}{\textit{Creation}} \\[2pt]
Small/micro entity & \shortstack{1.25$^{*}$ [0.96, 1.63]\\{\scriptsize $p=.092;\ q=.405$}} & \shortstack{1.01 [0.80, 1.29]\\{\scriptsize $p=.904;\ q=.904$}} & \shortstack{1.05 [0.86, 1.29]\\{\scriptsize $p=.607;\ q=.815$}} & \shortstack{0.95 [0.76, 1.18]\\{\scriptsize $p=.630;\ q=.815$}} \\
U.S. assignee & \shortstack{1.20$^{*}$ [0.97, 1.47]\\{\scriptsize $p=.095;\ q=.405$}} & \shortstack{1.03 [0.85, 1.24]\\{\scriptsize $p=.796;\ q=.898$}} & \shortstack{1.19$^{**}$ [1.01, 1.41]\\{\scriptsize $p=.040;\ q=.146$}} & \shortstack{0.85 [0.71, 1.03]\\{\scriptsize $p=.102;\ q=.281$}} \\
Above avg. maturity & \shortstack{0.87 [0.70, 1.07]\\{\scriptsize $p=.190;\ q=.569$}} & \shortstack{0.95 [0.78, 1.16]\\{\scriptsize $p=.624;\ q=.898$}} & \shortstack{1.04 [0.87, 1.24]\\{\scriptsize $p=.667;\ q=.815$}} & \shortstack{0.88 [0.71, 1.09]\\{\scriptsize $p=.245;\ q=.539$}} \\
Broad CPC scope & \shortstack{2.67 [0.29, 24.67]\\{\scriptsize $p=.388;\ q=.841$}} & \shortstack{0.89 [0.32, 2.46]\\{\scriptsize $p=.823;\ q=.898$}} & \shortstack{\textit{unstable}\\{\scriptsize excluded from BH}} & \shortstack{1.08 [0.31, 3.78]\\{\scriptsize $p=.905;\ q=.905$}} \\
Has child & \shortstack{1.21 [0.96, 1.51]\\{\scriptsize $p=.101;\ q=.405$}} & \shortstack{1.09 [0.88, 1.35]\\{\scriptsize $p=.420;\ q=.841$}} & \shortstack{1.26$^{***}$ [1.06, 1.50]\\{\scriptsize $p=.008;\ q=.088$}} & \shortstack{0.80$^{**}$ [0.66, 0.97]\\{\scriptsize $p=.026;\ q=.144$}} \\
Continuation & \shortstack{1.06 [0.74, 1.52]\\{\scriptsize $p=.756;\ q=.898$}} & \shortstack{1.12 [0.80, 1.55]\\{\scriptsize $p=.509;\ q=.873$}} & \shortstack{0.98 [0.74, 1.29]\\{\scriptsize $p=.870;\ q=.905$}} & \shortstack{1.10 [0.81, 1.50]\\{\scriptsize $p=.528;\ q=.815$}} \\
\addlinespace\multicolumn{5}{l}{\textit{Addition}} \\[2pt]
Small/micro entity & \shortstack{0.92 [0.65, 1.32]\\{\scriptsize $p=.661;\ q=.848$}} & \shortstack{1.34$^{**}$ [1.02, 1.77]\\{\scriptsize $p=.035;\ q=.396$}} & \shortstack{0.95 [0.71, 1.26]\\{\scriptsize $p=.712;\ q=.869$}} & \shortstack{1.09 [0.81, 1.46]\\{\scriptsize $p=.572;\ q=.869$}} \\
U.S. assignee & \shortstack{1.30$^{*}$ [0.99, 1.71]\\{\scriptsize $p=.058;\ q=.524$}} & \shortstack{1.18 [0.93, 1.50]\\{\scriptsize $p=.178;\ q=.524$}} & \shortstack{1.23$^{*}$ [0.98, 1.53]\\{\scriptsize $p=.071;\ q=.248$}} & \shortstack{1.10 [0.87, 1.38]\\{\scriptsize $p=.432;\ q=.851$}} \\
Above avg. maturity & \shortstack{1.00 [0.73, 1.37]\\{\scriptsize $p=.992;\ q=.999$}} & \shortstack{0.93 [0.72, 1.19]\\{\scriptsize $p=.560;\ q=.848$}} & \shortstack{0.84 [0.65, 1.08]\\{\scriptsize $p=.169;\ q=.483$}} & \shortstack{1.04 [0.83, 1.30]\\{\scriptsize $p=.744;\ q=.869$}} \\
Broad CPC scope & \shortstack{0.77$^{*}$ [0.56, 1.05]\\{\scriptsize $p=.099;\ q=.524$}} & \shortstack{0.91 [0.70, 1.20]\\{\scriptsize $p=.514;\ q=.847$}} & \shortstack{0.79$^{*}$ [0.62, 1.02]\\{\scriptsize $p=.071;\ q=.248$}} & \shortstack{1.16 [0.83, 1.61]\\{\scriptsize $p=.396;\ q=.821$}} \\
Has child & \shortstack{1.27$^{*}$ [0.96, 1.69]\\{\scriptsize $p=.093;\ q=.524$}} & \shortstack{1.07 [0.84, 1.36]\\{\scriptsize $p=.602;\ q=.848$}} & \shortstack{1.04 [0.84, 1.30]\\{\scriptsize $p=.709;\ q=.869$}} & \shortstack{1.27$^{**}$ [1.01, 1.59]\\{\scriptsize $p=.038;\ q=.221$}} \\
Continuation & \shortstack{0.70$^{**}$ [0.50, 0.97]\\{\scriptsize $p=.032;\ q=.396$}} & \shortstack{0.89 [0.67, 1.17]\\{\scriptsize $p=.402;\ q=.726$}} & \shortstack{0.90 [0.66, 1.21]\\{\scriptsize $p=.476;\ q=.869$}} & \shortstack{1.04 [0.82, 1.32]\\{\scriptsize $p=.736;\ q=.869$}} \\
Persistent green & \shortstack{0.62$^{**}$ [0.40, 0.95]\\{\scriptsize $p=.027;\ q=.396$}} & \shortstack{1.24 [0.89, 1.72]\\{\scriptsize $p=.208;\ q=.524$}} & \shortstack{0.73$^{*}$ [0.53, 1.01]\\{\scriptsize $p=.054;\ q=.231$}} & \shortstack{0.93 [0.70, 1.25]\\{\scriptsize $p=.646;\ q=.869$}} \\
\addlinespace\multicolumn{5}{l}{\textit{Strict addition}} \\[2pt]
Small/micro entity & \shortstack{1.34 [0.90, 1.99]\\{\scriptsize $p=.144;\ q=.524$}} & \shortstack{1.14 [0.82, 1.57]\\{\scriptsize $p=.441;\ q=.772$}} & \shortstack{0.91 [0.66, 1.25]\\{\scriptsize $p=.558;\ q=.869$}} & \shortstack{1.04 [0.75, 1.45]\\{\scriptsize $p=.811;\ q=.873$}} \\
U.S. assignee & \shortstack{1.18 [0.88, 1.60]\\{\scriptsize $p=.265;\ q=.549$}} & \shortstack{1.29$^{*}$ [0.98, 1.69]\\{\scriptsize $p=.072;\ q=.524$}} & \shortstack{1.07 [0.84, 1.37]\\{\scriptsize $p=.568;\ q=.869$}} & \shortstack{1.04 [0.81, 1.34]\\{\scriptsize $p=.768;\ q=.873$}} \\
Above avg. maturity & \shortstack{1.00 [0.73, 1.37]\\{\scriptsize $p=.999;\ q=.999$}} & \shortstack{0.81 [0.61, 1.08]\\{\scriptsize $p=.146;\ q=.524$}} & \shortstack{0.75$^{**}$ [0.57, 0.99]\\{\scriptsize $p=.039;\ q=.221$}} & \shortstack{1.19 [0.94, 1.50]\\{\scriptsize $p=.145;\ q=.450$}} \\
Broad CPC scope & \shortstack{0.77 [0.55, 1.08]\\{\scriptsize $p=.129;\ q=.524$}} & \shortstack{0.92 [0.68, 1.24]\\{\scriptsize $p=.569;\ q=.848$}} & \shortstack{0.69$^{***}$ [0.51, 0.91]\\{\scriptsize $p=.010;\ q=.138$}} & \shortstack{0.94 [0.66, 1.33]\\{\scriptsize $p=.726;\ q=.869$}} \\
Has child & \shortstack{1.05 [0.77, 1.43]\\{\scriptsize $p=.742;\ q=.848$}} & \shortstack{1.02 [0.77, 1.35]\\{\scriptsize $p=.900;\ q=.988$}} & \shortstack{0.97 [0.76, 1.25]\\{\scriptsize $p=.827;\ q=.873$}} & \shortstack{1.14 [0.91, 1.43]\\{\scriptsize $p=.247;\ q=.629$}} \\
Continuation & \shortstack{0.82 [0.57, 1.18]\\{\scriptsize $p=.280;\ q=.561$}} & \shortstack{0.93 [0.67, 1.29]\\{\scriptsize $p=.665;\ q=.848$}} & \shortstack{0.79 [0.57, 1.11]\\{\scriptsize $p=.173;\ q=.483$}} & \shortstack{1.21 [0.94, 1.55]\\{\scriptsize $p=.130;\ q=.430$}} \\
Persistent green & \shortstack{0.71 [0.44, 1.17]\\{\scriptsize $p=.181;\ q=.524$}} & \shortstack{1.27 [0.85, 1.90]\\{\scriptsize $p=.246;\ q=.549$}} & \shortstack{0.69$^{**}$ [0.48, 1.00]\\{\scriptsize $p=.047;\ q=.221$}} & \shortstack{1.01 [0.71, 1.44]\\{\scriptsize $p=.954;\ q=.954$}} \\
\addlinespace\multicolumn{5}{l}{\textit{Replacement}} \\[2pt]
Small/micro entity & \shortstack{1.41 [0.71, 2.81]\\{\scriptsize $p=.329;\ q=.635$}} & \shortstack{1.07 [0.55, 2.12]\\{\scriptsize $p=.836;\ q=.936$}} & \shortstack{0.96 [0.55, 1.70]\\{\scriptsize $p=.896;\ q=.929$}} & \shortstack{0.97 [0.50, 1.89]\\{\scriptsize $p=.925;\ q=.942$}} \\
U.S. assignee & \shortstack{0.69 [0.40, 1.18]\\{\scriptsize $p=.170;\ q=.524$}} & \shortstack{1.01 [0.58, 1.75]\\{\scriptsize $p=.974;\ q=.999$}} & \shortstack{0.50$^{***}$ [0.31, 0.81]\\{\scriptsize $p=.005;\ q=.094$}} & \shortstack{0.83 [0.49, 1.40]\\{\scriptsize $p=.485;\ q=.869$}} \\
Above avg. maturity & \shortstack{1.59 [0.89, 2.85]\\{\scriptsize $p=.120;\ q=.524$}} & \shortstack{0.64 [0.36, 1.16]\\{\scriptsize $p=.142;\ q=.524$}} & \shortstack{1.12 [0.65, 1.95]\\{\scriptsize $p=.681;\ q=.869$}} & \shortstack{0.54$^{**}$ [0.30, 0.99]\\{\scriptsize $p=.047;\ q=.221$}} \\
Broad CPC scope & \shortstack{1.33 [0.71, 2.47]\\{\scriptsize $p=.373;\ q=.695$}} & \shortstack{0.87 [0.49, 1.58]\\{\scriptsize $p=.656;\ q=.848$}} & \shortstack{1.34 [0.80, 2.25]\\{\scriptsize $p=.262;\ q=.638$}} & \shortstack{0.49$^{**}$ [0.26, 0.93]\\{\scriptsize $p=.028;\ q=.221$}} \\
Has child & \shortstack{1.17 [0.66, 2.08]\\{\scriptsize $p=.592;\ q=.848$}} & \shortstack{0.70 [0.40, 1.23]\\{\scriptsize $p=.215;\ q=.524$}} & \shortstack{0.79 [0.49, 1.26]\\{\scriptsize $p=.314;\ q=.676$}} & \shortstack{0.90 [0.53, 1.52]\\{\scriptsize $p=.683;\ q=.869$}} \\
Continuation & \shortstack{1.02 [0.52, 2.01]\\{\scriptsize $p=.945;\ q=.999$}} & \shortstack{1.64 [0.83, 3.25]\\{\scriptsize $p=.156;\ q=.524$}} & \shortstack{1.15 [0.65, 2.07]\\{\scriptsize $p=.628;\ q=.869$}} & \shortstack{0.62$^{*}$ [0.37, 1.03]\\{\scriptsize $p=.063;\ q=.248$}} \\
Persistent green & \shortstack{0.80 [0.43, 1.49]\\{\scriptsize $p=.485;\ q=.823$}} & \shortstack{1.16 [0.62, 2.17]\\{\scriptsize $p=.635;\ q=.848$}} & \shortstack{1.36 [0.84, 2.21]\\{\scriptsize $p=.217;\ q=.579$}} & \shortstack{0.88 [0.51, 1.51]\\{\scriptsize $p=.639;\ q=.869$}} \\
\addlinespace\multicolumn{5}{l}{\textit{Cleanup}} \\[2pt]
Small/micro entity & \shortstack{1.02 [0.90, 1.16]\\{\scriptsize $p=.727;\ q=.848$}} & \shortstack{1.09 [0.95, 1.24]\\{\scriptsize $p=.207;\ q=.524$}} & \shortstack{0.97 [0.87, 1.07]\\{\scriptsize $p=.510;\ q=.869$}} & \shortstack{1.07 [0.94, 1.22]\\{\scriptsize $p=.308;\ q=.676$}} \\
U.S. assignee & \shortstack{1.07 [0.97, 1.19]\\{\scriptsize $p=.194;\ q=.524$}} & \shortstack{1.03 [0.92, 1.15]\\{\scriptsize $p=.598;\ q=.848$}} & \shortstack{0.98 [0.90, 1.07]\\{\scriptsize $p=.625;\ q=.869$}} & \shortstack{0.98 [0.87, 1.11]\\{\scriptsize $p=.810;\ q=.873$}} \\
Above avg. maturity & \shortstack{0.98 [0.87, 1.10]\\{\scriptsize $p=.708;\ q=.848$}} & \shortstack{0.91 [0.81, 1.03]\\{\scriptsize $p=.135;\ q=.524$}} & \shortstack{0.91$^{**}$ [0.83, 1.00]\\{\scriptsize $p=.045;\ q=.221$}} & \shortstack{1.04 [0.92, 1.18]\\{\scriptsize $p=.515;\ q=.869$}} \\
Broad CPC scope & \shortstack{0.97 [0.86, 1.10]\\{\scriptsize $p=.676;\ q=.848$}} & \shortstack{0.98 [0.86, 1.11]\\{\scriptsize $p=.730;\ q=.848$}} & \shortstack{0.85$^{***}$ [0.76, 0.94]\\{\scriptsize $p=.001;\ q=.038$}} & \shortstack{1.03 [0.88, 1.20]\\{\scriptsize $p=.745;\ q=.869$}} \\
Has child & \shortstack{1.02 [0.91, 1.15]\\{\scriptsize $p=.697;\ q=.848$}} & \shortstack{0.93 [0.83, 1.05]\\{\scriptsize $p=.257;\ q=.549$}} & \shortstack{0.95 [0.87, 1.05]\\{\scriptsize $p=.312;\ q=.676$}} & \shortstack{0.86$^{**}$ [0.75, 0.97]\\{\scriptsize $p=.016;\ q=.177$}} \\
Continuation & \shortstack{1.00 [0.85, 1.16]\\{\scriptsize $p=.957;\ q=.999$}} & \shortstack{0.92 [0.79, 1.06]\\{\scriptsize $p=.243;\ q=.549$}} & \shortstack{0.88$^{**}$ [0.77, 0.99]\\{\scriptsize $p=.041;\ q=.221$}} & \shortstack{1.01 [0.89, 1.16]\\{\scriptsize $p=.826;\ q=.873$}} \\
Persistent green & \shortstack{0.82$^{***}$ [0.72, 0.94]\\{\scriptsize $p=.004;\ q=.238$}} & \shortstack{1.17$^{**}$ [1.02, 1.34]\\{\scriptsize $p=.028;\ q=.396$}} & \shortstack{0.96 [0.85, 1.07]\\{\scriptsize $p=.440;\ q=.851$}} & \shortstack{1.50$^{***}$ [1.28, 1.75]\\{\scriptsize $p=<.001;\ q=<.001$}} \\
\end{longtable}
\end{ThreePartTable}
\normalsize

}{\placeholder{Run the R script to generate the joint moderator table.}}

\FloatBarrier
\subsection{Breadth of subsequent examiner search}
\label{app:breadth}
Table~\ref{tab:breadth-deep} shows the results for the same matched samples as used in our main results. Absolute reach counts distinct citing examiners, AU4s, and TC2s and is estimated by fixed-effect Poisson DiD.

\IfFileExists{generated/tables/table_breadth_deep.tex}{
  \scriptsize
\setlength{\tabcolsep}{3pt}
\begin{ThreePartTable}
\begin{TableNotes}[flushleft]\footnotesize
\item[]
The sample and matching design correspond to the baseline design for each event and treatment branch. Counts of unique citing examiners, AU4s, and TC2s are estimated using matched-pair Poisson DiD models and reported as percentage changes (\%). Cross-AU4 and cross-TC2 citation shares are estimated using linear DiD models and reported as percentage-point changes (pp). Share models condition on observations for which the corresponding citation share is defined; their sample sizes are therefore smaller. Standard errors are clustered by matched pair. SEs for exponentiated Poisson effects use the delta method; linear-model SEs are multiplied by 100 to obtain percentage points. $^{***}p<.01$, $^{**}p<.05$, $^{*}p<.10$.
\end{TableNotes}
\begin{longtable}{llcrrrr}
\caption{Category revision effects on the breadth of examiner citations}\label{tab:breadth-deep}\\
\toprule
\shortstack{Breadth\\outcome} & Estimator & Unit & Effect & SE & $p$ & $N$ \\
\midrule
\endfirsthead
\multicolumn{7}{c}{\tablename\ \thetable\ -- continued} \\
\toprule
\shortstack{Breadth\\outcome} & Estimator & Unit & Effect & SE & $p$ & $N$ \\
\midrule
\endhead
\midrule
\multicolumn{7}{r}{\footnotesize Continued on next page} \\
\endfoot
\bottomrule
\insertTableNotes
\endlastfoot
\addlinespace\multicolumn{7}{l}{\textit{Creation}} \\[2pt]
Unique citing examiners & Poisson DiD & \% & $6.6$ & 4.2 & $.101$ & 15,068 \\
Unique citing AU4s & Poisson DiD & \% & $8.3^{**}$ & 4.1 & $.035$ & 15,068 \\
Unique citing TC2s & Poisson DiD & \% & $6.8^{*}$ & 3.9 & $.071$ & 15,060 \\
Cross-AU4 citation share & Linear DiD & pp & $0.0$ & 1.9 & $.987$ & 2,906 \\
Cross-TC2 citation share & Linear DiD & pp & $0.9$ & 2.5 & $.730$ & 2,892 \\
\addlinespace\multicolumn{7}{l}{\textit{Addition}} \\[2pt]
Unique citing examiners & Poisson DiD & \% & $-8.8^{*}$ & 4.4 & $.056$ & 11,004 \\
Unique citing AU4s & Poisson DiD & \% & $-8.4^{*}$ & 4.4 & $.070$ & 11,004 \\
Unique citing TC2s & Poisson DiD & \% & $-4.4$ & 4.4 & $.327$ & 11,000 \\
Cross-AU4 citation share & Linear DiD & pp & $-3.8$ & 2.3 & $.106$ & 1,204 \\
Cross-TC2 citation share & Linear DiD & pp & $4.1$ & 3.8 & $.278$ & 1,204 \\
\addlinespace\multicolumn{7}{l}{\textit{Strict addition}} \\[2pt]
Unique citing examiners & Poisson DiD & \% & $-0.2$ & 5.5 & $.975$ & 8,948 \\
Unique citing AU4s & Poisson DiD & \% & $-0.3$ & 5.5 & $.953$ & 8,948 \\
Unique citing TC2s & Poisson DiD & \% & $2.7$ & 5.4 & $.606$ & 8,948 \\
Cross-AU4 citation share & Linear DiD & pp & $0.5$ & 2.6 & $.853$ & 1,008 \\
Cross-TC2 citation share & Linear DiD & pp & $4.6$ & 4.1 & $.258$ & 1,008 \\
\addlinespace\multicolumn{7}{l}{\textit{Replacement}} \\[2pt]
Unique citing examiners & Poisson DiD & \% & $-20.4^{**}$ & 8.4 & $.032$ & 2,784 \\
Unique citing AU4s & Poisson DiD & \% & $-20.6^{**}$ & 8.3 & $.027$ & 2,784 \\
Unique citing TC2s & Poisson DiD & \% & $-19.0^{**}$ & 8.0 & $.032$ & 2,784 \\
Cross-AU4 citation share & Linear DiD & pp & $2.2$ & 7.6 & $.776$ & 256 \\
Cross-TC2 citation share & Linear DiD & pp & $20.7^{**}$ & 8.5 & $.016$ & 256 \\
\addlinespace\multicolumn{7}{l}{\textit{Cleanup}} \\[2pt]
Unique citing examiners & Poisson DiD & \% & $12.2^{***}$ & 2.1 & $<.001$ & 61,728 \\
Unique citing AU4s & Poisson DiD & \% & $11.0^{***}$ & 2.0 & $<.001$ & 61,728 \\
Unique citing TC2s & Poisson DiD & \% & $10.6^{***}$ & 2.0 & $<.001$ & 61,720 \\
Cross-AU4 citation share & Linear DiD & pp & $-0.6$ & 1.3 & $.660$ & 8,604 \\
Cross-TC2 citation share & Linear DiD & pp & $0.3$ & 1.3 & $.841$ & 8,602 \\
\end{longtable}
\end{ThreePartTable}
\normalsize

}{\placeholder{Run the R script to generate the citation-breadth table.}}

\clearpage
\putbib
\end{bibunit}

\end{document}